\documentclass[a4paper,12pt]{article}

\usepackage{amsmath}
\usepackage{amssymb}
\usepackage{color}
\usepackage{mathrsfs}

\usepackage[dvipdfmx]{graphicx}
\usepackage{graphicx}

\usepackage{tikz}

\DeclareMathOperator{\Ai}{Ai}
\DeclareMathOperator{\Det}{Det}

\DeclareMathOperator{\Res}{Res}
\DeclareMathOperator{\sgn}{sgn}
\DeclareMathOperator{\Tr}{Tr}

\newcommand{\llangle}{\langle\!\langle}
\newcommand{\rrangle}{\rangle\!\rangle}

\newcommand{\eq} {equation}
\newcommand{\eqa} {eqnarray}
\newcommand{\NN} {\nonumber}

\usepackage{geometry}
\numberwithin{equation}{section}
\allowdisplaybreaks

\begin{document}

\newcommand{\aff}[1]{${}^{#1}$}
\renewcommand{\thefootnote}{\fnsymbol{footnote}}

\begin{titlepage}
\begin{flushright}
{\footnotesize KIAS-P18012}\\
{\footnotesize YITP-18-76}
\end{flushright}
\begin{center}
{\Large\bf
Supersymmetry Breaking in a Large $N$ Gauge Theory\\\vspace{0.5em}
with Gravity Dual
}\\
\bigskip\bigskip
{\large Masazumi Honda,\footnote{\tt masazumi.honda@weizmann.ac.il}}\aff{1}
{\large Tomoki Nosaka,\footnote{\tt nosaka@yukawa.kyoto-u.ac.jp}}\aff{2}
{\large Kazuma Shimizu\footnote{\tt kazuma.shimizu@yukawa.kyoto-u.ac.jp}}\aff{3}\\
\bigskip
{\large and Seiji Terashima\footnote{\tt terasima@yukawa.kyoto-u.ac.jp}}\aff{3}\\
\bigskip\bigskip
\aff{1}: {\small
\it Department of Particle Physics and Astrophysics, Weizmann Institute of Science\\
Rehovot 7610001, Israel
}\\
\bigskip
\aff{2}: {\small
\it School of Physics, Korea Institute for Advanced Study\\
85 Hoegiro Dongdaemun-gu, Seoul 02455, Republic of Korea
}\\
\bigskip
\aff{3}: {\small
\it Yukawa Institute for Theoretical Physics, Kyoto University\\
Kitashirakawa-Oiwakecho, Sakyo, Kyoto 606-8502, Japan
}
\end{center}

\begin{abstract}
We study phase structure of mass-deformed ABJM theory which is a three dimensional $\mathcal{N}=6$ superconformal theory deformed by mass parameters and has the gauge group $\text{U}(N)\times \text{U}(N)$ with Chern-Simons levels $(k,-k)$ which may have a gravity dual.
We discuss that the mass deformed ABJM theory on $S^3$ breaks supersymmetry in a large-$N$ limit if the mass is larger than a critical value.
To see some evidence for this conjecture, we compute the partition function exactly, and numerically by using the Monte Carlo Simulation for small $N$.
We discover that the partition function has zeroes as a function of the mass deformation parameters if $N\ge k$, which supports the large-$N$ supersymmetry breaking.
We also find a solution to the large-$N$ saddle point equations, where the free energy is consistent with the finite $N$ result.
\end{abstract}

\end{titlepage}

\renewcommand{\thefootnote}{\dag\arabic{footnote}}
\setcounter{footnote}{0}

\tableofcontents

\section{Introduction}
Spontaneous supersymmetry (SUSY) breaking in string/M-theory is one of the most important subjects and has been discussed intensively in the context of phenomenology and cosmology.
The SUSY breaking in string/M-theory is the super-Higgs phenomenon in general since the theory has gauged SUSY rather than global one.
In contrast, various situations in string/M-theory are expected to have holographic descriptions by SUSY quantum field theories (QFT), whose SUSY are global.
Therefore it is interesting to discuss SUSY breaking in QFT with a gravity dual which is typically large-$N$ and strongly coupled.
As far as we know, the only explicit examples of such problem are the models discussed in \cite{KS1,KS2,KS3,EKSS} in which the SUSY is kinematical, i.e. the SUSY algebra does not includes the Hamiltonian.\footnote{
See \cite{MN,Massai} for related works in gravity side.
}
Main reason for the existence of the very few examples is that it is technically hard since we typically need non-perturbative analysis in this type of problem.

In this paper we study the SUSY breaking problem in so-called massive ABJM theory \cite{HLLLP,GRVV} which is a three dimensional $\mathcal{N}=6$ superconformal theory known as ABJM theory \cite{ABJM} deformed by two mass parameters and has the gauge group $\text{U}(N)\times \text{U}(N)$ with Chern-Simons levels $(k,-k)$.
It is expected that the ABJM theory without the masses is the low-energy effective theory of $N$ coincident M2-branes and dual to the M-theory on $AdS_4 \times S^7 /\mathbb{Z}_k$.
The mass deformation in this situation corresponds to the introduction of the background flux.
Then the massive ABJM theory in the large-$N$ limit is holographically dual to M-theory on asymptotic $AdS_4$ geometry \cite{Lin:2004nb,CKK}.

We argue that the mass deformed ABJM theory on $S^3$ breaks supersymmetry in the large-$N$ limit with $k$ fixed if the mass is larger than a critical value.
We can adress this because the partition function of the theory on $S^3$ can be exactly computed by the localization technique \cite{KWY,J,HHL}.
Note that the localization technique can be applied to our theory with finite $N$ on $S^3$, like the Witten index on $S^1 \times M$ where $M$ is a compact manifold, even if it will break the supersymmetry spontaneously in the large-$N$ or large volume limit.

The arguments are based on the existences of the zeroes of the partition function which will be related to the SUSY breaking and the phase transition at the critical mass which is expected from the large-$N$ saddle point solution found in \cite{NST2}.
A summary of our arguments for the SUSY breaking of the theory will be explained in sec.~\ref{susybreaking}.

In the previous work \cite{NST2} a part of the authors considered this theory with the two equal mass parameters, which enjoys the ${\cal N}=6$ supersymmetry.
We studied the partition function in the M-theory limit ($N\rightarrow\infty$ with $k$ kept finite) by using the saddle point approximation, and found that the saddle point solution which gives the free energy $F\sim N^{3/2}$ disappears as we increase the mass deformation parameter to some critical value.
Although this would suggest that a phase transition occurs at that point, the whole phase structure is still unclear.
\begin{figure}[t]
\begin{center}
\includegraphics[width=7.5cm]{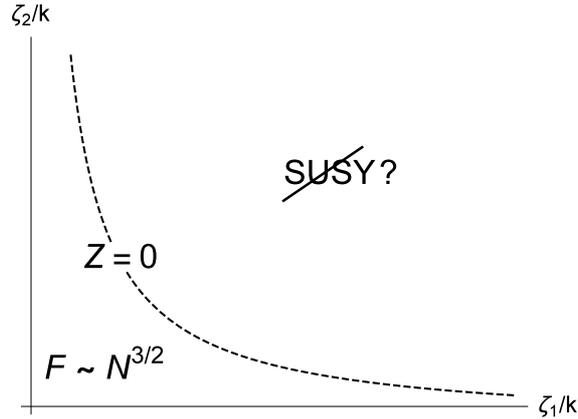}
\end{center}
\caption{
Our proposal on phase structure of the massive ABJM theory on $S^3$ in the large-$N$ limit.
The dashed line denotes expected zeroes of the sphere partition function: $\frac{\zeta_1 \zeta_2}{k^2}=\frac{1}{16}$.
}
\label{fig:proposal}
\end{figure}

The rest of this paper is organized as follows.
After introducing the mass deformed ABJM theory with two mass parameters $\zeta_1$, $\zeta_2$ in the next section, in sec.~\ref{susybreaking} we argue that this theory with $\zeta_1=\zeta_2$ breaks the supersymmetry for $\zeta_1/k>1/4$.
In the same section we also argue that the supersymmetry breaking does not occur if $\zeta_1=0$ or $\zeta_2=0$.
In sec.~\ref{zeta10} we first study the latter cases with $\zeta_1=0$ in detail, and indeed find the large-$N$ free energy obeys $N^{3/2}$-law for any value of $\zeta_2$.
Then, in sec.~\ref{bothnonzero} we consider the case with both $\zeta_1$ and $\zeta_2$ are non-zero.
In this case the partition function $Z(k,N,\zeta_1,\zeta_2)$ for $N\ge k$ can have zeroes at some finite $\zeta_1,\zeta_2$, this was explicitly shown for $N=2$.
We provide positive evidence for the existence of zeroes from the Monte Carlo simulation.
We also argue the physical interpretation for the zeroes, and estimate how the partition function behaves in the large-$N$ limit.
Our proposal on the phase structure in the large-$N$ limit is summarized in fig.~\ref{fig:proposal}.
We expect that the partition function vanishes when $\frac{\zeta_1 \zeta_2}{k^2}=\frac{1}{16}$ and the theory is in the SUSY breaking phase for $\frac{\zeta_1 \zeta_2}{k^2}\geq \frac{1}{16}$.
In sec.~\ref{discuss} we summarize our analysis and propose future directions.

We summarize technical details in appendices.
In app.~\ref{Sdualsection}, starting from the localization formula \eqref{Original} we rewrite the partition function to a simpler matrix model \eqref{Sdual}, which we use in the subsequent sections.
In particular, in the new expression the integration is absolutely convergent, hence we can evaluate the partition function numerically by applying the Monte Carlo method.
In app.~\ref{app:detail_exact} we display the exact computation of the partition function with $\zeta_1=0$ with $k$ and $N$ being small integers, and summarize the results in app.~\ref{Z1exact_results}.
As $N$ increases these results match with the saddle point approximation in sec.~\ref{saddle}, which support the validity of the saddle point approximation.
We also compare the exact partition function with the partition function of the linear quiver theory with single hypermultiplet obtained in \cite{NY1}.

\section{Review on Mass deformed ABJM theory}
In this section we review some basic facts on the mass-deformed ABJM theory on $S^3$.
The field content of the ABJM theory consists of, in the 3d ${\cal N}=2$ SUSY notation, a $\text{U}(N)_k$ vector multiplet ${\cal V}=(A_\mu,\sigma,\chi,D)$, a $\text{U}(N)_{-k}$ vector multiplet ${\widetilde {\cal V}}=({\widetilde A}_\mu,{\widetilde\sigma},{\widetilde\chi},{\widetilde D})$, two chiral multiplets ${\cal Z}_\alpha=(A_\alpha,\phi_\alpha,F_\alpha)$ in $(\Box,\bar{\Box})$ representation under $\text{U}(N)_k\times \text{U}(N)_{-k}$ and two chiral multiplets ${\cal W}_{\dot{\alpha}}=(B_{\dot{\alpha}},\psi_{\dot{\alpha}},G_{\dot{\alpha}})$ in $(\bar{\Box},\Box)$ representation.\footnote{
The (anti-)bi-fundamental chiral multiplets have $\text{U}(1)_\text{R}$ charges $1/2$.
}
Here the vector multiplets obey the Chen-Simons action with level $\pm k$, while the action for the chiral multiplets consists of the superpotential together with the following minimal coupling to the vector multiplets
\begin{align}
S_\text{kin}
&=\int d^3x\sqrt{g}\Tr\Bigl[|D_\mu A_a|^2+|D_\mu W_{\dot{a}}|^2
+\frac{3}{4r_{S^3}^2}(|A_a|^2+|W_{\dot{a}}|^2)\nonumber \\
&\quad +\frac{1}{r_{S^3}}|\sigma A_a-A_a{\widetilde\sigma}|^2
+i(\bar{A}^aDA_a-A_a{\widetilde D}\bar{A}^a)\nonumber \\
&\quad +\frac{1}{r_{S^3}} |{\widetilde \sigma}B_{\dot{a}} -B_{\dot{a}}\sigma |^2
+i(\bar{B}^{\dot{a}}{\widetilde D}B_{\dot{a}}-B_{\dot{a}}D\bar{B}^{\dot{a}})
\Bigr]+(fermions) .
\label{minimalcoupling}
\end{align}

We can introduce a mass by turning on a background vector multiplet ${\cal V}^{\text{(bgd)}}=(A_\mu^{\text{(bgd)}},\sigma^{\text{(bgd)}}$, $\chi^{\text{(bgd)}},D^{\text{(bgd)}})$ of a global symmetry in the following supersymmetric configuration\footnote{
This type of mass is usually called real mass.
We can also add ``complex mass'' by adding quadratic terms in superpotential but it is known that $S^3$ partition function of general 3d $\mathcal{N}=2$ theory is independent of complex mass. 
}
\cite{FP}
\begin{align}
A_\mu^{\text{(bgd)}}=0,\quad
\sigma^{\text{(bgd)}}=\delta,\quad
\chi^{\text{(bgd)}}=0,\quad
D^{\text{(bgd)}}=-\delta.
\label{background}
\end{align}
where we have set the radius of $S^3$ to $r_{S^3}=1$. 
Here we turn on the background multiplets of the flavor symmetries $\text{U}(1)_1\times\text{U}(1)_2\times\text{U}(1)_3$ commuting with the ${\cal N}=2$ supersymmetry under which the chiral multiplets are charged\footnote{
These charges are denoted as $h_4$, $h_1$, $h_2$ in \cite{Kim}, respectively.
These $\text{U}(1)$ symmetries are a part of non-Abelian R-symmetry in higher SUSY language.
}
as in table~\ref{tab:charges}.
The background gauge fields also minimally couples to the chiral multiplets in the same way as \eqref{minimalcoupling}, hence it modifies the action as
\begin{align}
S
&\rightarrow S+\int \sqrt{g}\Tr\Biggl[
\Bigl(\frac{\delta^1}{2}+\frac{\delta^2}{2}+\frac{\delta^3}{2}\Bigr)
(-i|A_1|^2
-2(\bar{A}^1\sigma A_1-\bar{A}^1A_1{\widetilde\sigma})
)
+\Bigl(\frac{\delta^1}{2}+\frac{\delta^2}{2}+\frac{\delta^3}{2}\Bigr)^2|A_1|^2\nonumber \\
&\quad+\Bigl(\frac{\delta^1}{2}-\frac{\delta^2}{2}-\frac{\delta^3}{2}\Bigr)
(-i|A_2|^2
-2(\bar{A}^2\sigma A_2-\bar{A}^2A_2{\widetilde\sigma})
)
+\Bigl(\frac{\delta^1}{2}-\frac{\delta^2}{2}-\frac{\delta^3}{2}\Bigr)^2|A_2|^2\nonumber \\
&\quad+\Bigl(-\frac{\delta^1}{2}+\frac{\delta^2}{2}-\frac{\delta^3}{2}\Bigr)
(
-i|B_{\dot{1}}|^2
-2({\bar B}^{\dot{1}}{\widetilde\sigma}B_{\dot{1}}-{\bar B}^{\dot{1}}B_{\dot{1}}\sigma)
)
+\Bigl(-\frac{\delta^1}{2}+\frac{\delta^2}{2}-\frac{\delta^3}{2}\Bigr)^2|B_{\dot{1}}|^2\nonumber \\
&\quad+\Bigl(-\frac{\delta^1}{2}-\frac{\delta^2}{2}+\frac{\delta^3}{2}\Bigr)
(
-i|B_{\dot{2}}|^2
-2({\bar B}^{\dot{2}}{\widetilde\sigma}B_{\dot{2}}-{\bar B}^{\dot{2}}B_{\dot{2}}\sigma)
)
+\Bigl(-\frac{\delta^1}{2}-\frac{\delta^2}{2}+\frac{\delta^3}{2}\Bigr)^2|B_{\dot{2}}|^2
\Biggr].
\label{massterms}
\end{align}
Here $\delta^{i}$ $(i=1,2,3)$ are the vacuum expectation values \eqref{background} of the background vector multiplets ${\cal V}^{(\text{bgd},i)}$ for $\text{U}(1)_i$.
In this paper we choose $\delta^i$ as\footnote{
We are using the notation different from \cite{NST2}.
For $\delta^2=0$ case, the background gauge fields couple uniformly to $A_\alpha$, $\bar{B}^{\dot{\alpha}}$ in \eqref{massterms} and hence can be absorbed into the shift of 
$(\sigma,D,{\widetilde\sigma},{\widetilde D})\rightarrow(\sigma,D,{\widetilde\sigma},{\widetilde D})+(-\pi\zeta/k,\pi\zeta/k,\pi\zeta/k,-\pi\zeta/k)$ with $\zeta =(\zeta_1 +\zeta_2 )/2$.
These field redefinitions generate the Fayet-Illiopoulos terms out of the Chern-Simons term instead.
}
\begin{align}
\delta^1=\frac{2(\zeta_1+\zeta_2)}{k},\quad
\delta^2=\frac{2(\zeta_1-\zeta_2)}{k},\quad
\delta^3=0,
\end{align}
so that $\zeta_1$, $\zeta_2$ are interpreted as the mass parameters for the chiral multiplets as $m_1=2\zeta_1/k$ for ${\cal Z}_1$, ${\cal W}_{\dot{2}}$ and $m_2=2\zeta_2/k$ for ${\cal Z}_2$, ${\cal W}_{\dot{1}}$.
\begin{table}[t]
\begin{center}
\begin{tabular}{|c|c|c|c|}
\hline
             &$\text{U}(1)_1$&$\text{U}(1)_2$&$\text{U}(1)_3$\\ \hline
$\mathcal{Z}_1$ &$\frac{1}{2}$ &$\frac{1}{2}$  &$\frac{1}{2}$\\ \hline
$\mathcal{Z}_2$  &$\frac{1}{2}$ &$-\frac{1}{2}$ &$-\frac{1}{2}$\\ \hline
$\mathcal{W}_{\dot{1}}$  &$-\frac{1}{2}$  &$\frac{1}{2}$  &$-\frac{1}{2}$\\ \hline
$\mathcal{W}_{\dot{2}}$  &$-\frac{1}{2}$  &$-\frac{1}{2}$ &$\frac{1}{2}$\\ \hline
\end{tabular}
\caption{
Charges of the $\text{U}(1)_1\times\text{U}(1)_2\times\text{U}(1)_3$ flavor symmetry.
}
\label{tab:charges}
\end{center}
\end{table}

Applying the localization method  \cite{KWY,J,HHL}, the sphere partition function of the massive ABJM theory is given by the following matrix model \cite{JKPS}
\begin{align}
Z
=\frac{1}{(N!)^2}\int\frac{d^N\lambda}{(2\pi)^N}\frac{d^N{\widetilde\lambda}}{(2\pi )^N}
e^{\frac{ik}{4\pi}\sum_i(\lambda_i^2-{\widetilde\lambda}_i^2)}
\frac{\prod_{i\neq j }^N  2\sinh\frac{\lambda_i-\lambda_j}{2} 
\cdot 2\sinh\frac{{\widetilde\lambda}_i-{\widetilde\lambda}_j}{2} }
{\prod_{i,j=1}^N
 2\cosh\frac{\lambda_i-{\widetilde\lambda}_j -4\pi \zeta_1 /k}{2} 
\cdot 2\cosh\frac{\lambda_i-{\widetilde\lambda}_j -4\pi \zeta_2 /k}{2} 
} .
\label{Original}
\end{align}
In the rest of sections, we practically analyze another equivalent representation for $Z$:
\begin{align}
Z(N,k,\zeta_1,\zeta_2)=\frac{1}{N!}\int\frac{d^Nx}{(2\pi k)^N}\prod_{i=1}^N\frac{e^{\frac{2i\zeta_1}{k}x_i}}{2\cosh\frac{x_i}{2}}
\frac{
\prod_{i<j}^N(2\sinh\frac{x_i-x_j}{2k})^2
}{
\prod_{i,j=1}^N2\cosh\frac{x_i-x_j+4\pi\zeta_2}{2k}
},
\label{Sdual}
\end{align}
which we derive in app.~\ref{Sdualsection}.
For $k=1$ and $\zeta_1=\zeta_2=0$, this latter expression coincides with the partition function of the ${\cal N}=8$ $\text{U}(N)$ Yang-Mills theory coupled with a fundamental chiral multiplet, which is dual to the ABJM theory under the $\text{SL}(2,\mathbb{Z})$ transformation in the type IIB brane setup.
Because of this reason we simply refer to \eqref{Sdual} as the S-dual representation even for general $(k,\zeta_1,\zeta_2)$.

Note that the integration in the S-dual representation \eqref{Sdual} is absolutely convergent in contrast to the representation \eqref{Original} where the convergence is achieved by the rapidly oscillating factors.
Because of this fact, it is much easier to apply the Monte Carlo simulation of the partition function to \eqref{Sdual} than \eqref{Original}.
With the help of the Monte Carlo simulation of \eqref{Sdual} we will observe a novel behavior of the partition function: the partition function vanishes at some finite values of $\zeta_1,\zeta_2$, which was not encountered in the undeformed case or the case of the R-charge deformation ($\zeta_1,\zeta_2\in i\mathbb{R}$).

\section{Evidence for SUSY breaking}
\label{susybreaking}
In this section, we discuss why we expect the SUSY breaking of the mass deformed ABJM theory on $S^3$ in the large-$N$ limit at some finite $(\zeta_1 ,\zeta_2 )$ and explain our criterion for the SUSY breaking which we will examine in the following sections. 

First, in the case of $\zeta_1=\zeta_2 =\zeta$, there is a large-$N$ saddle point solution for the original matrix model \eqref{Original} which exist only for $0 \leq \frac{\zeta}{k} < \frac14$ \cite{NST2}.
This solution becomes the saddle point solution of the massless ABJM theory \cite{HKPT} in the $\zeta \rightarrow 0$ limit and gives the $N^{3/2}$-law of the free energy: 
\begin{\eq}
-\log{Z} = \frac{\pi\sqrt{2k}}{3} \left( 1+\frac{16\zeta^2}{k^2} \right) N^{3/2} .
\end{\eq}
However, this saddle point solution becomes singular in the $\frac{\zeta}{k} \rightarrow \frac14$ limit.
There is another large-$N$ solution for any value of $\zeta$.
The free energy of this solution is proportional to $N^2$ and this solution may correspond to a confinement vacuum.\footnote{
This statement is not precise because the Chern-Simons interaction remains and theory may be in a gapped phase.
Nevertheless we will call the confinement phase for such case also.
Note that here we take $k/N \rightarrow 0$ limit, thus the Chern-Simons interaction will be ignored for the leading order in the large-$N$ limit and the Yang-Mills term always induced by the renormalization flow.
We also note that the ${\cal N}=2$ SUSY pure Yang-Mills theory does not have SUSY vacua.
}
Although it would be possible that there are other solutions,\footnote{
With some numerical methods, we can not find any solution other than the two solutions.
}
these results strongly indicate a phase transition at $\frac{\zeta}{k} \rightarrow \frac14$.

We expect that this phase transition comes from SUSY breaking as follows.
Let us  take the mass very large, i.e. $\frac{\zeta}{k} \gg 1$, then, at least naively, the hypermultiplets become heavy and decouple from the vector multiplets.
The remaining ${\cal N}=2$ SUSY pure Chern-Simons theory will spontaneously break SUSY as shown in \cite{Witten, Ohta}, and becomes the confinement phase in the large-$N$ limit.
This expectation is consistent with the above large-$N$ solutions.
However, for the mass deformed ABJM theory, the SUSY index was computed to be non-zero in \cite{KK,CKK}.
In this theory, there are infinitely many discrete classical vacua which are characterized by the fuzzy $S^3$ solutions given in \cite{Terashima,GRVV}, which represent M5-branes.
Although the contribution to the index for the trivial vacuum, where all the scalar fields are zero, vanishes as in the pure SUSY Chern-Simons theory, other vacua give the non-zero contributions to the index if there are no coincident M5-branes.
This result seems to contradict with the above argument of the SUSY breaking.
However, this results is for the theory on $T^3$, not on $S^3$.
For the ${\cal N}=2$ SUSY theory on $S^3$, there are mass terms for the hypermultiplets proportional to the curvature of $S^3$.
The mass term will lift all of the vacua except the trivial vacuum at the origin classically.\footnote{
We expect that the energy of the possible metastable SUSY breaking vacuum is proportional to $\zeta$ and the free energy will be proportional to $\zeta r_{S^3}$.
The extra contribution by the curvature induced mass term to the free energy for the fuzzy sphere solutions will also proportional to $\zeta r_{S^3}$ because the size of the 
fuzzy sphere grows as $\zeta$ grows.
Of course, this is not valid except the weak coupling limit and the phase of the theory can be non-trivial.
}
Thus, the result of \cite{KK,CKK} on the SUSY index does not exclude the possibility that the mass deformed ABJM theory on $S^3$ has the SUSY breaking phase.\footnote{
Here, we assume that the theory is regarded as a deformation of the ABJM theory on $S^3$ for a small $\zeta/k$ case.
For a enough large $\zeta/k$ case, we think that the curvature effect of $S^3$ is almost negligible, but still remains.
This picture will lead the SUSY breaking scenario explained here.
}
Here note that we do not take the large volume limit.

\subsection{Criterion for SUSY breaking}
By now, we have not defined what the spontaneous SUSY breaking on $S^3$ is.
Usually, the SUSY breaking means that there is no states with zero energy in the theory.
For $S^3$, we can not define states with an appropriate Hamiltonian and time, thus it is difficult to use this definition.
Instead of this definition for the SUSY breaking, the spontaneous breaking of a symmetry $\hat{Q}$ can be defined as ${}^\exists{\cal \hat{O}}$ s.t. $\langle 0|[\hat{Q},{\cal \hat{O}}]|0\rangle \neq 0$.
In the path-integral formalism, this corresponds to
\begin{align}
Q\text{ is spontaneously broken}\ \mathop{\Longleftrightarrow}^{\text{def}}\ 
{}^\exists{\cal O}\text{ s.t. }\langle Q{\cal O}\rangle \neq 0,
\label{susybreaking180212a}
\end{align}
where the condensation is the order parameter.
Note that this correspondence is valid for the theory with enough number of non-compact space directions, in which the notion of vacuum is meaningful, otherwise, $\langle Q{\cal O} \rangle$ corresponds to $\Tr[\hat{Q},{\cal \hat{O}}]$,\footnote{
For the SUSY, it corresponds to $\Tr(-1)^{\hat{F}} \{ \hat{Q},{\cal \hat{O}} \}= \Tr[ (-1)^{\hat{F}} \hat{Q},{\cal \hat{O}}]$.
}
not to $\langle 0|[\hat{Q},{\cal O}]|0\rangle$.
Because $Q$ is a symmetry generator which behaves well, we expect that $\langle Q{\cal O}\rangle=0$ ($\Tr[Q,{\cal O}]=0$) is trivial identity due to the invariance of path integral measure (cyclic invariance of $\Tr$).\footnote{
In the case of $Q=$SUSY, $Q{\cal O}$ is such as $F$-term and $D$-term.
Unfortunately we cannot compute $\langle F\rangle$ or $\langle D\rangle$ by using the supersymmetry localization.
We can compute $\langle \int F\rangle$ and $\langle \int D\rangle$, but they are trivially zero.
This is consistent with the fact that there is no SUSY breaking for the theory on $S^3$ with $N$ finite. 
}
For example, for SUSY quantum mechanics case, the invariance of the Witten index means $\Tr[(-1)^{\hat{F}} \hat{Q},{ \hat{Q}^\dagger}] \sim \Tr (-1)^{\hat{F}} \hat{H} = 0$.
Thus, the definition of (\ref{susybreaking180212a}) is meaningful for the theory with some space with enough number of non-compact directions.
Since $S^3$ is compact, we need to take the large volume limit or large-$N$ limit which can effectively gives extra dimension.
If this happens, there should exist a massless Goldstone fermion in the theory which makes the (SUSY) partition function $Z$ vanished.

Instead of the large volume limit, we take a large-$N$ limit in which the SUSY breaking is meaningful.
Thus, we need a criterion of the SUSY breaking in the large-$N$ limit from a finite $N$ result.
For the theory in which we can define the Witten index, the vanishing of it, i.e. $Z=0$, is the necessary condition for the SUSY breaking for the finite volume.
For the other theories also, we expect that the massless Goldstone fermion makes $Z=0$.
Indeed, for a superconformal theory on $S^3$, the theory can break the SUSY if $Z=0$ because the radius of $S^3$ is not physical.
Such theories were discussed in \cite{MN,KWY3,Suyama1,Suyama2}.
For our case, the theory is not conformal, but we take the large-$N$ limit.
Thus, we regard $Z=0$ as a criterion of the SUSY breaking.\footnote{
In the gravity dual, the SUSY is gauged and the theory is described by a supergravity.
In the supergravity, there are massless fermions, however, there are no zero modes around the SUSY vacuum which is an asymptotic $AdS_4$ background.
In a SUSY breaking vacuum, some fermions near the boundary have zero modes.
}
\footnote{
From the analogy to the case with bosonic zero mode, an appropriate analysis would be to add an explicit-susy-breaking deformation to kill the zero mode and see what happens in the limit of zero deformation.
In this approach, however, we cannot use the result of the localization.
}
It is worth to note that $Z=0$ does not necessarily mean SUSY breaking as in Witten index.
However, for our case, interpreting $Z=0$ as SUSY breaking is the most natural possibility because the mass deformed ABJM theory on $S^3$ will be smoothly connected to the pure SUSY CS theory in the large mass limit whose SUSY is broken for $k\leq N$.

In the following sections, we will give further supporting arguments for the above picture of the SUSY breaking phase for the mass deformed ABJM theory on $S^3$ using the S-dual representation of the matrix model.
Here we will summarize these argument for the SUSY breaking shortly.
The S-dual representation of the matrix model (for $k=1$) is obtained from the $\text{U}(N)$ Yang-Mills theory with an adjoint and fundamental matter fields where $\zeta_1$ and $\zeta_2$ corresponds to the FI term and the mass for the adjoint matter, respectively.
Because of the FI term (and the mass term), this theory will break the SUSY at the origin of the Coulomb branch moduli space which will be favored by the mass terms induced from the curvature of $S^3$.
This picture will be right for a generic large value of $\zeta_1$ and $\zeta_2$.
However,  for $\zeta_1=0$, the FI term vanishes and the SUSY will not break.\footnote{
The mass deformed ABJM also will not break the SUSY for this case because a half of the hypermultiplets remain massless and do not decouple.
}
For this case, as we will see later, we can construct a large-$N$ solution for any value of $\zeta_2$, thus there are no critical mass for this case.
This is consistent with the above picture.

In order to investigate further, we will compute the partition function $Z$ for finite $N$ exactly and numerically using the Monte Carlo method for various points of $(\zeta_1,\zeta_2 )$.
We expect that some values of $N$ for which we computed $Z$ are not very large, but enough large for the large-$N$ expansion.
Indeed, the computed values of $Z$ are consistent with the large-$N$ solutions for $\zeta_1=\zeta_2<k/4$ and $\zeta_1=0$.
These actual computations of $Z$ for finite $N$ shows that as increasing $\zeta_i$, $Z$ is decreasing and oscillating, thus $Z=0$ for some values of $\zeta_i$.
We expect that this zero corresponds to the SUSY breaking in the large-$N$ limit.
Furthermore, if we increase $N$ with other parameters fixed, the smallest value of $\zeta_i$ which gives $Z=0$ decreasingly approaches to the critical point of the large-$N$ solution.
Therefore, the extrapolation of this to the large-$N$ limit may be consistent with the SUSY breaking picture above.

\section{The case with one massless hypermultiplet ($\zeta_1=0$)}
\label{zeta10}
In this section we consider the case with $\zeta_1=0$.
In this case we find a solution to the saddle point equation for the partition function in the S-dual representation \eqref{Sdual}.
We can also compute the exact values of the partition function for finite $(N,k)$ by a slight generalization \cite{Nosaka,OZ} of the technique used in the ABJM theory \cite{TW,PY}.
We will see a good agreement of the both results.

\subsection{Saddle point analysis in the large-$N$ limit}
\label{saddle}
In this subsection we compute the partition function in the large-$N$ limit
\begin{\eq}
N\rightarrow\infty,\quad {\rm with\ fixed}\ (k,\zeta_2 ).
\end{\eq}
In this limit, we can evaluate the partition function by the saddle point method.
To perform the saddle point analysis, we first introduce the effective action $S_\text{eff}$ by
\begin{align}
Z=\frac{1}{N!} \int  \frac{d^Nx}{(2\pi k)^N}\ e^{-S_\text{eff}(x)}
\label{ZinS}
\end{align}
where
\begin{eqnarray}
S_\text{eff}(x)
&=& -\frac{2i\zeta_1}{k}\sum_{i=1}^N x_i
+\sum_{i=1}^N\log \left( 2\cosh\frac{x_i}{2} \right) \nonumber\\
&& -\sum_{i<j}^N\log\Bigl(2\sinh\frac{x_i-x_j}{2k}\Bigr)^2 
 +\sum_{i,j=1}^N\log \left( 2\cosh\frac{x_i-x_j+4\pi\zeta_2}{2k} \right).
\label{S}
\end{eqnarray}
We rearrange the eigenvalues $x_i$ such that $x_{i+1}\geq x_i$
by the permutation symmetry
and regard $x_i$ as a function of $s=i/N -1/2$,
which becomes the continuous variable in the large-$N$ limit:
\begin{\eq}
x_i \quad\rightarrow\quad x( s ),\quad {\rm with}\ s\in [-1/2 ,1/2]
\ \   {\rm and}\ \  \frac{dx}{ds}\geq 0 .
\end{\eq}
Then the summations over $i$ are replaced by the integral over $s$
\begin{align}
\sum_{i} \rightarrow N\int_{-\frac{1}{2}}^{\frac{1}{2}}ds.
\end{align}

We look for saddle point solutions by the approach taken in \cite{HKPT} which has been used to derive $\mathcal{O}(N^{\frac{3}{2}})$ behaviors of free energies, rather than the traditional approach often applied for matrix models in the planar limit.\footnote{
The traditional approach was taken in \cite{Grassi:2014vwa,Hatsuda:2015lpa} for $\zeta_1 =0=\zeta_2$ identifying 't Hooft coupling with $N/N_f$ where $N_f$ denotes an additional power put on the $\cosh$ (For our case, $N_f =1$).
}
This is achieved by taking the following ansatz
\begin{align}
x(s) =\sqrt{N}z (s) ,
\label{eq:ansatz}
\end{align}
with an ${\cal O}(1)$ real\footnote{
In our actual analysis, we have looked for solutions with complex $z(s)$ under the ansatz \eqref{eq:ansatz} but we have found only a real solution as a result.
Because of this, we take $z(s)$ to be real for simplicity of explanations in the main text.
Precisely speaking, we should first take the variation $\frac{\delta S_\text{eff}}{\delta z(s)}$ with $z(s)\in\mathbb{C}$ before assuming $z(s)\in\mathbb{R}$.
This induces a new constraint $\frac{\delta S_\text{eff}}{\delta \text{Im}(z(s))}=0$ in addition to \eqref{saddleeq} and \eqref{bc}; nevertheless the final result \eqref{zsfinal} remains the same.
}
function $z(s)$, and perform large-$N$ expansion of $S_\text{eff}(x)$ to simplify the saddle point equation.
It is easy to write down the leading part for the first and second terms in \eqref{S}:
\begin{align}
-\frac{2i\zeta_1}{k}\sum_{i=1}^N x_i
+\sum_{i=1}^N\log \left( 2\cosh\frac{x_i}{2} \right)
=N^{\frac{3}{2}}\int_{-\frac{1}{2}}^{\frac{1}{2}} ds\Bigl(-\frac{2i\zeta_1z}{k}+\frac{|z|}{2}\Bigr)+{\cal O}(N).
\end{align}
We can also expand the third and fourth terms in \eqref{S} respectively by using the techniques of \cite{NST2} to see that the leading part of $S_\text{eff}$ in the large-$N$ limit is proportional to $N^{\frac{3}{2}}$.
First we rewrite these terms as
\begin{align}
&-\sum_{i<j}^N\log\Bigl(2\sinh\frac{x_i-x_j}{2k}\Bigr)^2
=-\frac{N^2}{2}\int ds ds'\log\Bigl(2\sinh\frac{\sqrt{N}(z-z')}{2k}\Bigr)^2\nonumber \\
=& \frac{N^2}{2}\int dsds' \Biggl[\sgn(z-z')\frac{\sqrt{N}(z-z')}{k}
+
\log \left( 1-e^{-\sqrt{N}\mid\frac{z-z^{\prime}}{k}\mid}\right)^2
\Biggr],
\label{sinh}
\end{align}
and\footnote{
Note that odd functions of $z-z'$ do not contribute.
}
\begin{align}
&\sum_{i,j}^N\log \left( 2\cosh\frac{x_i-x_j+4\pi\zeta_2}{2k} \right)\nonumber \\
=& N^2\int dsds' 
\Biggl[ \sgn(z-z')\frac{\sqrt{N}(z-z')}{2k}
+\log \left( 1+e^{-\sgn(z-z^{\prime})\frac{\sqrt{N}(z-z^{\prime})+4\pi\zeta_{2}}{k}}\right)
\Biggr] .
\label{cosh}
\end{align}
where $z'$ is the abbreviation for $z(s')$.
Note that the $\mathcal{O}(N^{5/2})$ terms, which are the first terms in \eqref{sinh} and \eqref{cosh}, are canceled and only the second terms remain.  
Here we use the following formula in the large-$N$ limit:\footnote{
The condition that this evaluation is valid is following \cite{NST2}:
\[
 -\frac{1}{4} < {\rm Im} (w)- {\rm Re} (w) 
\frac{{\rm Im}(\dot{v}) }{{\rm Re} (\dot{v})} < \frac{1}{4}.
\]
In this case this condition is satisfied.
}
\begin{align}
 \int_{s_0}^{1/2} ds \ln (1 \pm e^{-2y(s)}) &\sim
\frac{1}{\sqrt{N} \dot{v}(s_{0})}
\int_{w(s_0 )}^\infty dt \ln (1 \pm e^{-2t}), 
\label{logcoshapprox1} \\
 \int_{-1/2}^{s_0} ds \ln ( 1 \pm e^{2y(s)}) &\sim
\frac{1}{\sqrt{N} \dot{v}(s_{0})}
\int_{-w(s_0 )}^\infty dt \ln (1 \pm e^{-2t}),
\label{logcoshapprox2}
\end{align}
where $y(s)=\sqrt{N}v(s)+w(s)$ and $s_{0}$ is the zero of $v(s)$.
These formulas are obtained by changing the integration variable and reflected with the fact that the contribution to the integral in l.h.s of \eqref{logcoshapprox1} and \eqref{logcoshapprox2} is coming from only $s\sim s_{0}$ region in the large-$N$ limit.
Using these formulas the second terms in \eqref{sinh} and \eqref{cosh} can be evaluated as 
\begin{\eqa}
&& 2kN^{\frac{3}{2}}\int  \frac{ds'}{\dot{z}_1(s')}
 \Biggl[  -2 \int_0^{\infty} dt \log (1-e^{-2t}) 
+ \int_{\frac{2\pi\zeta_{2}}{k}}^{\infty} dt \log (1+e^{-2t})
+ \int_{-\frac{2\pi\zeta_{2}}{k}}^{\infty} dt \log (1+e^{-2t})
\Biggr]\nonumber \\
&&= \frac{\pi^2kN^{\frac{3}{2}}}{2}\left(1+ \frac{16\zeta_{2}^2}{k^2}
\right)\int \frac{ds'}{\dot{z}_1(s')}.
\end{\eqa}
Putting the above computations together, we find the following large-$N$ expansion for the effective action
\begin{align}
S_\text{eff}=N^{3/2}\int ds F(z,\dot{z})+{\cal O}(N),
\label{SinlargeN}
\end{align}
with
\begin{align}
F(z,\dot{z})=\Bigl[-\frac{2i\zeta_1z}{k}+\frac{|z|}{2}+\frac{\pi^2k}{2}\Bigl(1+\frac{16\zeta_2^2}{k^2}\Bigr)\frac{1}{\dot{z}}\Bigr].
\end{align}
The overall scaling $N^{3/2}$ in \eqref{SinlargeN} implies that the integration \eqref{ZinS} is dominated in the large-$N$ limit by the saddle point configuration satisfying the following equation of motion
\begin{align}
0=\frac{\partial F}{\partial z}-\frac{d}{dt}\frac{\partial F}{\partial\dot{z}}=-\frac{2i\zeta_1}{k}+\frac{\sgn(z)}{2}-\frac{d}{ds}\Bigl[-\frac{\pi^2k}{2}\Bigl(1+\frac{16\zeta_2^2}{k^2}\Bigr)\frac{1}{\dot{z}^2}\Bigr]
\label{saddleeq}
\end{align}
together with the boundary condition
\begin{align}
0=\frac{\partial F}{\partial \dot{z}}=-\frac{\pi^2k}{2}\bigg[\bigl(1+\frac{16\zeta_2^2}{k^2}\Bigr)\frac{1}{\dot{z}^2}\bigg]_{\text{boundary}} .
\label{bc}
\end{align}

First let us consider the case for $\zeta_1=0$.
First of all the equation of motion \eqref{saddleeq} has the following two local solutions depending on $\sgn(z)$
\begin{align}
z^{(+)}(s)&=\sqrt{2\pi^2k\Bigl(1+\frac{16\zeta_2^2}{k^2}\Bigr)}(z_b-\sqrt{2(s_b-s)}),\quad (\sgn(z)=+1)\nonumber \\
z^{(-)}(s)&=-\sqrt{2\pi^2k\Bigl(1+\frac{16\zeta_2^2}{k^2}\Bigr)}(z_b-\sqrt{2(-s_b+s)}),\quad (\sgn(z)=-1)
\label{solntosaddleeq}
\end{align}
where $s_b$ and $z_b$ are the integration constants.\footnote{
We have excluded the other two solutions by the condition $\dot{z}(s)\geq 0$. 
}
The bulk solution would be obtained by connecting these solutions appropriately and determining the integration constants so that $z(s)$ satisfies the boundary condition $\dot{z}(s)=\pm \infty$ \eqref{bc} at every point where $z(s)$ or $\dot{z}(s)$ is discontinuous.
Notice that both of $z^{(\pm)}(s)$ satisfies $\dot{z}^{(\pm)}(s)=\infty$ at only a single point $s=s_b$.
Therefore, if we split the support $-1/2<s<1/2$ into segments by the points of discontinuity, $z(s)$ on each segment must be given as a smooth junction of $z^{(-)}(s)$ and $z^{(+)}(s)$.
Since $z^{(+)}(s)$ cannot be followed by $z^{(-)}(s)$ due to the assumption that $z(s)$ is monotonically increasing, we conclude that the solution is given by a single junction of $z^{(-)}(s)$ with $s_b=-1/2$ ($-1/2<s<s_0$) and $z^{(+)}(s)$ with $s_b=1/2$ ($s_0<s<1/2$) with some $s_0$.
The remaining constants $s_0,s_b$ are determined from $z^{(-)}(s_0)=z^{(+)}(s_0)$ and $\dot{z}^{(-)}(s_0)=\dot{z}^{(+)}(s_0)$ as $s_0=0$, $z_b=1$ (for both domain).
In summary we obtain the following unique solution as the saddle point configuration:
\begin{align}
z(s)=\sgn(s)\sqrt{2\pi^2k\Bigl(1+\frac{16\zeta_2^2}{k^2}\Bigr)}(1-\sqrt{1-2|s|}).
\label{zsfinal}
\end{align}
In the language of the eigenvalue density, this solution corresponds to
\begin{\eq}
\rho (z) = \frac{ds}{dz} 
= \frac{1}{\sqrt{2\pi^2 k \left( 1 +16\zeta_2^2 /k^2 \right)}}
 \left( 1 -\frac{|z|}{\sqrt{2\pi^2 k \left( 1 +16\zeta_2^2 /k^2 \right)}}  \right) .
\end{\eq}

Substituting this solution to \eqref{SinlargeN}, we find that the partition function in the large-$N$ limit is given as
\begin{align}
\left. -\log Z \right|_{\zeta_1 =0}
\approx \frac{\pi\sqrt{2k}}{3}\sqrt{1+\frac{16\zeta_2^2}{k^2}}N^{\frac{3}{2}}.
\label{Zinsaddle}
\end{align}

For $\zeta_{1}\neq 0$ we could not solve the saddle point equation with the ansatz we used here because the solution can not satisfy the boundary condition \eqref{bc} due to the existence of the imaginary term $\frac{2i\zeta_{1}}{k}$ in \eqref{saddleeq}. 
However, the partition function with $\zeta_{1}=0,\ \zeta_{2}\neq0$ and that with $\zeta_1\neq 0, \ \zeta_{2}=0$ is the same because the partition function is invariant under exchanging $\zeta_{1}$ and $\zeta_{2}$. 
This fact suggests that even when $\zeta_{2}=0,\zeta_{1} \neq 0$, there exists the solution of the saddle point equation in large-$N$ limit and the free energy can be evaluated by the saddle point approximation.

\subsection{Exact partition function for finite $(N, k)$}
\label{TWPY}
Next we compute the partition function for some finite $(N,k)$ by the technique used in \cite{Nosaka}.
We start with the partition function written in the Fermi gas formalism
\begin{align}
Z(N,k,0,\zeta_2)=\frac{1}{N!}\int\frac{d^Nx}{(2\pi)^N}\det_{i,j}\langle x_i|{\widehat\rho} (\hat{q} ,\hat{p}) |x_j\rangle,
\end{align}
where $[\hat{q}, \hat{p}] = i\hbar$ with $\hbar =2\pi k$ and\footnote{
For a later convenience we have symmetrized the density matrix by another similarity transformation from \eqref{rhosimplified}.
}
\begin{align}
{\widehat\rho}=
\sqrt{\frac{1}{2\cosh\frac{\widehat q}{2}}}
\frac{e^{\frac{2i\zeta_2}{k}{\widehat p}}}{2\cosh\frac{\widehat p}{2}}
\sqrt{\frac{1}{2\cosh\frac{\widehat q}{2}}}.
\end{align}
If we consider the generating function of the partition function or equivalently the grand partition function $\sum_{N=0}^\infty z^NZ(N)$, we can show that it is written as the following Fredholm determinant
\begin{align}
\sum_{N=0}^\infty z^NZ(N)=\Det(1+z{\widehat\rho})\equiv \exp\Bigl[\sum_{n=1}^\infty \frac{(-1)^{n-1}}{n}z^n \Tr{\widehat\rho}^n\Bigr].
\label{eq:grand}
\end{align}
Comparing the coefficient of $z^N$ on the both sides, we find that the partition function $Z(N)$ is determined by $\Tr{\widehat\rho}^n$ with $n\le N$, as
\begin{align}
Z(1)=\Tr{\widehat\rho},\quad
Z(2)=\frac{1}{2}(\Tr{\widehat\rho})^2-\frac{1}{2}\Tr{\widehat\rho}^2,\quad
Z(3)=\frac{1}{6}(\Tr{\widehat\rho})^3-\frac{1}{2}\Tr{\widehat\rho}\Tr{\widehat\rho}^2+\frac{1}{3}\Tr{\widehat\rho}^3,\quad
\cdots.
\end{align}

We can compute $\Tr{\widehat\rho}^n$ by completely the same way as that in the case of R-charge deformation \cite{Nosaka}.
First we notice that the matrix element $\langle x|{\widehat \rho}|y\rangle$ has the following structure
\begin{align}
\langle x|{\widehat\rho}|y\rangle
=\frac{1}{2\cosh\frac{x}{2}}\frac{1}{2k\cosh\frac{x-y+4\pi \zeta_2}{2k}}
=\frac{E(x)E(y)}{k(\alpha M(x)+\alpha^{-1}M(y))},
\label{TWstructure}
\end{align}
with
\begin{align}
E(x)=\frac{e^{\frac{x}{2k}}}{\sqrt{2\cosh\frac{x}{2}}},\quad
M(x)=e^{\frac{x}{k}},\quad
\alpha=e^{\frac{2\pi\zeta_2}{k}}.
\end{align}
For $\alpha =1$, 
this form is in the range of application of Tracy-Widom's lemma \cite{Tracy:1995ax}
which has been very powerful tool to systematically compute ${\rm Tr}\hat{\rho}^n$
in various M2-brane theories without masses \cite{Hatsuda:2012hm,PY,Hatsuda:2012dt,Matsumoto:2013nya,Honda:2014npa,Hatsuda:2014vsa,Moriyama:2014gxa,Moriyama:2014nca,Moriyama:2017gye}.
We can easily extend it to general $\alpha$ as follows.
The structure \eqref{TWstructure} can be expressed as a quasi-commutation relation for ${\widehat\rho}$
\begin{align}
\alpha {\widehat M}{\widehat\rho}+\alpha^{-1}{\widehat\rho}{\widehat M}={\widehat E}|0\rrangle\llangle 0|{\widehat E},\quad\quad ({\widehat E}=E({\widehat q}),\quad {\widehat M}=M({\widehat q})) ,
\end{align}
where $|p \rrangle$ is momentum eigenstate satisfying
\begin{align}
\langle x|x'\rangle =2\pi\delta(x-x'),\quad
\llangle p|p'\rrangle =2\pi\delta(p-p'),\quad
\langle x|p\rrangle=\frac{1}{\sqrt{k}}e^{\frac{ixp}{\hbar}},\quad
\llangle p|x\rangle=\frac{1}{\sqrt{k}}e^{-\frac{ixp}{\hbar}}.
\label{eq:braket}
\end{align}
This relation can be generalized straightforwardly for ${\widehat\rho}^n$ as
\begin{align}
\alpha^n{\widehat M}{\widehat\rho}^n-(-1)^n\alpha^{-n}{\widehat\rho}{\widehat M}
=\sum_{\ell=0}^{n-1}(-1)^\ell \alpha^{n-1-2\ell}{\widehat\rho}^\ell {\widehat E}|0\rrangle\llangle 0| {\widehat E} {\widehat\rho}^{n-1-\ell}.
\end{align}
This implies that we can compute the matrix element of ${\widehat\rho}^n$ from two sets of functions $\phi_\ell(x)$ and $\psi_\ell(x)$
as
\begin{align}
\langle x|{\widehat \rho}^n|y\rangle=\frac{E(x)E(y)}{\alpha^nM(x)-(-1)^n\alpha^{-n}M(y)}\sum_{\ell=0}^{n-1}(-1)^\ell\phi_\ell(x)\psi_{n-1-\ell}(y),
\end{align}
where
\begin{align}
\phi_\ell(x)=\alpha^{-\ell}\langle x|{\widehat E}^{-1}{\widehat\rho}^\ell {\widehat E}|0\rrangle,\quad
\psi_\ell(x)=\alpha^\ell\llangle 0|{\widehat E}^{-1}{\widehat\rho}^\ell {\widehat E}|x\rrangle=\phi_\ell(x)|_{\alpha\rightarrow\alpha^{-1}} .
\end{align}
We can show that the function $\phi_\ell(x)$ satisfies the following recursion relation
\begin{align}
\phi_{\ell+1}(x)&=\int\frac{dy}{2\pi}\frac{1}{E(x)}\alpha^{-1}\rho(x,y)E(y)\phi_\ell(y)\nonumber \\
&=\int \frac{dy}{2\pi k}\frac{1}{e^{\frac{y}{k}}+\alpha^2e^{\frac{x}{k}}}\frac{e^{\frac{y}{k}}}{e^{\frac{y}{2}}+e^{-\frac{y}{2}}}\phi_\ell(y),
\label{recursive_int}
\end{align}
as well as $\psi_\ell (x)$.
In app.~\ref{app:detail_exact}, we explain how to practically solve the recursion relation for integer $k$ while their details are slightly different between odd $k$ and even $k$ cases. 
According to the algorithm, we have computed $Z(N,k,1,\zeta_2)$ by Mathematica for $(k=1 ,N\le 12)$, $(k=2,N\le 9 )$, $(k=3 ,N\le 5)$, $(k=4 ,N\le 5 )$ and $(k=6 ,N\le 4 )$.
In app.~\ref{Z1exact_results}, we explicitly write down a part of the results and also compare them with the result of saddle point approximation \eqref{Zinsaddle}.

\section{General deformation with $\zeta_1,\zeta_2\neq 0$}
\label{bothnonzero}
In this section we consider the case for $\zeta_1, \zeta_2\neq 0$.
Note that this may affect the sign of the partition function because the integrand of \eqref{Sdual} for $\zeta_1\neq 0$ has the oscillation factor $e^{\frac{2i \zeta_1}{k}\sum_i x_i}$ in contrast to the $\zeta_1=0$ case, where the integrand was positive semi-definite.
Therefore the partition function may be negative or zero depending on the parameters $(N,k,\zeta_1 ,\zeta_2 )$.
For large $\zeta_1,\zeta_2$, we can easily see that this actually happens as follows.
In this limit, the hypermultiplets become very massive and integrating them out leads us to the $\mathcal{N}=2$ SUSY $\text{U}(N)_k\times \text{U}(N)_{-k}$ pure Chern-Simons theory schematically.\footnote{
More precisely, integrating out the matter fields induces level shifts of all the possible mixed CS terms which are among the gauge symmetry $\text{U}(N)\times \text{U}(N)$, flavor symmetry $\text{U}(1)_1 \times \text{U}(1)_2$ and $\text{U}(1)_\text{R}$ symmetry.
In the case of the massive ABJM theory, most of the shifts are canceled and we have only contributions from the gauge-$\text{U}(1)_\text{R}$ and flavor $\text{U}(1)_\text{R}$ CS terms but these terms do not affect the zeroes of the partition function.
Here the integration of the matter fields is assumed to be at the origin of the Coulomb moduli space.
Thus, the decoupling of the matter fields in the large mass limit is possible.
}
It is known that the sphere partition function of the pure Chern-Simons theory vanishes for $k<N$.
In this section, we will see that the zeroes appear also for finite $(\zeta_1 ,\zeta_2 )$.

In this case we could not find a solution to the saddle point equation.
The technique for small integers $k,N$ in sec.~\ref{TWPY} is not applicable either.
Nevertheless we can evaluate the partition function exactly for $N=1,2$, which suggest the partition function has zeroes as a function of $\zeta_1,\zeta_2$ only for $(N,k)=(2,1),(2,2)$.
We argue a possible interpretation for this zeroes.
We further conjecture the zeroes for general $k,N$, and provide positive evidence from the numerical computation of the partition function for $N\ge 3$.

\subsection{Exact expression for $N=1,2$}
\label{sec:exactN2}
In this subsection we review 
the exact results for $N=1,2$ obtained in \cite{RS}.\footnote{
The notation in \cite{RS} is related to ours by
\[
Z_{\rm ours}(N,k,\zeta_1 ,\zeta_2 ) 
=2^{-2N}Z_{\rm Russo-Silva}( N,k, m_1 =-4\pi\zeta_2/k ,\zeta_2 =-4\pi\zeta_1 /k ). 
\] 
}
The relation \eqref{eq:grand} between the partition function and ${\rm Tr}{\widehat\rho}^n $ is correct also for general $\zeta_1$ if we take ${\widehat\rho}$ as
\begin{align}
\langle x|{\widehat\rho}|y\rangle
=\frac{e^{\frac{2i\zeta_1}{k}x}}{2\cosh\frac{x}{2}}\frac{1}{2k\cosh\frac{x-y+4\pi \zeta_2}{2k}}.
\end{align}
For $N=1$, the partition function is simply given by $Z(1,k,\zeta_1,\zeta_2)=\Tr{\widehat\rho}$, which can be exactly computed as
\begin{align}
Z(1,k,\zeta_1,\zeta_2)
=\int_{-\infty}^\infty\frac{dx}{2\pi}
\frac{e^{\frac{2i\zeta_1x}{k}}}{2\cosh\frac{x}{2}}\frac{1}{2k\cosh\frac{2\pi\zeta_2}{k}}
=\frac{1}{4k\cosh\frac{2\pi\zeta_1}{k}\cosh\frac{2\pi\zeta_2}{k}}.
\label{eq:exactN1}
\end{align}
For $N=2$, we need to compute $\Tr{\widehat\rho}^2$, which is given by the following two dimensional integration
\begin{align}
\Tr{\widehat\rho}^2
=\int_{-\infty}^\infty \frac{dx}{2\pi}\frac{dy}{2\pi} \frac{e^{\frac{2i\zeta_1(x+y)}{k}}}{16k^2\cosh\frac{x}{2}\cosh\frac{y}{2}\cosh\frac{x-y+4\pi\zeta_2}{2k}\cosh\frac{x-y-4\pi\zeta_2}{2k}}.
\end{align}
After changing the integration variables to $x_\pm=x\pm y$, we can easily perform the $x_+$-integration, which leads to
\begin{align}
\Tr{\widehat\rho}^2
=\frac{1}{16\pi k^2\sin\frac{4\pi\zeta_1}{k}}
\int_{-\infty}^\infty dx_- \frac{\sin\frac{2\zeta_1x_-}{k}}
{\sin\frac{x_-}{2}\cosh\frac{x_-+4\pi\zeta_2}{2k}\cosh\frac{x_--4\pi\zeta_2}{2k}} .
\end{align}
For $k\in \mathbb{Z}_+$, this integral can be evaluated by considering an integral with the same integrand along a rectangular whose corners are $x_- = (-\infty,\infty ,\infty +2\pi ik ,-\infty +2\pi ik)$ \cite{Okuyama,RS}, and we obtain
\begin{\eq}
\Tr{\widehat\rho}^2
= \frac{1}{8k^2 \sinh{\frac{4\pi\zeta_1}{k}}\cosh^2{\frac{2\pi\zeta_2}{k}}
\left( 1-(-1)^k \cosh{4\pi\zeta_1} \right)} 
 \Biggl[ \sum_{n=1}^{k-1}  
\frac{(-1)^n \sin^2{\frac{\pi n}{k}}\sinh{\frac{4\pi\zeta_1 n}{k}}}
 {\cosh{\frac{2\pi\zeta_2 +i\pi n}{k}} \cosh{\frac{2\pi\zeta_2 -i\pi n}{k}}}
+R_k \Biggr] ,
\end{\eq}
where
\begin{\eq}
R_k 
=\begin{cases}
(-1)^{\frac{k -1}{2}} \frac{k \coth{\frac{2\pi\zeta_2}{k}}\cosh{\frac{2\pi\zeta_1}{k}}}
{\cosh{2\pi\zeta_2}} \sin{\frac{8\pi\zeta_1 \zeta_2}{k}} & 
{\rm for\ odd}\ k \cr
(-1)^{\frac{k}{2}+1} \frac{k \coth{\frac{2\pi\zeta_2}{k}}\sinh{\frac{2\pi\zeta_1}{k}}}
{\sinh{2\pi\zeta_2}} \cos{\frac{8\pi\zeta_1 \zeta_2}{k}} & 
{\rm for\ even}\ k 
\end{cases} .
\end{\eq}
For example, the final results for $k=1,2,3,4$ are explicitly given by 
\begin{align}
Z(2,1,\zeta_1,\zeta_2)
&=\frac{\sin 8\pi\zeta_1\zeta_2}{8\sinh 4\pi\zeta_1\sinh 4\pi\zeta_2 \cosh 2\pi\zeta_1 \cosh 2\pi\zeta_2},\nonumber \\
Z(2,2,\zeta_1,\zeta_2)
&=\frac{\sin^2 2\pi\zeta_1\zeta_2}{8\sinh^2 2\pi\zeta_1 \sinh^2 2\pi\zeta_2},\nonumber \\
Z(2,3,\zeta_1,\zeta_2)
&=\frac{1}{24(\cosh\frac{4\pi\zeta_1}{3}+\cosh\frac{8\pi\zeta_1}{3})(\cosh\frac{4\pi\zeta_2}{3}+\cosh\frac{8\pi\zeta_2}{3})}
\left( 2-\frac{\sin\frac{8\pi\zeta_1\zeta_2}{3}}{\sinh\frac{2\pi\zeta_1}{3}\sinh\frac{2\pi\zeta_2}{3}}\right),\nonumber \\
Z(2,4,\zeta_1,\zeta_2)
&=\frac{1}{128\sinh^2\pi\zeta_1\sinh^2\pi\zeta_2}
\left(1-\frac{1}{\cosh\pi\zeta_1}-\frac{1}{\cosh\pi\zeta_2}+\frac{\cos 2\pi\zeta_1\zeta_2}{\cosh\pi\zeta_1\cosh\pi\zeta_2}\right) .
\label{ZN2exact}
\end{align}
We easily see from these results that
the partition function for $(N,k)=(2,1),(2,2)$ has zeroes at finite $(\zeta_1 /k ,\zeta_2 /k)$.

\subsection{$N\ge 3$ from Monte Carlo Simulation}
\label{sec:MC}
In this subsection we provide numerical evidence that the partition function has zeroes at finite $(\zeta_1 /k ,\zeta_2 /k)$ also for $N\geq 3$.
For this purpose, we apply (Markov chain) Monte Carlo method to the partition function in the $S$-dual representation \eqref{Sdual}:
\begin{\eq}
Z(k,N,\zeta_1 ,\zeta_2 )
= \frac{1}{N!}\int \frac{d^N x}{(2\pi k )^N}\ e^{-S(k,N,\zeta_1 ,\zeta_2 ;x)} ,
\end{\eq} 
where
\begin{\eqa}
S(k,N,\zeta_1 ,\zeta_2 ;x)
&=& -\sum_{i<j}^N \log{\left( 2\sinh\frac{x_i-x_j}{2k} \right)^2}
 +\sum_{i,j=1}^N \log{\left(2\cosh\frac{x_i-x_j+4\pi\zeta_2}{2k}\right)} \NN\\
&& +\sum_{i=1}^N \log{\left( 2\cosh\frac{x_i}{2} \right)}
   -\log{\cos\left(\frac{2\zeta_1}{k}\sum_{i=1}^N x_i\right)} .
\end{\eqa}

\subsubsection{Algorithm}
First we explain our algorithm.
There are two subtleties in applying the Monte Carlo method to our problem.
The first subtlety, which will not be problematic as explained below, is that Monte Carlo simulation can directly calculate only ``expectation values" or equivalently ratio of two functions rather than $Z$ itself.
The second one is that the Boltzmann weight $e^{-S}$ is not positive semi-definite for $\zeta_1 \neq 0$ and hence cannot be regarded as probability.
This problem appears in many contexts such as finite density QCD, real time systems and theories with CS terms.

We take care of these subtleties as follows.
Instead of $Z$ itself, we consider the ratio
\begin{align}
Z_{\rm MC}(N,k,\zeta_1 ,\zeta_2 )
=\frac{Z(N,k,\zeta_1,\zeta_2)}{Z(N,k,0,\zeta_2)}
=\Biggl\langle \cos\left(\frac{2\zeta_1}{k}\sum_{i=1}^Nx_i \right) 
\Biggr\rangle_{\zeta_1 =0},
\label{ZcMC}
\end{align}
where $\langle \mathcal{O}(x) \rangle_{\zeta_1 =0}$ denotes the expectation value of $\mathcal{O}(x)$ under the action $S(N,k,\zeta_1 =0 ,\zeta_2 )$.
Then we approximate the ratio\footnote{
This is so-called reweighting method.
}
by Hybrid Monte Carlo simulation\footnote{
The application to a similar system is explained in app.~A of \cite{KEK}.
}
by taking samples generated with the probability $\sim \left. e^{-S} \right|_{\zeta_1 =0}$.
Note that studying only the ratio is sufficient for our purpose since $Z(N,k,0,\zeta_2)$ is real positive and we are interested in the sign of the partition function.\footnote{
Of course we can also compute $Z(N,k,\zeta_1 ,\zeta_2 )$ itself by combining $Z_{\rm MC}(N,k,\zeta_1 ,\zeta_2 )$ with $Z(N,k,0 ,\zeta_2 )$ computed in another way.
For example, we know the exact values of $Z(N,k,0 ,\zeta_2 )$ for various $(N,k,\zeta_2)$ obtained in sec.~\eqref{TWPY} and Monte Carlo simulation of $Z(N,k,0 ,\zeta_2 )$ is much easier than the $\zeta_1 \neq 0$ case if we use the algorithm in \cite{KEK}.
}
Since we are taking samples of the oscillating function, whose oscillation is controlled by $\zeta_1 /k$, we typically need more statistics for larger $\zeta_1 /k$ to obtain precise approximations.
Note also that the $S$-dual representation \eqref{Sdual} has much milder oscillation than the original matrix model \eqref{Original}.
This is why we are using the $S$-dual representation as in \cite{KEK}.

\subsubsection{Results}
\begin{figure}[t]
\begin{center}
\includegraphics[width=7.8cm]{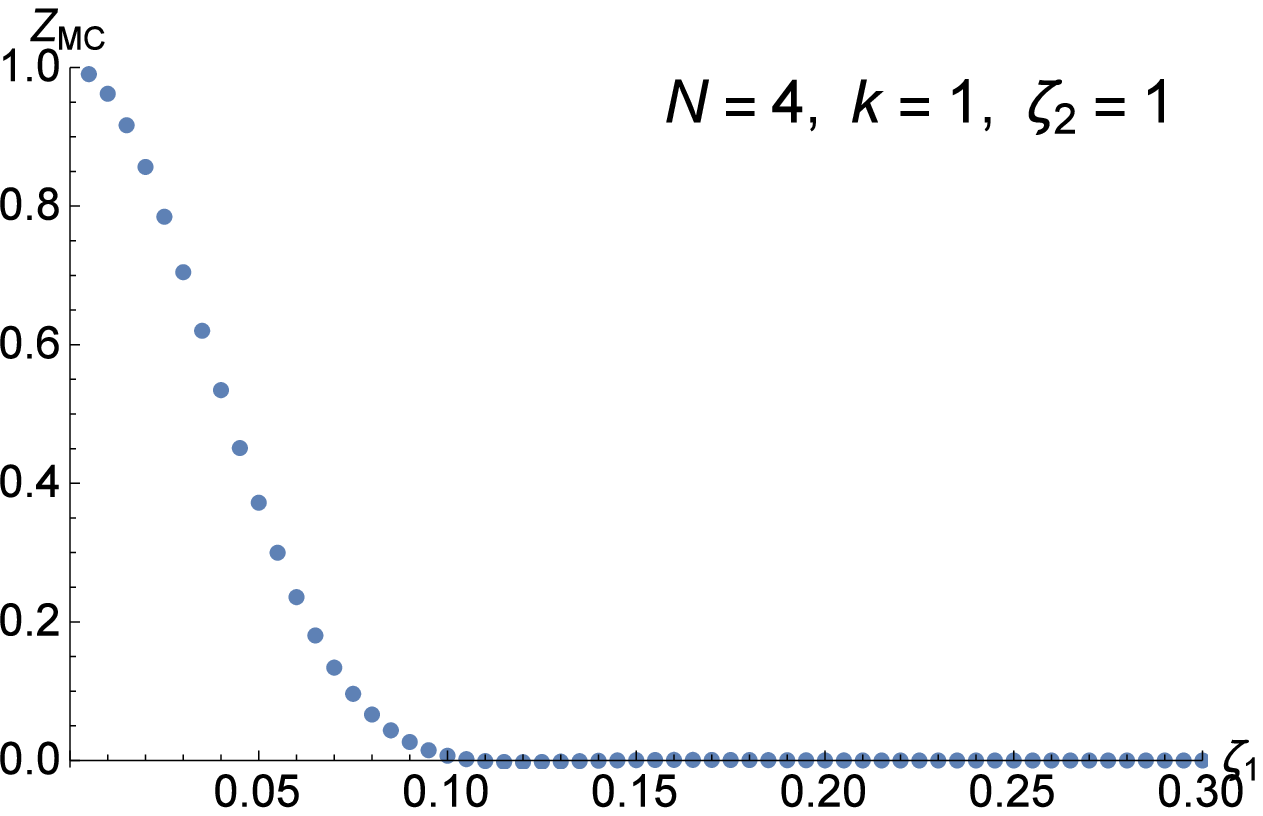}
\includegraphics[width=7.8cm]{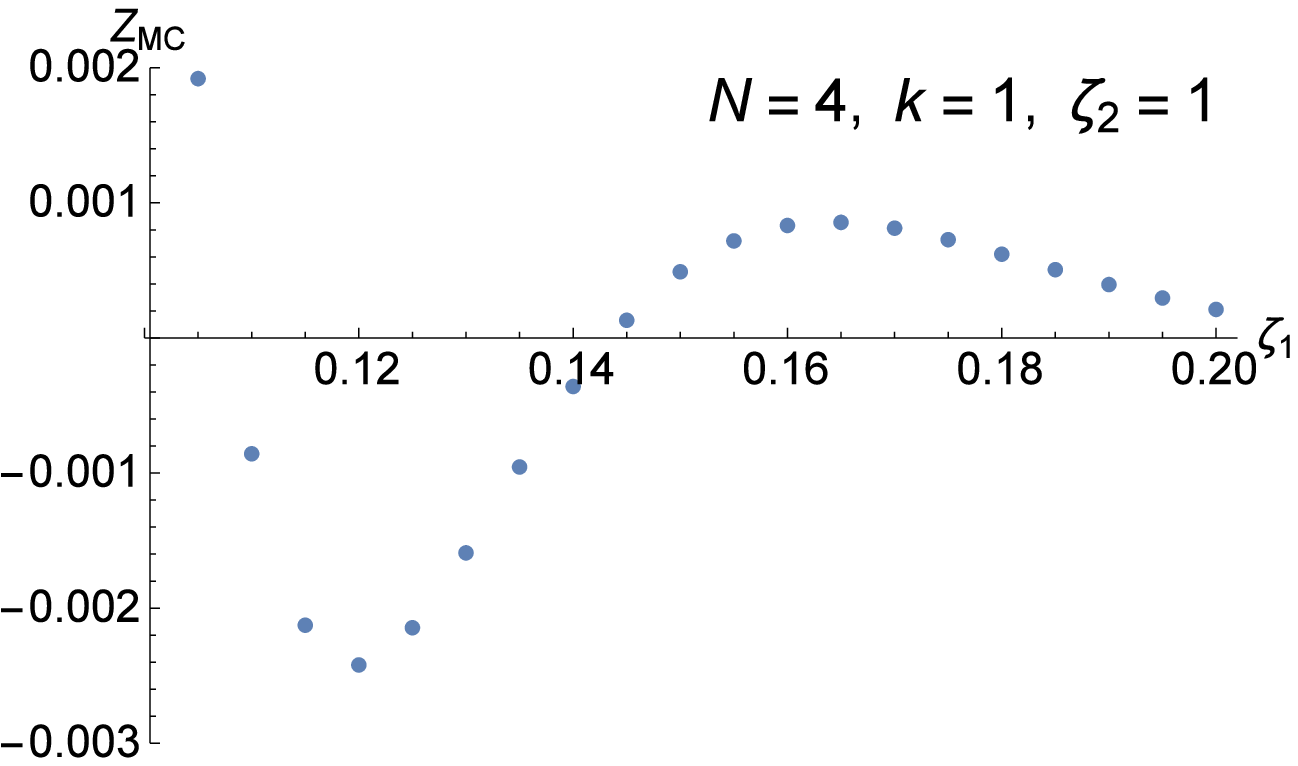}
\caption{
The ratio \eqref{ZcMC} computed by Monte Carlo simulation
is plotted against $\zeta_1$ for $(N,k,\zeta_2 )=(4,1,1)$.
The right panel is the zoomup of the left panel 
around the negative peak of the partition function.
\label{fig:N4k1zeta2_1}
}
\end{center}
\end{figure}
Now we present numerical results for the ratio $Z_{\rm MC}$ \eqref{ZcMC}, which has the same sign as the partition function $Z(N,k,\zeta_1 ,\zeta_2 )$ itself.
Fig.~\ref{fig:N4k1zeta2_1} plots $Z_{\rm MC}$ for $(N,k,\zeta_2 )=(4, 1,1)$ as a function of $\zeta_1$.
The statistical errors are estimated by Jackknife method although they are practically almost invisible in the figures.
The right panel of fig.~\ref{fig:N4k1zeta2_1} is the zoomup of the left figure in the range $\zeta_1 \in [0.1,0.2]$.
From the right figure, we easily see that the partition function takes negative values when $\zeta_1$ $=$ $0.110$, $0.115$, $\cdots ,0.135$ even if we take into account the errors.
Therefore there must be a zero of the partition function for $\zeta_1 \leq 0.110$ and the plot indicates that the zero is located at $0.105 <\zeta_1 <0.110$. 

\begin{figure}[p]
\begin{center}
\includegraphics[width=7.8cm]{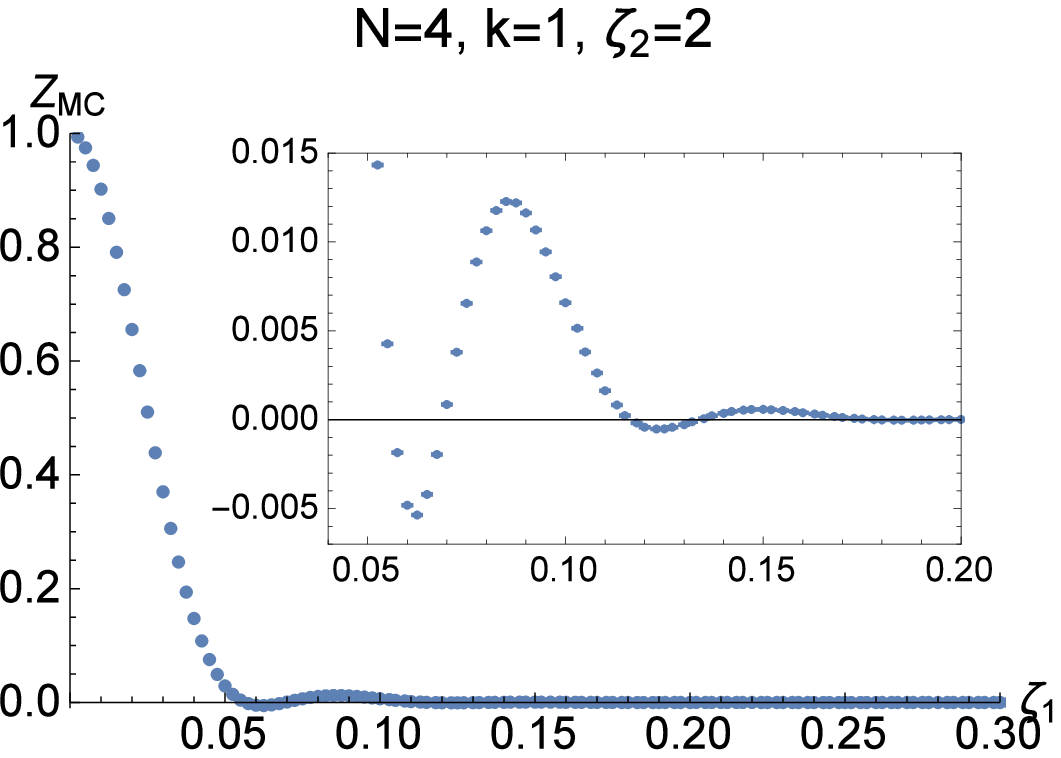}
\includegraphics[width=7.8cm]{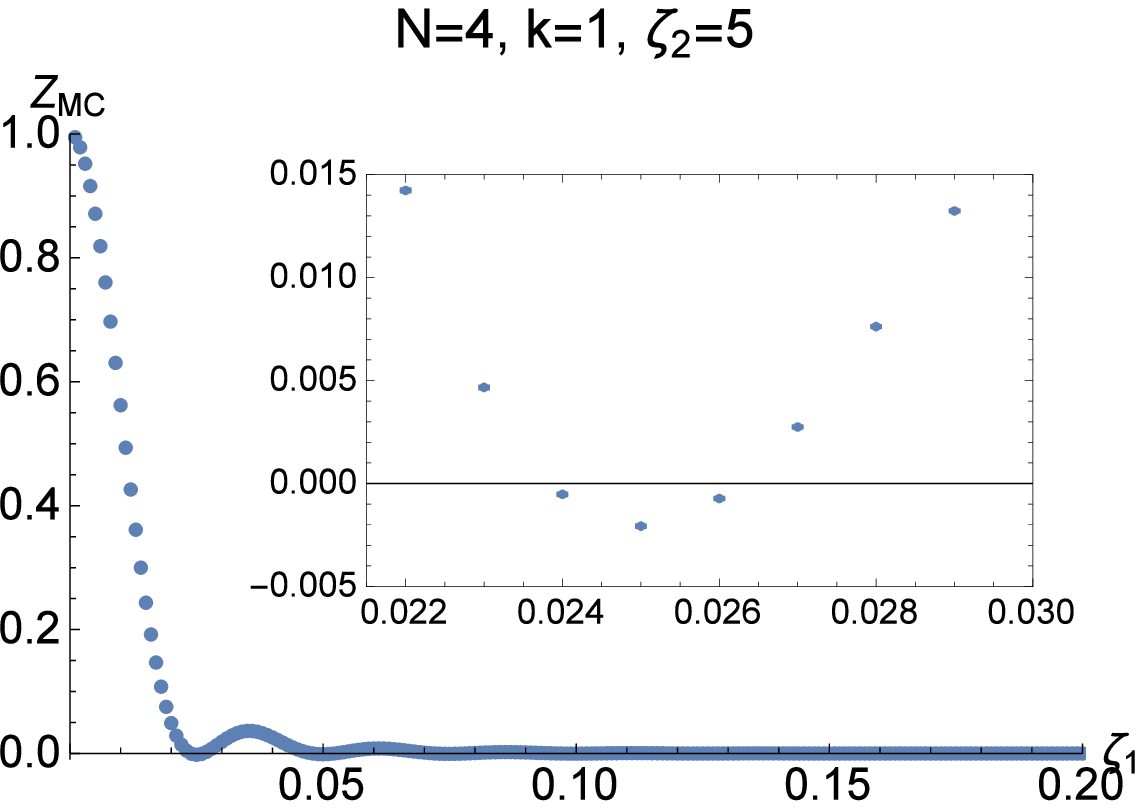}\\ \vspace{1.0em}
\includegraphics[width=7.8cm]{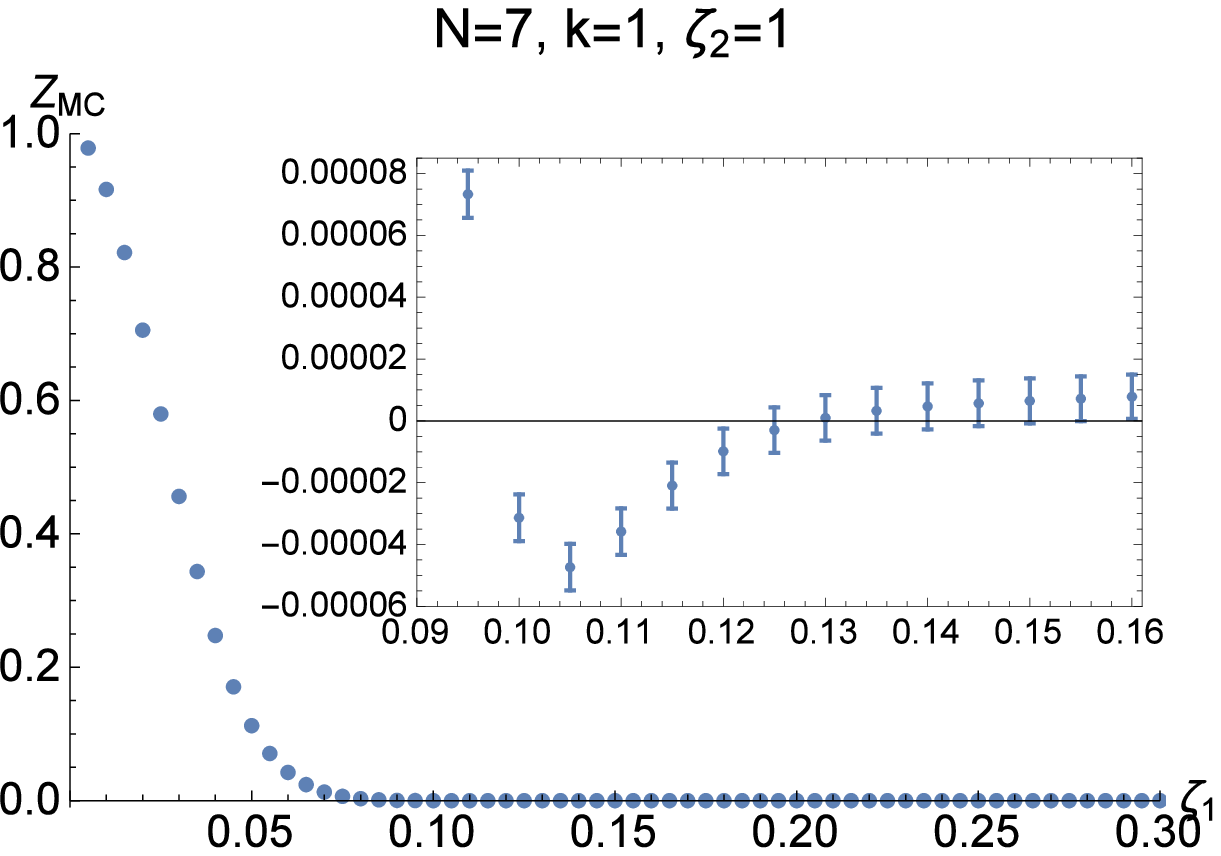}
\includegraphics[width=7.8cm]{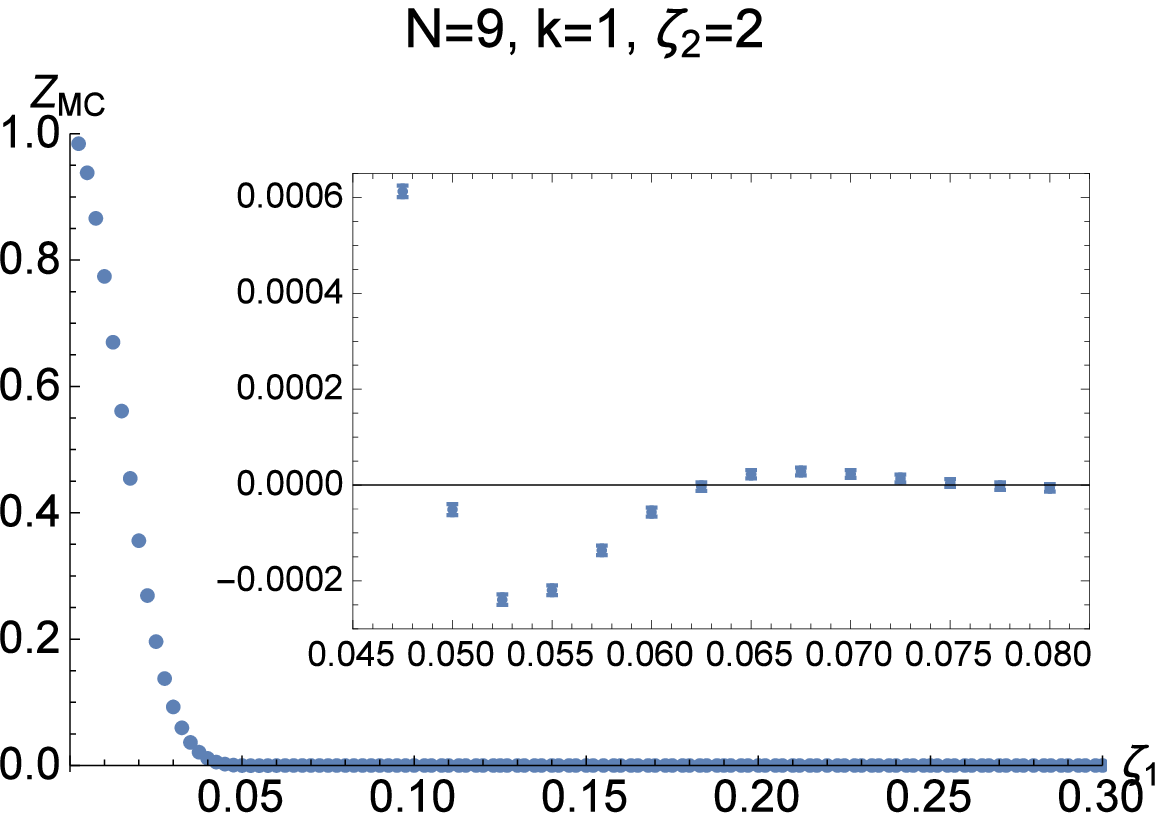}\\ \vspace{1.0em}
\includegraphics[width=7.8cm]{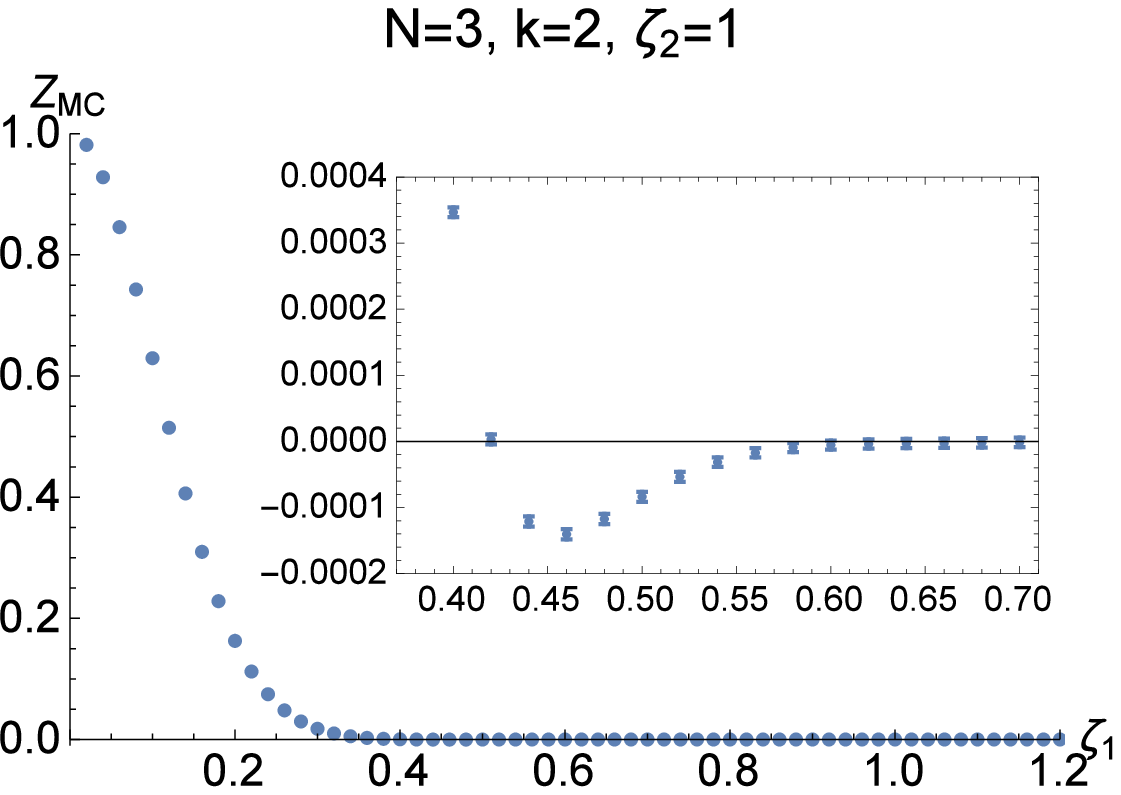}
\includegraphics[width=7.8cm]{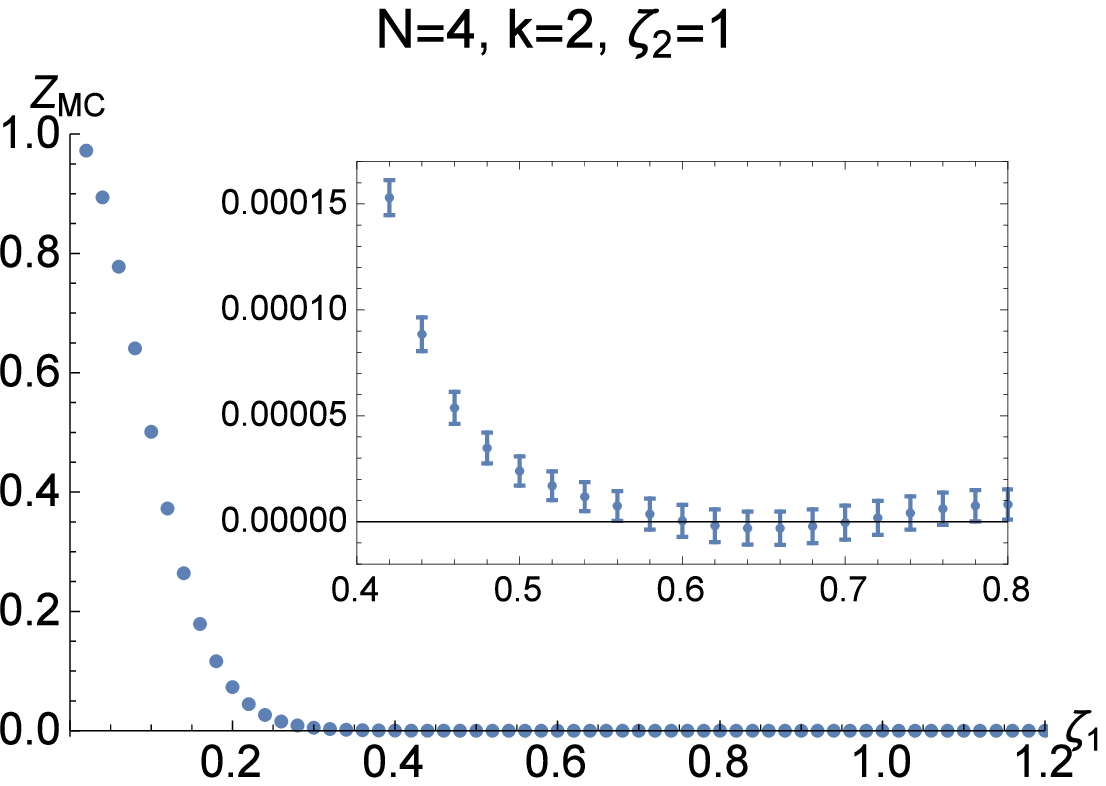}
\caption{
The numerical plots of $Z_{\rm MC}$ as functions of $\zeta_1$ 
for various $(N,k,\zeta_2 )$
with their zoomups around the minima.
\label{fig:various_cases}
}
\end{center}
\end{figure}

We have found similar results for other values of $(N,k,\zeta_2 )$ whose samples are shown in fig.~\ref{fig:various_cases}.
These figures indicate that the partition function has the zeroes at finite $\zeta_1 /k$ for various $(N,k,\zeta_2 )$.
Note also that we sometimes encounter subtle cases.
For example, in the case of $(N,k,\zeta_2 )=(4,2,1)$ shown in the right-bottom of fig.~\ref{fig:various_cases}, the minimum is consistent with both positive and negative $Z$
within the numerical errors.\footnote{
Similar behaviors have been observed for $(N,k,\zeta_2 )$ $=$ $(3,1,1)$, $(4,2,2)$, $(5,2,1)$, $(5,2,2)$.
}
We expect that this type of behavior appear when the partition function is positive semidefinite but has zeroes as in the case of $(N,k)=(2,2)$ whose analytic result is given in the second line of \eqref{ZN2exact}.
For this type of cases, any numerical simulation with nonzero errors cannot establish existence of zeroes since numerical values at the zeroes must be consistent with all the possible signs of $Z$ within errors.
Therefore, for this type of cases, the best thing we can do by numerical simulation is to check existence of points consistent with $Z=0$.
For all values of $(N\geq 2,k,\zeta_2 )$ which we have analyzed, we have checked that there exists at least one value of $\zeta_1$ consistent with $Z=0$ within errors.
For the cases where we have established existence of first zeroes of $Z$, we write down bounds on the zeroes in tables \ref{tab:zeta1}, \ref{tab:zeta2} and \ref{tab:zeta5} for fixed $(k,\zeta_2 )$ (see tab.~\ref{negativeextrema} for $Z_{\rm MC}$ at the first negative peaks and their errors).
We also estimate their precise locations by constructing\footnote{
This is done by the command ``Interpolation" in Mathematica.
The values without ``$\pm$" are (first) zeroes of the interpolating functions for the average values of $Z_{\rm MC}$.
The values including ``$\pm$" denotes zeroes of interpolating functions for the average values plus/minus the errors.
}
interpolating functions of all the data points of $Z_{\rm MC}$ for fixed $(N,k,\zeta_2 )$ and finding zeroes of the interpolating functions.
We will discuss implications of these values in sec.~\ref{conjlargeN}.
\begin{table}[t]
\begin{center}
\begin{tabular}{|c|c|c|}
\hline
$N$& Bounds on the zeroes  & Estimate of the zeroes \\ \hline
$2$&$\zeta_1=0.125$                  &  $\zeta_1=0.125$ \\ \hline
$4$&$0.105<\zeta_1<0.11$          & $\zeta_1 =0.108084 \pm 0.000016 $\\ \hline
$5$&$0.105<\zeta_1<0.11$          & $\zeta_1 =0.105249 \pm 0.000041$\\ \hline
$7$&$0.095<\zeta_1<0.1$            & $\zeta_1 =0.0975822^{+0.0004201}_{-0.0003715}$\\ \hline
$9$&$0.085<\zeta_1<0.095$&$\zeta_1=0.0898839^{+0.0003752}_{-0.0004039}$\\ \hline
\end{tabular}
\end{center}
\caption{
Bounds on first zeroes of the partition function and estimate of their precise locations by interpolating functions for $(k,\zeta_2 )=(1,1)$.
The value for $N=2$ is the exact value.
}
\label{tab:zeta1}
\end{table}
\begin{table}[t]
\begin{center}
\begin{tabular}{|c|c|c|}
\hline
$N$& Bounds on the zeroes  & Estimate of the zeroes \\ \hline
$2$&$\zeta_1=0.0625$          &  $\zeta_1=0.0625$\\ \hline
$4$&$0.055<\zeta_1<0.0575$ & $\zeta_1 = 0.0565766 \pm 0.0000060$\\ \hline
$5$&$0.0525<\zeta_1<0.055$ & $\zeta_1 =0.0543974\pm 0.0000068 $\\ \hline
$7$&$0.05<\zeta_1<0.0525$   & $\zeta_1 =0.0518753^{+0.0000324}_{-0.0000320}$\\ \hline
$9$&$0.0475<\zeta_1<0.05$   & $\zeta_1 =0.0496673^{+0.0000700}_{-0.0000677}$\\ \hline
\end{tabular}
\end{center}
\caption{
Bounds and estimate of first zeroes of the partition function for $(k,\zeta_2 )=(1,2)$.
}
\label{tab:zeta2}
\end{table}
\begin{table}[t]
\begin{center}
\begin{tabular}{|c|c|c|}
\hline
$N$& Bounds on the zeroes  & Estimate of the zeroes \\ \hline
$2$&$\zeta_1=0.025$   & $\zeta_1=0.025$ \\ \hline
$4$&$0.0023<\zeta_1<0.024$ & $\zeta_1 =0.0238516 \pm 0.0000102$\\ \hline
$5$&$0.023<\zeta_1<0.024$   & $\zeta_1 =0.023177 \pm 0.000007$ \\ \hline
$7$&$0.022<\zeta_1<0.023$   & $\zeta_1 =0.022638^{+0.000042}_{-0.000039}$\\ \hline
$9$&$0.021<\zeta_1<0.023$   & $\zeta_1 =0.0218204^{+0.0000247}_{-0.0000241} $ \\ \hline
\end{tabular}
\end{center}
\caption{
Bounds and estimate of first zeroes of $Z$ for $(k,\zeta_2 )=(1,5)$.
}
\label{tab:zeta5}
\end{table}
\begin{table}[t]
\begin{center}
\begin{tabular}{|c|c|c|c|c|}
\hline
$N$&$\zeta_2$&$\zeta_1$&$Z_{\text{MC}}$&   Errors               \\ \hline
$4$&$1$      &$0.12$   &$-0.00242055$      &$7.70257\times 10^{-6}$\\ \cline{2-5}
   &$2$      &$0.0625$ &$-0.00536328$      &$0.0000115859$         \\ \cline{2-5}
   &$5$      &$0.025$  &$-0.00206855$      &$0.000035207$          \\
    \hline\hline
$5$&$1$      &$0.115$  &$-0.000807839$     &$7.76448\times 10^{-6}$\\ \cline{2-5}
   &$2$      &$0.06$   &$-0.00579732$      &$0.0000125459$       \\ \cline{2-5}
   &$5$      &$0.025$  &$-0.00481262$      &$0.000036708$        \\ \hline\hline
$7$&$1$      &$0.105$  &$-0.0000473376$    &$7.52208\times 10^{-6}$\\ \cline{2-5}
   &$2$      &$0.055$  &$-0.000501799$     &$0.0000102382$         \\ \cline{2-5}
   &$5$      &$0.035$  &$-0.00411775$      &$0.000018267$          \\ \hline\hline
$9$&$1$      &$0.1$ &$-0.0000167033$     &$2.0722\times 10^{-6}$         \\ \cline{2-5}
   &$2$      &$0.0525$ &$-0.000239292$     &$0.0000109368$         \\ \cline{2-5}
   &$5$      &$0.035$  &$-0.00176051$      &$0.0000142852$         \\ \hline
\end{tabular}
\end{center}
\caption{
$Z_{\rm MC}$ at the first negative peaks and their statistical errors for various $(N,k,\zeta_2 )$.
}
\label{negativeextrema}
\end{table}

\subsection{Physical origins of the zeroes and Fermi gas formalism}
\label{sec:origin}
In this subsection we discuss physical origins of the zeroes of the partition function.
For this purpose, we apply Fermi gas formalism and identify which effects trigger the change of the sign of $Z$.
Note that some techniques in the Fermi gas formalism are not available for $\zeta_1 \neq 0$ since the Hamiltonian is not hermitian.
However, there is a technique which is still available.
This is a formal $\hbar$-expansion of ${\rm Tr}{\widehat\rho}^n$ via Wigner transformation where ${\rm Tr}{\widehat \rho}^n$ is expressed as a phase space integral of a function whose explicit representation can be obtained by acting differential operators on $\rho (q,p)$.
In this technique, it does not matter whether or not the Hamiltonian is hermitian since the problem is reduced to compute a perturbative series of the explicit two dimensional integral with respect to $\hbar$.
Fortunately, this analysis has been already done in \cite{Nosaka} for imaginary $(\zeta_1 ,\zeta_2 )$ in the context of the R-charge deformation and hence we can obtain the $\hbar$-expansion simply by analytic continuation of the result in \cite{Nosaka} up to a subtlety discussed below.\footnote{
This analysis was done in sec.~4 of \cite{Nosaka}.
The result in our notation can be obtained by taking $p\rightarrow 1$, $q\rightarrow 1$, $\xi \rightarrow \frac{4i}{k}\zeta_1$ and $\eta \rightarrow \frac{4i}{k}\zeta_2$.
}
Once we find ${\rm Tr}{\widehat \rho}^n$ approximated in this way, one can compute the grand potential $J(\mu )$ by the following Mellin-Barnes expression
\begin{\eq}
J(\mu )
=-\int_{\epsilon -i\infty}^{\epsilon +i\infty} \frac{dt}{2\pi i}
 \Gamma (t) \Gamma(-t) \mathcal{Z}(t) e^{t\mu} 
\quad (0<\epsilon <1 ),
\end{\eq}
where $\mathcal{Z}(t) = {\rm Tr}{\widehat \rho}^{t}$ and the canonical partition function can be obtained from $J(\mu )$ by
\begin{\eq}
Z(N) =\int d\mu\ e^{J(\mu ) -\mu N } .
\end{\eq}
The $\hbar$-expansion of $\mathcal{Z}(n)$ takes the form
\begin{\eq}
\mathcal{Z}(n) 
= \sum_{s=0}^\infty \hbar^{2s-1} \mathcal{Z}_{2s}(n)
+\mathcal{O}(e^{-\frac{\sharp}{\hbar}})
\label{eq:small_h}
\end{\eq}
where the second term denotes non-perturbative effects of the $\hbar$-expansion which we are ignoring.
The work \cite{Nosaka} computed the first four coefficients $\mathcal{Z}_0$, $\mathcal{Z}_2$, $\mathcal{Z}_4$ and $\mathcal{Z}_6$ which are explicitly written down in app.~A of \cite{Nosaka}.
For example, the leading order coefficient $\mathcal{Z}_0$ is given by
\begin{\eq}
\mathcal{Z}_0 (n)
=\frac{1}{2\pi} B\Bigl[ \frac{1+4i\zeta_1 /k}{2}n ,\frac{1-4i\zeta_1 /k}{2}n \Bigr]
   B\Bigl[ \frac{1+4i\zeta_2 /k}{2}n ,\frac{1-4i\zeta_2 /k}{2}n \Bigr] ,
\end{\eq}
where we are keeping $(\zeta_1 /k ,\zeta_2 /k)$ fixed and
\begin{\eq}
B(x,y) = \frac{\Gamma (x) \Gamma (y)}{\Gamma (x+y)} .
\end{\eq}
The large-$N$ behavior of $Z(N)$ can be easily derived by the large-$\mu$ behavior of $J(\mu )$ which has the following structure
\begin{\eq}
J(\mu ) 
=J^{\rm pert}(\mu ) +\mathcal{O}
\left( e^{-\frac{2\mu}{1 \pm 4i\zeta_1 /k}} ,
e^{-\frac{2\mu}{1 \pm 4i\zeta_2 /k}},e^{-\mu} \right)
+\mathcal{O}\left( e^{-\frac{\sharp}{\hbar}\mu}\right) ,
\label{eq:J}
\end{\eq}
where
\begin{\eq}
J_{\rm pert}(\mu )
= \frac{C(\zeta_1 ,\zeta_2 ,k)}{3}\mu^3 +B(\zeta_1 ,\zeta_2 ,k)\mu
 +A(\zeta_1 ,\zeta_2 ,k) .
\label{eq:large_mu}
\end{\eq}
Several comments are in order.
First, the $\hbar$-expansions for the coefficients $C$ and $B$ are terminated at leading and sub-leading orders respectively:
\begin{\eqa}
&& C= \frac{2}{\pi^2 k(1+16\zeta_1^2 /k^2)(1+16\zeta_2^2 /k^2)} ,\NN\\
&& B=\frac{\pi^2 C}{3} 
-\frac{1}{6k}\left( \frac{1}{1+16\zeta_1^2 /k^2} +\frac{1}{1+16\zeta_2^2 /k^2} \right)
+\frac{k}{24} .
\end{\eqa}
The coefficient $A$ receives all order corrections and it has been conjectured in \cite{Nosaka} that the exact answer for $A$ is given by
\begin{\eq}
A
=\frac{1}{4}\Biggl[ 
 A_{\rm ABJM}(  k +4i\zeta_1 ) +A_{\rm ABJM}(  k -4i\zeta_1 )
+A_{\rm ABJM}(  k +4i\zeta_2 ) +A_{\rm ABJM}(  k -4i\zeta_2 )
\Biggr] ,
\end{\eq}
where \cite{KEK,Hatsuda:2014vsa}
\begin{align}
A_\text{ABJM}(k)
=\frac{2\zeta(3)}{\pi^2k}\Bigl(1-\frac{k^3}{16}\Bigr)+\frac{k^2}{\pi^2}\int_0^\infty dx\frac{x}{e^{kx}-1}\log(1-e^{-2x}) .
\end{align}
If the approximation by $J_{\rm pert}(\mu )$ is reliable, then the canonical partition function is approximated by
\begin{align}
Z \simeq Z_{\rm pert} ,\quad
Z_{\rm pert}
=\int d\mu\ e^{J_{\rm pert}(\mu ) -\mu N }
=e^A C^{-\frac{1}{3}}\Ai\left[ C^{-\frac{1}{3}}(N-B)\right] .
\label{eq:Zpert}
\end{align}
The large-$N$ limit of this formula  exhibits the $N^{3/2}$-law:\footnote{
In the large-$N$ limit, the $\mu$-integral is dominated by $\mu =\sqrt{\frac{N-B}{C}}$.
Therefore the non-perturbative effects in \eqref{eq:J} contribute to $Z$ like $\sim \mathcal{O}(e^{-\sqrt{kN}})$, $\mathcal{O}(e^{-\sqrt{N/k}})$.
}
\begin{\eq}
-\log{Z} 
= \frac{2}{3}C^{-1/2} N^{3/2} +\mathcal{O}(N^{1/2})
= \frac{\pi \sqrt{2k}}{3}
\sqrt{\left( 1+\frac{16\zeta_1^2}{k^2}\right)
\left( 1+\frac{16\zeta_2^2}{k^2}\right)} N^{3/2} 
+\mathcal{O}(N^{1/2}) ,
\label{eq:FlargeN}
\end{\eq}
which agrees with \eqref{Zinsaddle} for $\zeta_1 =0$ and the result of \cite{NST2} for $\zeta_1 =\zeta_2 =\zeta$.

The second term in \eqref{eq:J} is non-perturbative corrections of the large-$\mu$ expansion whose exponents can be explicitly derived by the $\hbar$-expansion \eqref{eq:small_h}.
These corrections for the massless case have been identified with membrane instanton effects whose type IIA picture is D2-branes wrapping (warped) $\mathbb{RP}^3$ in $AdS_4 \times \mathbb{CP}^3$ \cite{Drukker:2011zy}.
The third term in \eqref{eq:J} takes over the non-perturbative correction of the $\hbar$-expansion in \eqref{eq:small_h} whose exponents cannot be determined by the above arguments.
It has been conjectured in \cite{Nosaka} that the exponent for imaginary $(\zeta_1 ,\zeta_2 )$ is given by $\mathcal{O}(e^{-\frac{4\mu}{k(1\pm 4i\zeta_1 /k)(1\pm 4i\zeta_2 /k)}} )$.
These corrections for the massless case have been identified with worldsheet instanton effects coming from fundamental strings wrapping $\mathbb{CP}^1$ \cite{Cagnazzo:2009zh}. 

Let us estimate when we can trust the approximation by the perturbative part $J_{\rm pert}(\mu )$ in the large-$\mu$ expansion \eqref{eq:large_mu}, or equivalently when the canonical partition function is well approximated by \eqref{eq:Zpert}.
We easily see that the second term in \eqref{eq:large_mu} is exponentially suppressed for any $(\zeta_1 ,\zeta_2 )$ and therefore we can ignore the second term in the large-$N$ limit.
Then let us focus on the third term which comes from non-perturbative effects of the $\hbar$-expansion.
We have not estimated the exponent of the third term for real $(\zeta_1 ,\zeta_2 )$ precisely.
However, the exponent for real $(\zeta_1 ,\zeta_2 )$ should be the same as the naive analytic continuation of the one for imaginary $(\zeta_1 ,\zeta_2 )$ in the domain where the partition function is holomorphic with respect to $(\zeta_1 ,\zeta_2 )$.
Therefore, if we assume the above conjecture on the exponent for imaginary $(\zeta_1 ,\zeta_2 )$ in \cite{Nosaka}, then we should have the following correction in \eqref{eq:large_mu} for real $(\zeta_1 ,\zeta_2 )$: 
\begin{\eq}
\mathcal{O}\Bigl( e^{-\frac{4\mu}{k(1\pm 4i\zeta_1 /k)(1\pm 4i\zeta_2 /k)}} \Bigr)
= \mathcal{O}\Bigl( e^{-\frac{4\mu}{k(1 +16\zeta_1^2 /k^2)(1 +16\zeta_2^2 /k^2)}
\left[ 1-\frac{16\zeta_1 \zeta_2}{k^2} \mp \frac{4i(\zeta_1 +\zeta_2 )}{k} \right] } \Bigr) .
\label{eq:NPhbar}
\end{\eq}
Note that this correction is no longer exponentially suppressed for 
\begin{\eq}
\frac{\zeta_1 \zeta_2}{k^2}\geq \frac{1}{16},
\end{\eq}
and we cannot approximate the grand potential $J(\mu )$ by $J_{\rm pert}(\mu )$ in this region.
This also implies that the holomorphy of the partition function with respect to $(\zeta_1 ,\zeta_2 )$ is broken at $\frac{16\zeta_1 \zeta_2}{k^2} = \frac{1}{4}$ because if we start with imaginary $(\zeta_1 ,\zeta_2 )$, then the naive analytic continuation to real $(\zeta_1 ,\zeta_2 )$ does not commute with the large-$N$ limit.
Namely, if we take the large-$N$ limit first, then the free energy behaves as $\sim N^{3/2}$ and its continuation to real $(\zeta_1 ,\zeta_2 )$ is also described by the same formula for any $(\zeta_1 ,\zeta_2 )$ which is very likely different from the large-$N$ limit after the continuation in the domain $\frac{16\zeta_1 \zeta_2}{k^2}\geq \frac{1}{4}$.
\begin{figure}[t]
\begin{center}
\includegraphics[width=7.8cm]{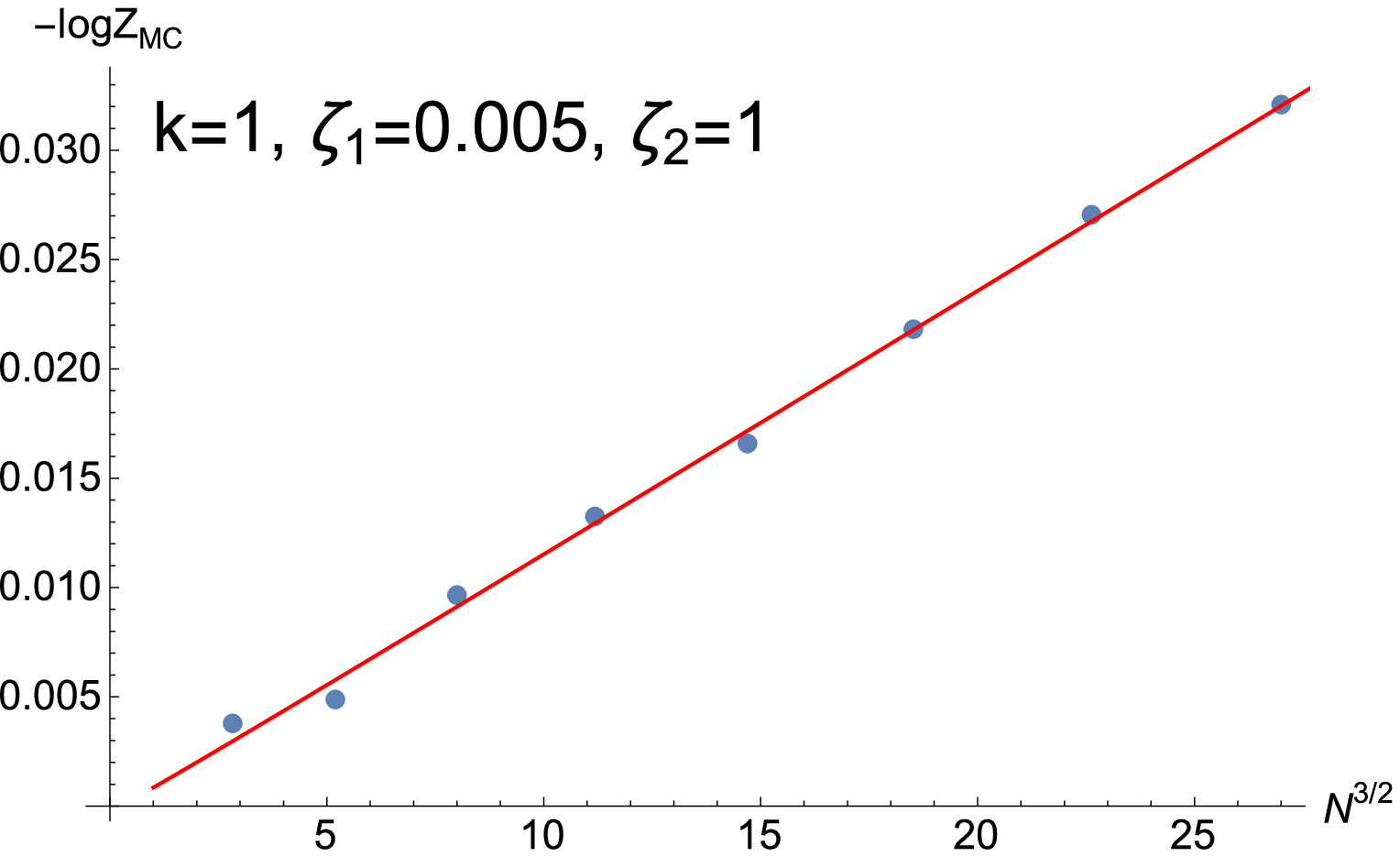}
\includegraphics[width=7.8cm]{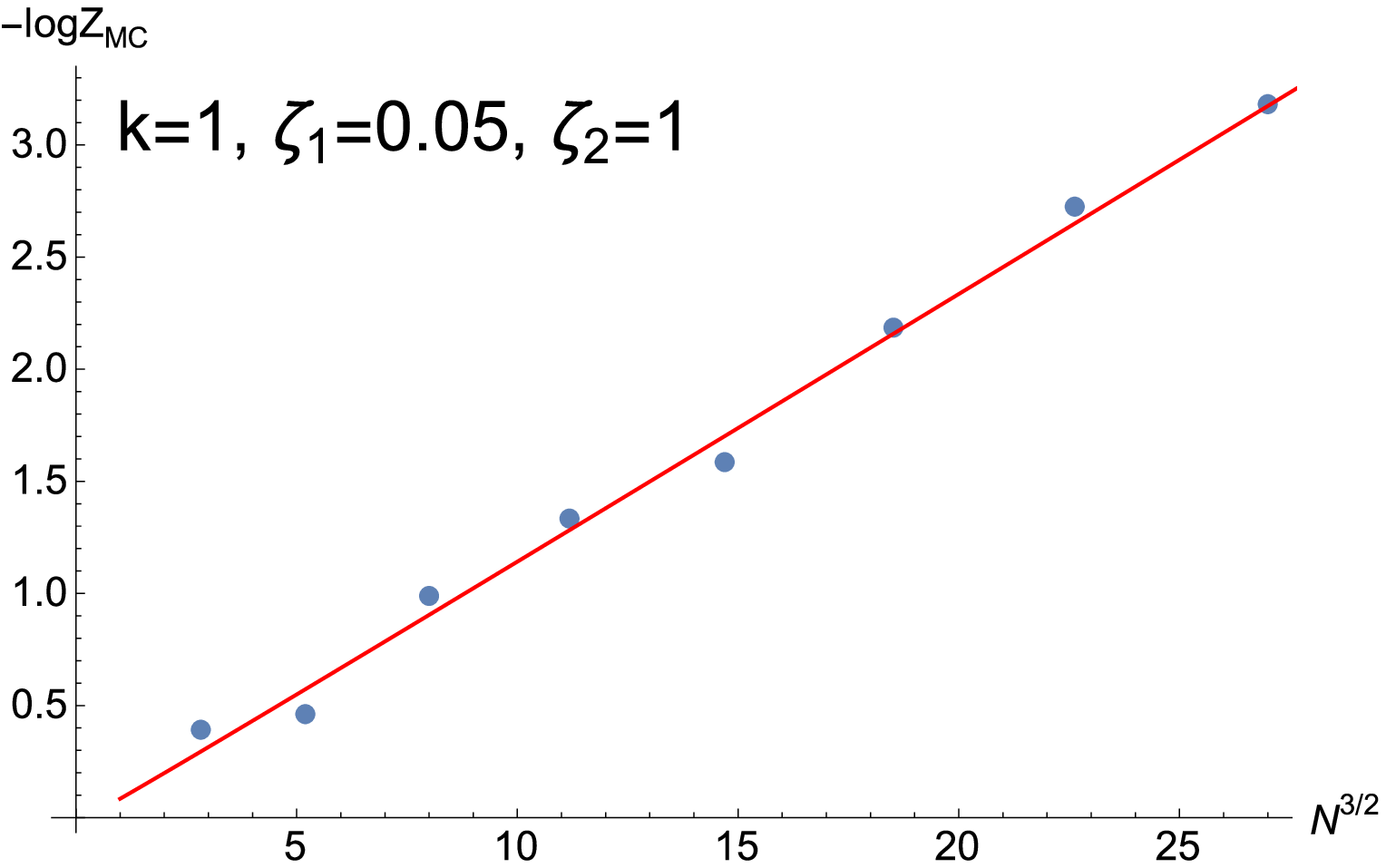}\\
\includegraphics[width=7.8cm]{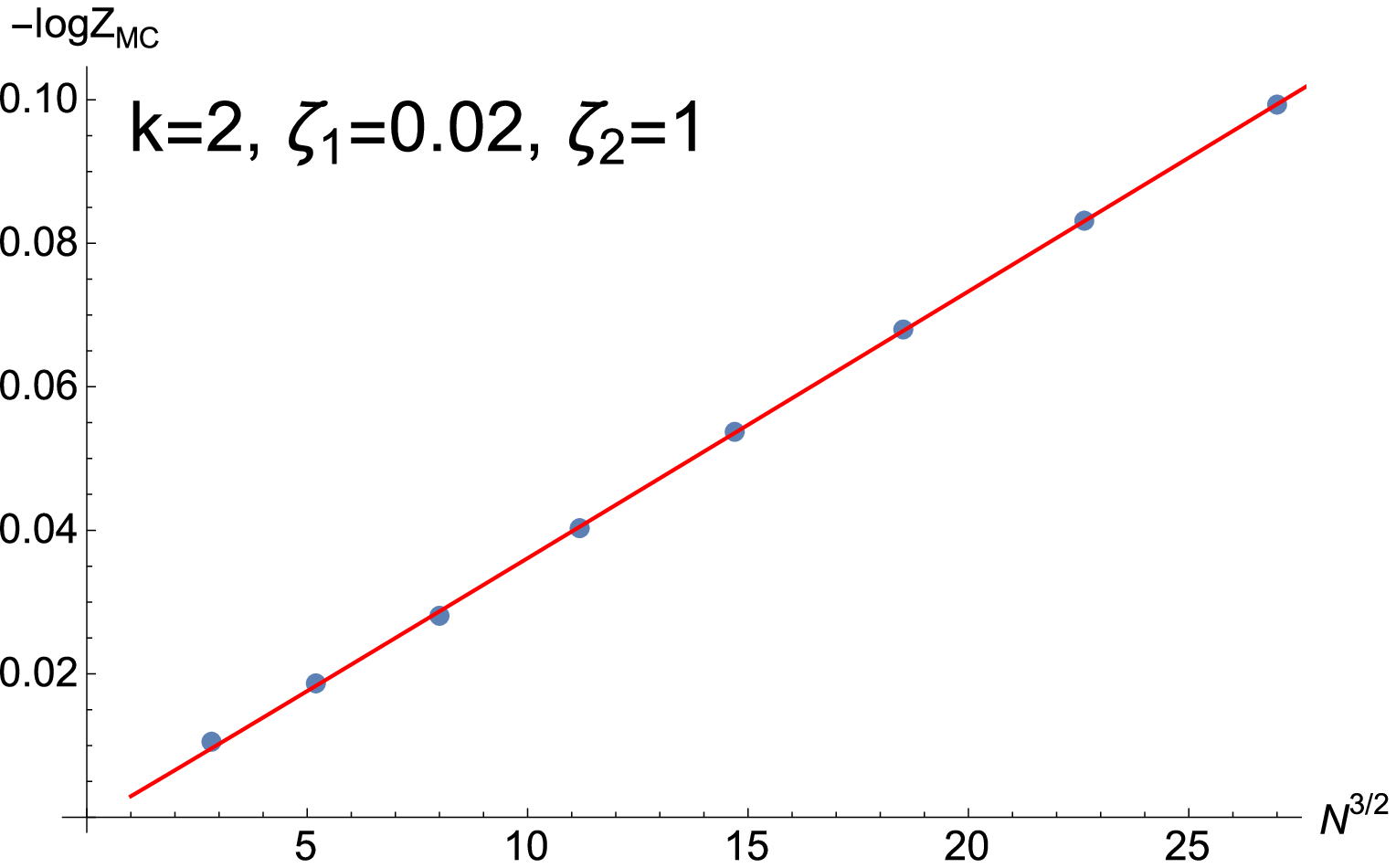}
\includegraphics[width=7.8cm]{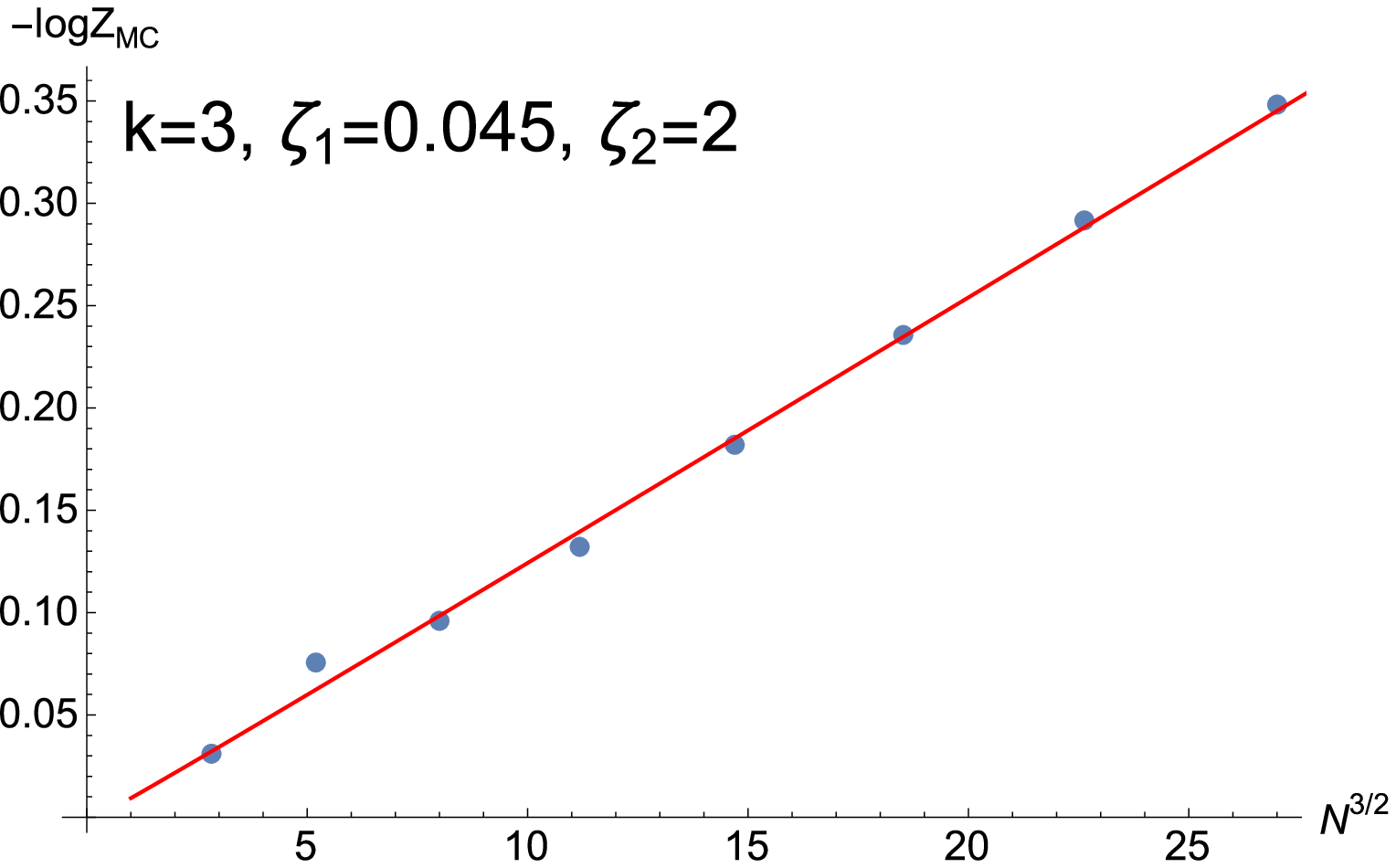}
\caption{
The quantity $-\log{Z_{\rm MC}}$ with $Z_{\rm MC}(N,k,\zeta_1 ,\zeta_2 )$  $=$ $\frac{Z(N,k,\zeta_1,\zeta_2)}{Z(N,k,0,\zeta_2)}$ is plotted against $N^{3/2}$ for $\frac{\zeta_1 \zeta_2}{k^2} <\frac{1}{16}$.
The symbols are the numerical results obtained by the Monte Carlo simulation.
The red line denotes the result computed by the Airy function formula \eqref{eq:Zpert}, namely $-\log{\frac{Z_{\rm pert}(N,k,\zeta_1,\zeta_2)}{Z_{\rm pert}(N,k,0,\zeta_2)} }$.
\label{fig:VS_Airy}
}
\end{center}
\end{figure}

\begin{figure}[t]
\begin{center}
\includegraphics[width=7.8cm]{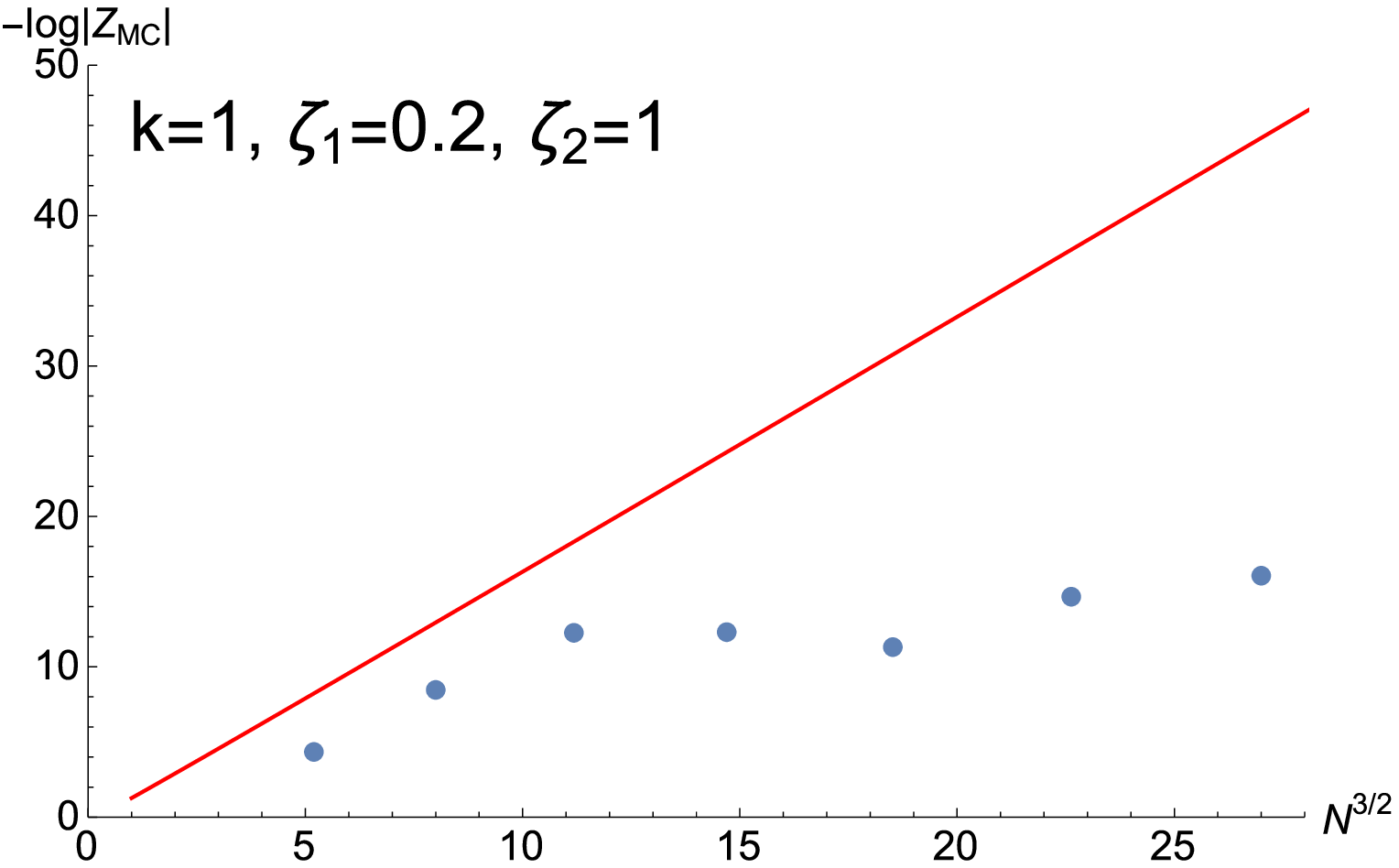}
\includegraphics[width=7.8cm]{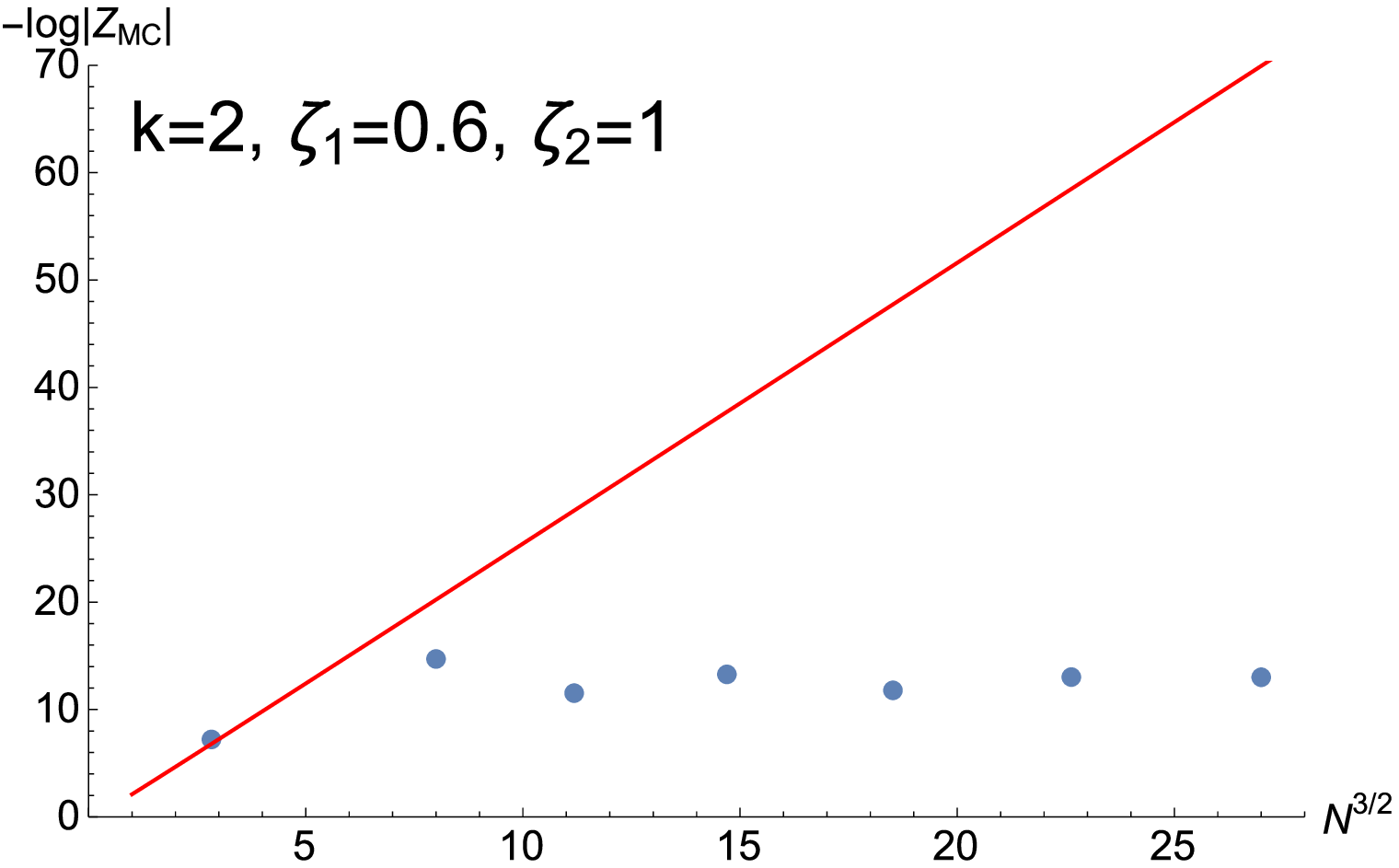}
\caption{
Similar plots to fig.~\ref{fig:VS_Airy} for $\frac{\zeta_1 \zeta_2}{k^2} \geq \frac{1}{16}$.
\label{fig:VS_Airy2}
}
\end{center}
\end{figure}

The above estimate is consistent with our numerical results obtained in sec.~\ref{sec:MC}.
In fig.~\ref{fig:VS_Airy} we compare the ratio $Z_{\rm MC}(N,k,\zeta_1 ,\zeta_2 )$  $=$ $\frac{Z(N,k,\zeta_1,\zeta_2)}{Z(N,k,0,\zeta_2)}$ obtained by the Monte Carlo simulation with the one computed by the approximation \eqref{eq:Zpert} for some cases with $\frac{\zeta_1 \zeta_2}{k^2} <\frac{1}{16}$ where we expect \eqref{eq:Zpert} to be good approximation.
The plots show that our numerical results agree with the Airy function formula \eqref{eq:Zpert} and exhibit the $N^{3/2}$-law.
Although we explicitly present only the four cases, we have observed similar behaviors for various other values of $(k,\zeta_1 ,\zeta_2 )$ satisfying $\frac{\zeta_1 \zeta_2}{k^2} <\frac{1}{16}$.
Figure \ref{fig:VS_Airy2} shows similar plots for $\frac{\zeta_1 \zeta_2}{k^2} \geq \frac{1}{16}$ where we expect that we cannot trust \eqref{eq:Zpert} due to the correction \eqref{eq:NPhbar}.
In contrast to fig.~\ref{fig:VS_Airy}, we easily see that the numerical results do not agree with \eqref{eq:Zpert} and no longer exhibit the $N^{3/2}$-law.
We have also found similar behaviors for various other values of $(k,\zeta_1 ,\zeta_2 )$ with $\frac{\zeta_1 \zeta_2}{k^2} \geq \frac{1}{16}$.
Thus our numerical results support our expectation that the approximation by \eqref{eq:Zpert} is valid for $\frac{\zeta_1 \zeta_2}{k^2} < \frac{1}{16}$.

\subsection{Conjecture on phase structure in the large-$N$ limit}
\label{conjlargeN}
We discuss the phase structure of the mass deformed ABJM theory in the large-$N$ limit.
Let us first recall the results obtained so far by the various analyzes:
\begin{itemize}
\item In the case of $\zeta_1=\zeta_2=\zeta$, the partition function in the representation \eqref{Original} has been analyzed by the saddle point method in \cite{NST2}
as reviewed in sec.~\ref{susybreaking}.
The saddle point configuration in \cite{NST2} realizes the $\mathcal{O}(N^{3/2})$ free energy and becomes singular at $\zeta/k=1/4$.

\item In sec.~\ref{saddle}, we have constructed the saddle point solution for $\zeta_1 =0$ in the $S$-dual representation, which gives the $\mathcal{O}(N^{3/2})$ free energy \eqref{Zinsaddle}.
This behavior is consistent with the exact results for finite $N$ obtained in sec.~\ref{TWPY}.
Note also that we can obtain the result for $\zeta_1 \neq 0$, $\zeta_2 =0$ by the replacement $\zeta_2 \rightarrow \zeta_1$ in \eqref{Zinsaddle} since the partition function is symmetric under $\zeta_1 \leftrightarrow \zeta_2$.

\item In sec.~\ref{sec:exactN2}, we have written down the exact results for $N=1,2$ and arbitrary $(k,\zeta_1 ,\zeta_2)$ obtained in \cite{RS}.
It has turned out that the partition function for $N=2$ has the zeroes at finite $(\zeta_1 ,\zeta_2)$ while the one for $N=1$ does not.

\item In sec.~\ref{sec:MC}, we have performed the Monte Carlo simulation for higher $N$.
We have observed that the partition function has the zeroes at finite $(\zeta_1 ,\zeta_2 )$ given $(N,k)$.
The bounds and estimates on the zeroes given in tables \ref{tab:zeta1}, \ref{tab:zeta2} and \ref{tab:zeta5}, imply that the first zeroes do not increase by $N$.
It is natural to expect that the partition function becomes zero at some finite $(\zeta_1, \zeta_2 )$ also in the large-$N$ limit.

\item In sec.~\ref{sec:origin}, we have argued when one can trust the approximation in terms of the perturbative grand potential \eqref{eq:large_mu} in the Fermi gas formalism, which gives the $\mathcal{O}(N^{3/2})$ free energy in the large-$N$ limit.
We have found that the approximation is reliable for $\frac{\zeta_1 \zeta_2}{k^2}<\frac{1}{16}$ while in the other regime $\frac{\zeta_1 \zeta_2}{k^2}\geq \frac{1}{16}$, the expected non-perturbative effects \eqref{eq:NPhbar} of the $\hbar$-expansion are no longer exponentially suppressed.
Note that for $\zeta_1=\zeta_2=\zeta$, the approximation starts to be invalid at $\zeta /k =1/4$, which is the same as the condition that the saddle point in \cite{NST2} becomes singular.

\end{itemize}
Based on the above results, we propose the following scenario (see fig.~\ref{fig:proposal} for schematic picture):
\begin{itemize}
\item [(i)] For small $(\zeta_1,\zeta_2 )$, the large-$N$ free energy behaves as $F\sim N^{3/2}$ whose explicit form is given by \eqref{eq:FlargeN}.
We expect that this formula is valid for $\frac{\zeta_1 \zeta_2}{k^2}<\frac{1}{16}$, and becomes invalid for $\frac{\zeta_1 \zeta_2}{k^2}\geq \frac{1}{16}$.
\item [(ii)] The partition function vanishes at some finite values of $(\zeta_1,\zeta_2 )$.
We expect that this occurs at the boundary of the validity of \eqref{eq:FlargeN}: $\frac{\zeta_1 \zeta_2}{k^2}= \frac{1}{16}$.
We interpret this as the SUSY breaking at this point.
\end{itemize}
\begin{figure}[t]
\begin{center}
\includegraphics[width=8.2cm]{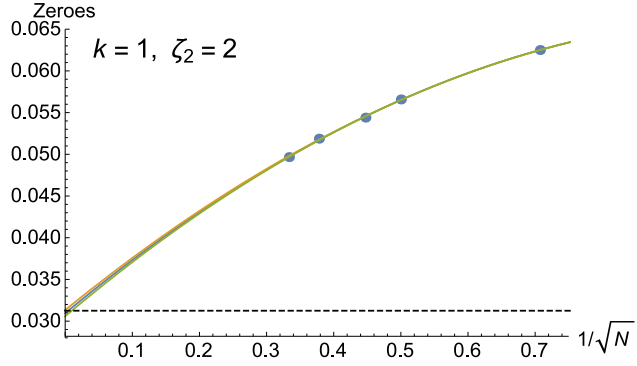}
\end{center}
\caption{
Fitting of the estimated zeroes for $(k,\zeta_2)=(1,2)$ given in table \ref{tab:zeta2} by the function $a(k,\zeta_2 )+\frac{b(k,\zeta_2 )}{\sqrt{N}}+\frac{c(k,\zeta_2 )}{N}$.
The three thick lines denote the fitting functions of the estimates from the interpolating functions of $Z_{\rm MC}$ and $Z_{\rm MC}$ plus/minus the errors.
The black dashed line denotes our expectation on the zero in the large-$N$ limit: $\zeta_1 =\frac{k^2}{16 \zeta_2}$, which is $\frac{1}{32}=0.03125$ for $(k,\zeta_2)=(1,2)$.
The result of the fitting is $a(k=1,\zeta_2 =2 ) = 0.0309507^{+0.0003777}_{-0.0003645}$.
}
\label{fig:fit}
\end{figure}

We already have strong evidence of the first point by the saddle point analysis for $\zeta_1 =\zeta_2$ in \cite{NST2} and Fermi gas analysis in sec.~\ref{sec:origin}.
Now we provide further evidence for the second point.
From tables \ref{tab:zeta1}, \ref{tab:zeta2} and \ref{tab:zeta5}, we observe that the locations of the first zeroes decrease slowly as $N$ increases.
Therefore it is plausible that the first zeroes in the large-$N$ limit are at some finite values of $(\zeta_1, \zeta_2 )$.
It would be nontrivial whether or not the first zeroes in the large-$N$ limit coincide with our expectation $\frac{\zeta_1 \zeta_2}{k^2}=\frac{1}{16}$.
We perform consistency checks of this by fitting analysis of our numerical data.
Note that the fitting analysis in the current situation is subtle in the following two reasons.
First we do not know asymptotic behaviors of the first zeroes for large-$N$.
In other words, we do not know what appropriate fitting functions are a priori.
Second, we do not have sufficient data of the first zeroes since it is only for the five values of $N$ ($N=2,3,5,7,9$).  
Nevertheless the fitting analysis provides quite nontrivial consistency checks as we will see soon.
We have constructed fitting functions for the first zeroes $\zeta_1 (N, k,\zeta_2 )$ with fixed $(k,\zeta_2 )$ and varied $N$.
As a conclusion, we have found that when we find a fitting function nicely interpolating numerical data, asymptotic value of the fitting function at $N\rightarrow\infty$ agrees with our expectation $\zeta_1 =\frac{k^2}{16\zeta_2}$.
\begin{figure}[t]
\begin{center}
\includegraphics[width=7.0cm]{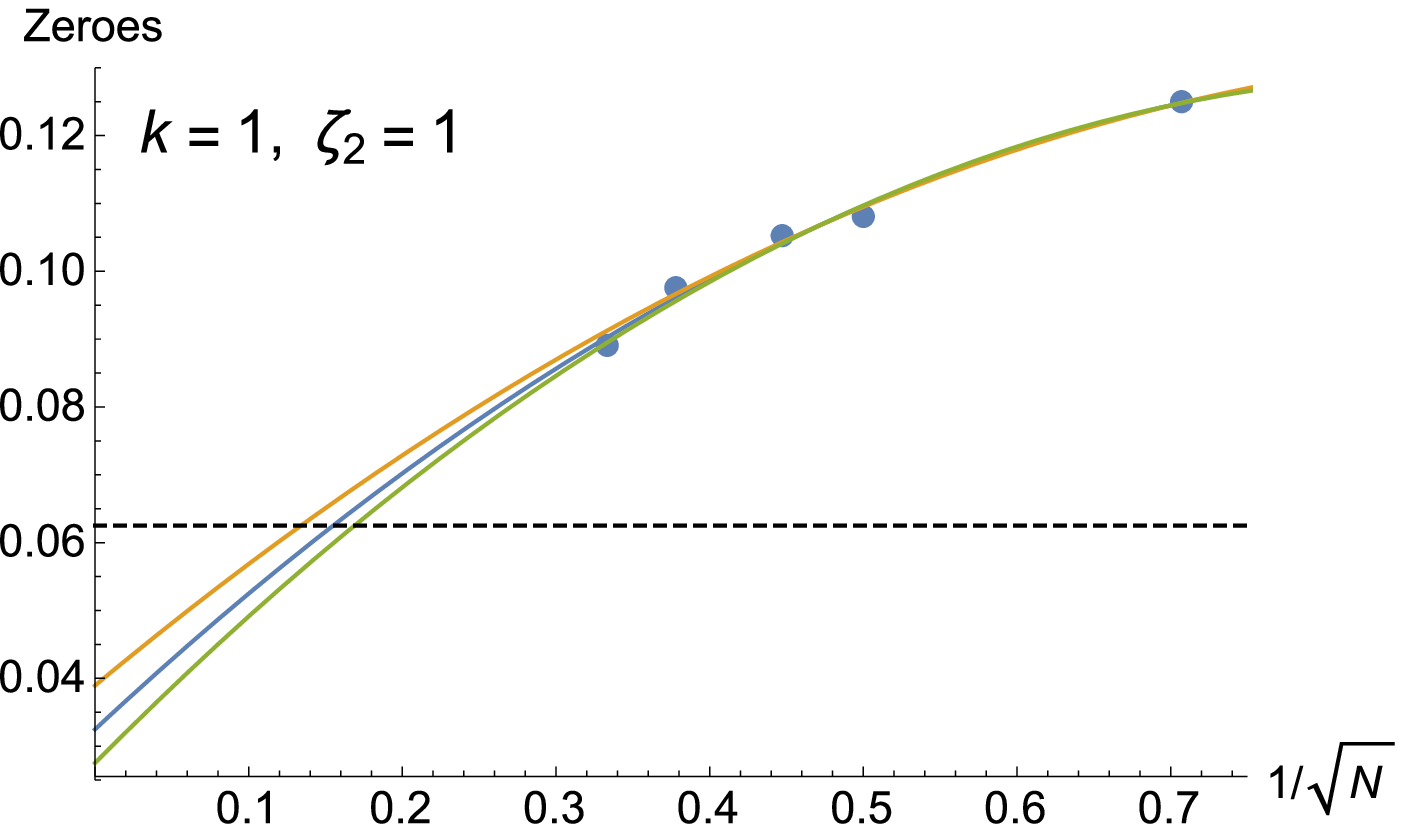}
\includegraphics[width=7.0cm]{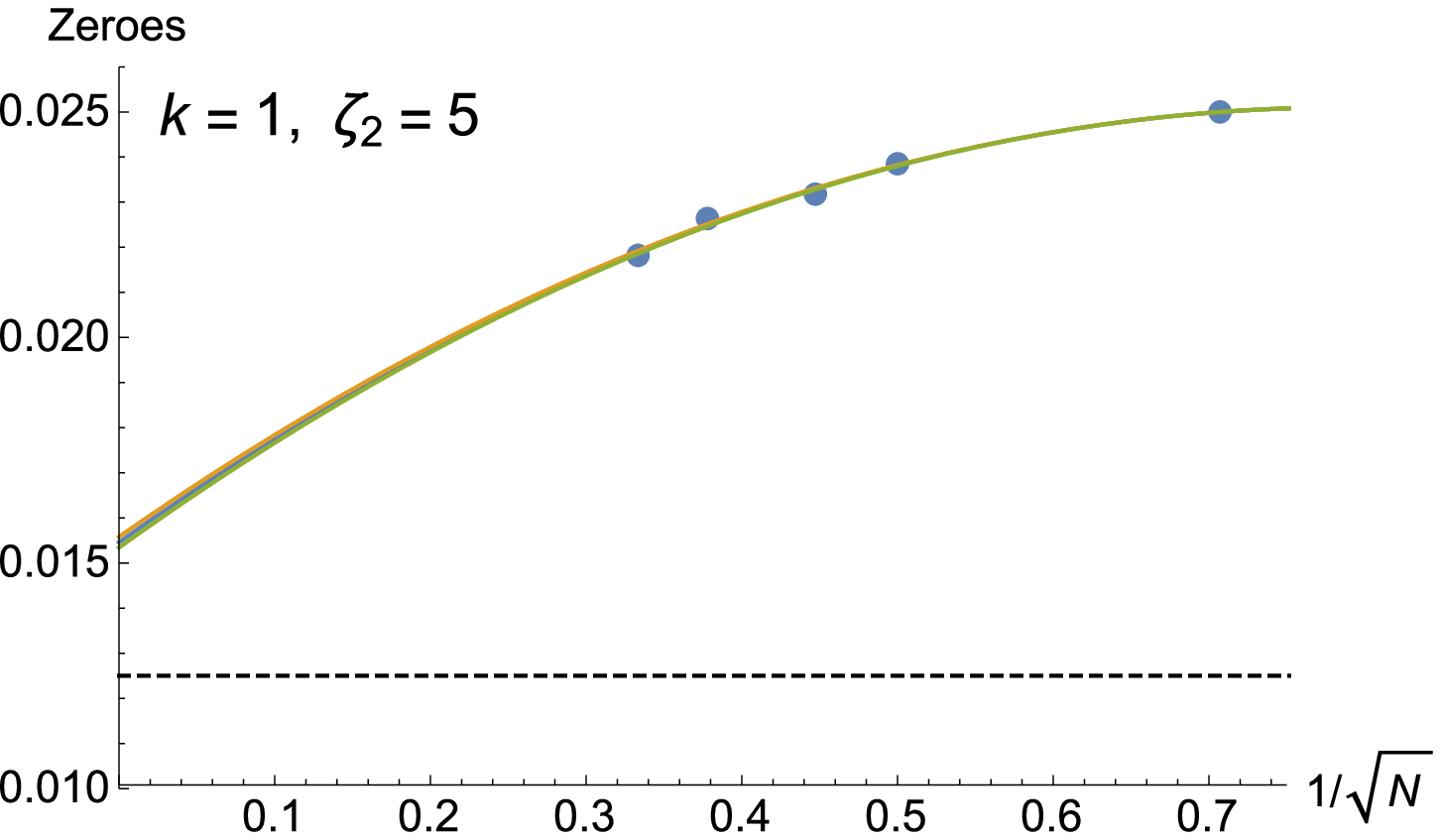}
\end{center}
\caption{
Similar plots to fig.~\ref{fig:fit} for $(k,\zeta_2)=(1,1)$ and $(1,5)$.
The intercepts of the fitting functions are $a(k=1,\zeta_2 =1 )$ $=$ $0.0620729^{+0.0014133}_{-0.0011213}$ and $a(k=1,\zeta_2 =5 )$ $=$ $0.0194694^{+0.0000579}_{-0.0000551}$.
}
\label{fig:fit2}
\end{figure}

\begin{figure}[t]
\begin{center}
\includegraphics[width=7.0cm]{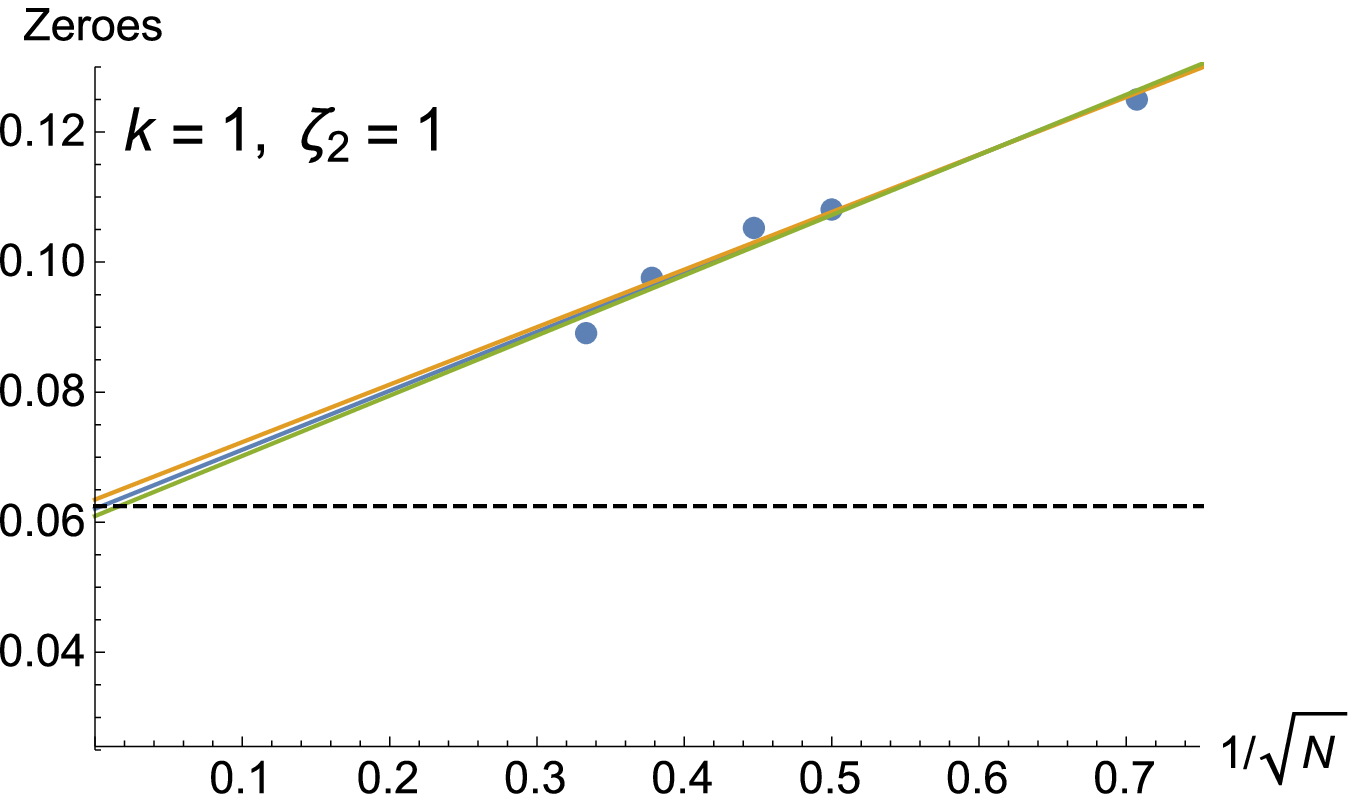}
\includegraphics[width=7.0cm]{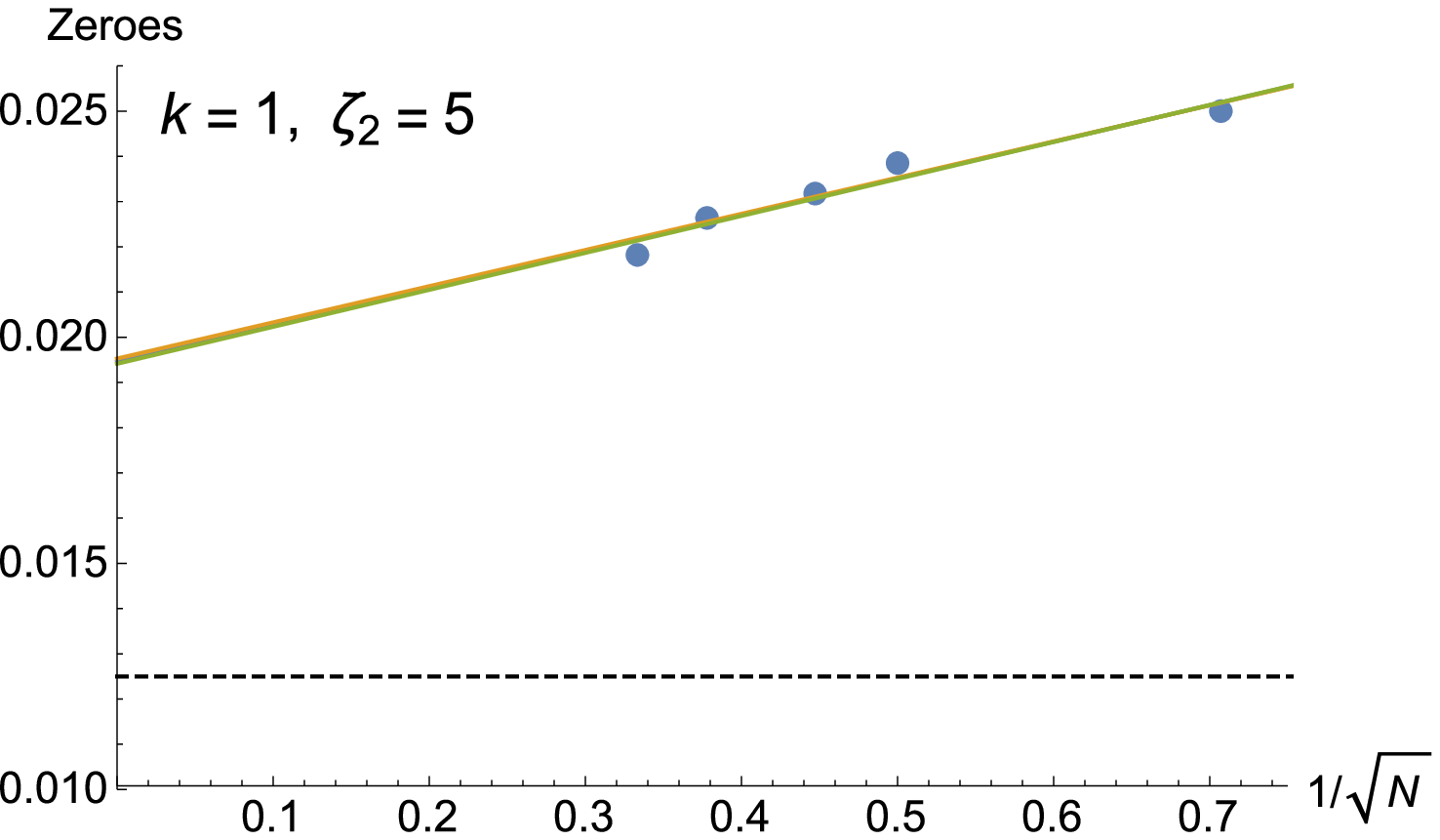}
\end{center}
\caption{
Similar analysis to fig.~\ref{fig:fit2} by using the different fitting function
$a(k,\zeta_2 )+\frac{b(k,\zeta_2 )}{\sqrt{N}}$.
The values of the intercepts are 
$a(k=1,\zeta_2 =1 )$ $=$ $0.0577778^{+0.0038557}_{-0.0033802}$
and $a(k=1,\zeta_2 =5 )$ $=$ $0.0154572^{+0.0001219}_{-0.000114526}$.
}
\label{fig:fit3}
\end{figure}

In fig.~\ref{fig:fit}, we plot the estimated zeroes for $(k,\zeta_2 )=(1,2)$ given in tab.~\ref{tab:zeta2} against $1/\sqrt{N}$.
We construct a fitting function for this data by the ansatz 
\begin{\eq}
a(k,\zeta_2 )+\frac{b(k,\zeta_2 )}{\sqrt{N}}+\frac{c(k,\zeta_2 )}{N} ,
\label{eq_fit1}
\end{\eq}
where $a(k,\zeta_2 )$ corresponds to the zero in the large-$N$ limit with respect to $\zeta_1$.
We easily see that the fitting function nicely interpolates the data points.
Therefore it is natural to compare the asymptotic value of the fitting function at large-$N$ with our expectation. 
As a result, we have found 
\begin{\eq}
a(k=1,\zeta_2 =2 ) = 0.0309507^{+0.0003777}_{-0.0003645} ,
\end{\eq}
which includes our expectation on the zero at large-$N$: 
$\zeta_1$ $=$ $\frac{k^2}{16 \zeta_2}|_{(k,\zeta_2)=(1,2)}$ $=$ $0.03125$.
This strongly supports our expectation on the large-$N$ phase structure.
Fig.~\ref{fig:fit2} shows results by the same fitting function \eqref{eq_fit1} for the other values of $\zeta_2$.
It is clear that the fitting functions \eqref{eq_fit1} for $(k,\zeta_2 )=(1,1)$ and $(1,5)$ do not nicely interpolate the data as much as for the case of $(k,\zeta_2 )=(1,2)$ although the fitting for $(k,\zeta_2 )=(1,5)$ is better than the one for $(k,\zeta_2 )=(1,1)$.
From the fitting functions, we have found $a(k=1,\zeta_2 =1 )$ $=$ $0.0577778^{+0.0038557}_{-0.0033802}$ and $a(k=1,\zeta_2 =5 )$ $=$ $0.0154572^{+0.0001219}_{-0.000114526}$ which do not include our expectation $\zeta_1 =\frac{k^2}{16 \zeta_2}$ although the result for $(k,\zeta_2 )=(1,5)$ is not far from the expectation.
We interpret that this does not mean invalidity of our expectation since the fitting ansatz \eqref{eq_fit1} does not exhibit very nice interpolations for $(k,\zeta_2 )=(1,1)$ and $(1,5)$, and we need another fitting functions or more data.
In fig.~\ref{fig:fit3} we also present similar plots to fig.~\ref{fig:fit2} by the different fitting function $a(k,\zeta_2 )+\frac{b(k,\zeta_2 )}{\sqrt{N}}$ for $(k,\zeta_2 )=(1,1)$ and $(1,5)$.
Again the fitting functions do not interpolate the data very nicely and have different intercepts from the ones obtained by the ansatz \eqref{eq_fit1} although the intercept for $(k,\zeta_2 )=(1,1)$ includes our expectation: $a(k=1,\zeta_2 =1 )$ $=$ $0.0577778^{+0.0038557}_{-0.0033802}$.
To summarize we need to find more appropriate fitting functions or data points for larger $N$ in order to further check our expectation except for $(k,\zeta_2 )=(1,2)$.
We leave this for future work.

Moreover, the correlation between the supersymmetry breaking and the singularity in the saddle point approximation was argued for the pure Chern-Simons theory \cite{MoNi}.
Hence it would be more than just a minimal scenario for our theory to relate the singularity in the saddle point approximation with the supersymmetry breaking.
It would be interesting to test this conjecture by studying the partition function for larger $N$ in future.

\section{Discussion}
\label{discuss}
In this paper we have studied the mass deformed ABJM theory on the three sphere.
Based on the argument in sec.~\ref{susybreaking}, we expect that this theory exhibits a spontaneous supersymmetry breaking in large-$N$ limit at $\zeta_1=\zeta_2=k/4$.
To gain an evidence for this conjecture we have analyzed the partition function of the mass deformed ABJM theory for finite $k$ and $N$ by using the Monte Carlo simulation.
As a result we have found that the partition function vanishes at some finite values of $\zeta_1,\zeta_2$.
The numerical results also indicate that the zeroes exist for general $N$, and that the locus of the first zero stays finite as $N$ increases.
These observations are consistent with the expectation in the end of sec.~\ref{susybreaking} from the large-$N$ supersymmetry breaking.
Our result would shed new light to the phase structure of the mass deformed ABJM theory in the M-theory limit, which was unclear in the previous works \cite{NST,NST2}.

Precisely speaking, the correct physical interpretation for the zeroes of the partition function for finite $N$ is not clear, since a spontaneous symmetry breaking can happens only in the limit of large degree of freedom.
To test our conjecture it is important to study the partition function in the large-$N$ limit.
One possible method would be the saddle point approximation.
In the previous work we found the saddle point solution only for the special case $\zeta_1=\zeta_2$.
In this paper we found a solution for a new slice $\zeta_1=0$ by rewriting the matrix model into the S-dual representation.

Another direction is to improve the algorithm of the numerical simulation.
In this paper, we have treated the oscillation factor of \eqref{Sdual} in the quite naive way where we just regard the factor as the observable in the system with $\zeta_1 =0$.
In this approach, we need much more statistics than simulations without oscillating factors so that the simulation at large-$N$ becomes harder. 
It is nice if one can find more appropriate algorithm such as complex Langevin method and Lefschetz thimble.

Lastly, it would be interesting to compare the exact partition functions for $\zeta_1=0$ with those in \cite{NY1}.
In that paper they computed the partition function of the $\text{U}(N)_k\times \text{U}(N+M)_{-k}$ linear quiver superconformal Chern-Simons theory.
Naively, for $M=0$ this theory can be obtained by taking the decoupling limit $\zeta_2\rightarrow\infty$ in the mass deformed ABJM theory.
Indeed we observe for $N<k/2$ that, if we take the limit $\zeta_2\rightarrow \infty$ in the exact expressions for the partition function $Z(N,k,0,\zeta_2)$ they precisely coincide with those in \cite{NY1} up to the contribution of decoupled hypermultiplet $e^{-2\pi N^2\zeta_2/k}$ (see app.~\ref{comarewithNY}).
On the other hand we also observe that for some cases with $N\ge k/2$ the decay is slower than $e^{-2\pi N^2\zeta_2/k}$.
Actually in these cases the corresponding linear quiver theory is a bad theory \cite{Nosaka:2018eip}, hence it should not be the right decoupling limit of the mass deformed ABJM theory.
Though it is still not clear, the correct description of the decoupling limit might be obtained by expanding the Coulomb branch moduli around a configuration which depends on $\zeta_2$ in a non-trivial way.

\subsection*{Acknowledgement}
We would like to thank Sungjay Lee, Sanefumi Moriyama and Shuichi Yokoyama for valuable discussions.
T.~N.~ would appreciate Jin-beom Bae, Joonho Kim, Takaya Miyamoto and Dario Rosa for several pieces of advice on numerical simulation.
The numerical analysis were performed on the ATOM server which is supported by Korea Institute for Advanced Study.
The work of S.~T.~ was supported by JSPS KAKENHI Grant Number 17K05414.
K.~S.~ is supported by JSPS fellowship.
K.~S.~ is also supported by Grant-in-Aid for JSPS Fellow No.~18J11714.

\appendix
\section{Partition function in S-dual representation}
\label{Sdualsection}
In this appendix we derive the S-dual representation \eqref{Sdual} of the partition function, which we practically use in the main text.
The computation is essentially the same as those in the Fermi gas formalism for the ABJM theory \cite{MP}.
Let us start with the integral \eqref{Original}.
Changing the integration variables as $\lambda\rightarrow \lambda /k+\pi (\zeta_1 +\zeta_2 )$ and ${\widetilde\lambda}\rightarrow{\widetilde\lambda}/k -\pi (\zeta_1 +\zeta_2 )$, we find
\begin{align}
Z=\frac{1}{(N!)^2}\int\frac{d^N\lambda}{(2\pi k)^N}\frac{d^N{\widetilde\lambda}}{(2\pi k)^N}e^{\frac{i}{4\pi k}\sum_i(\lambda_i^2-{\widetilde\lambda}_i^2)+\frac{i\zeta}{k}\sum_i(\lambda_i+{\widetilde\lambda}_i)}
\frac{
\prod_{i<j}^N(2\sinh\frac{\lambda_i-\lambda_j}{2k})^2
\prod_{i<j}^N(2\sinh\frac{{\widetilde\lambda}_i-{\widetilde\lambda}_j}{2k})^2
}{
\prod_{i,j=1}^N
\prod_\pm
2\cosh\frac{\lambda_i-{\widetilde\lambda}_j\pm \mu}{2k}
}.
\end{align}
where
\begin{align}
\zeta=\frac{\zeta_1+\zeta_2}{2},\quad
\mu=2\pi(\zeta_1-\zeta_2).
\end{align}
Next we rewrite the 1-loop determinant into pair of determinants of $N\times N$ matrices by using the Cauchy determinant formula
\begin{align}
\frac{
\prod_{i<j}^N2\sinh\frac{x_i-x_j}{2}
\prod_{i<j}^N2\sinh\frac{y_i-y_j}{2}
}{
\prod_{i,j=1}^N2\cosh\frac{x_i-y_j}{2}
}
=\det_{i,j}\frac{1}{2\cosh\frac{x_i-y_j}{2}},
\label{Cauchy}
\end{align}
and then combine them by using the formula
\begin{align}
\frac{1}{N!}\int d^Nx \det_{i,j}f_i(x_j) \det_{i,j}g_i(x_j)=\det_{i,j} \int dxf_i(x)g_j(x).
\label{MM}
\end{align}
After these manipulations we obtain the following expression for the partition function
\begin{align}
Z=\frac{1}{N!}\int\frac{d^N\lambda}{(2\pi)^N}\det_{i,j}f(\lambda_i,\lambda_j),
\end{align}
where
\begin{align}
f(x,y)=\int\frac{dz}{2\pi}e^{\frac{i}{4\pi k}x^2+\frac{i\zeta}{k}x}\frac{1}{2k\cosh\frac{x-z-\mu}{2k}}e^{-\frac{i}{4\pi k}z^2+\frac{i\zeta z}{k}}\frac{1}{2k\cosh\frac{z-y-\mu}{2k}}.
\end{align}
Hence the partition function takes the form of the partition function of 1d $N$ particle non-interacting Fermi gas if we regard $f(\lambda',\lambda'')$ as the matrix element of a one-particle density matrix $\langle \lambda|{\widehat\rho}|\lambda'\rangle$ with position eigenstates $|\cdot \rangle$
\begin{align}
Z=\frac{1}{N!}\int\frac{d^Nx}{(2\pi)^N}\det_{i,j}\langle x_i|{\widehat\rho}|x_j\rangle,
\label{Fermigas}
\end{align}
where\footnote{
See \eqref{eq:braket} for the notation for the 1d quantum mechanics.
}
\begin{align}
{\widehat\rho}=e^{\frac{i}{4\pi k}{\widehat q}^2+\frac{i\zeta}{k}{\widehat q}}\frac{e^{\frac{i\mu}{2\pi k}{\widehat p}}}{2\cosh\frac{\widehat p}{2}}e^{-\frac{i}{4\pi k}{\widehat q}^2+\frac{i\zeta}{k}{\widehat q}}\frac{e^{\frac{i\mu}{2\pi k}{\widehat p}}}{2\cosh\frac{\widehat p}{2}}.
\end{align}
Now it is obvious from \eqref{Fermigas} that the partition function is invariant under any similarity transformation of the density matrix ${\widehat\rho}$.
We can simplify ${\widehat\rho}$ by the similarity transformation ${\widehat\rho}\rightarrow e^{-\frac{i}{4\pi k}{\widehat p}^2-\frac{i\zeta}{k}{\widehat p}}{\widehat\rho}e^{\frac{i}{4\pi k}{\widehat p}^2+\frac{i\zeta}{k}{\widehat p}}$ as
\begin{align}
{\widehat\rho}=\frac{e^{\frac{2i\zeta_1}{k}{\widehat q}}}{2\cosh\frac{\widehat q}{2}}
\frac{e^{\frac{2i\zeta_2}{k}{\widehat p}}}{2\cosh\frac{\widehat p}{2}},
\label{rhosimplified}
\end{align}
whose matrix element is
\begin{align}
\langle x|{\widehat\rho}|y\rangle=\frac{e^{\frac{2i\zeta_1}{k}x}}{2\cosh\frac{x}{2}}\frac{1}{2k\cosh\frac{x-y+4\pi \zeta_2}{2k}}.
\label{rhosimplifiedME}
\end{align}
Applying the Cauchy determinant formula \eqref{Cauchy} reversely to $\det_{i,j}\langle x_i|{\widehat\rho}|x_j\rangle$ in the Fermi gas formalism \eqref{Fermigas} with this new ${\widehat\rho}$, we finally obtain the S-dual representation for the partition function \eqref{Sdual}.

\section{Technical details on exact computation of $Z(N,k,0,\zeta_2)$}
\label{app:detail_exact}
In this appendix we explain details on how to solve the recursion relation \eqref{recursive_int} for integer $k$.

\subsubsection*{\underline{Even $k$}}
\begin{figure}[t]
\begin{center}
\includegraphics[width=6cm]{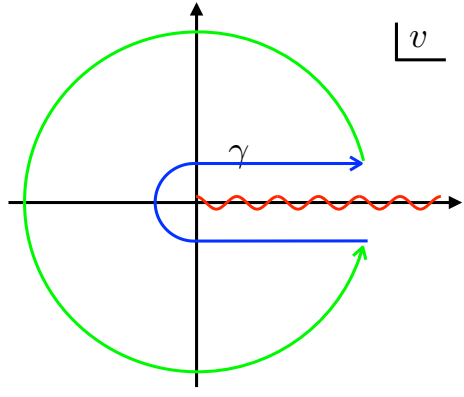}
\end{center}
\caption{The integration contour $\gamma$ in \eqref{TWPYRes} (blue) and the deformed contour to use the Cauchy theorem (green).
The cut of $\log^{(+)}$ is depicted by wavy red line.}
\label{TWPYpath}
\end{figure}

If $k\in 2\mathbb{N}$, we can introduce a new variable $u=e^{\frac{x}{k}}$ to rewrite the integration \eqref{recursive_int} as
\begin{align}
\phi_{\ell+1}(u)=\frac{1}{2\pi}\int_0^\infty dv\frac{v^{\frac{k}{2}}}{(v+\alpha^2u)(v^k+1)}\phi_\ell(v).
\label{int_rec_phitophi}
\end{align}
If we assume that $\phi_\ell(u)$ can be expanded as the following finite series (inductively correct) 
\begin{align}
\phi_\ell(u)=\sum_{j\ge 0}\phi_\ell^{(j)}(u) (\log u)^j,\quad\quad (\phi_\ell^{(j)}(u)\text{ are some rational functions of }u)
\end{align}
we can compute the integration \eqref{int_rec_phitophi} as \cite{PY}
\begin{align}
\phi_{\ell+1}(u)&=\frac{1}{2\pi}\sum_{j\ge 0}\Bigl[-\frac{(2\pi i)^j}{j+1}\int_\gamma\frac{v^{\frac{k}{2}}}{(v+\alpha^2u)(v^k+1)}\phi_\ell^{(j)}(v)B_{j+1}\Bigl(\frac{\log^{(+)}v}{2\pi i}\Bigr)\Bigr]\nonumber \\
&=\frac{1}{2\pi}\sum_{j\ge 0}\biggl[-\frac{(2\pi i)^{j+1}}{j+1}\sum_{w\in poles} \Res\Bigl[\frac{v^{\frac{k}{2}}}{(v+\alpha^2u)(v^k+1)}\phi_\ell^{(j)}(v)B_{j+1}\Bigl(\frac{\log^{(+)}v}{2\pi i}\Bigr),v\rightarrow w\Bigr]\biggr].
\label{TWPYRes}
\end{align}
Here $\log^{(+)}$ is logarithm function with the branch cat located on $\mathbb{R}^+$ and the integration contour $\gamma$ is as depicted in figure \ref{TWPYpath}.
The poles to be collected in the step  $\phi_\ell\rightarrow \phi_{\ell+1}$ are at most
\begin{align}
v&=-\alpha^2 u,\nonumber \\
v&=\alpha^{-2a}e^{\frac{\pi i(2b+1)}{k}},\quad (a=0,1,\cdots,\ell;\,\,b=0,1,\cdots,k-1),
\end{align}
which can be seen from the same argument as in \cite{Nosaka}.

After obtaining $\phi_\ell$ for $\ell=0,1,\cdots,n-1$, we can compute $\Tr{\widehat \rho}^n$ by
\begin{align}
\Tr{\widehat \rho}^n&=\frac{1}{2\pi(\alpha^n-(-1)^n\alpha^{-n})}\int_0^\infty du\frac{u^{\frac{k}{2}-1}}{u^k+1}\Psi_n(u)\quad\quad\Bigl(\Psi_n(u)=\sum_{\ell=0}^{n-1}(-1)^\ell\phi_\ell(u)\psi_{n-1-\ell}(u)\Bigr)\nonumber \\
&=\frac{1}{2\pi(\alpha^n-(-1)^n\alpha^{-n})}\sum_{j\ge 0}\biggl[-\frac{(2\pi i)^{j+1}}{j+1}\sum_{w\in poles}\Res\Bigl[\frac{u^{\frac{k}{2}-1}}{u^k+1}\Psi_n^{(j)}(u)B_{j+1}\Bigl(\frac{\log^{(+)}u}{2\pi i}\Bigr),u\rightarrow w\Bigr]\biggr],
\end{align}
where $\Psi_n=\sum_{j\ge 0}\Psi_n^{(j)}(u)(\log u)^j$ and $poles$ are (at most)
\begin{align}
u=\alpha^{-2a} e^{\frac{\pi i(2b+1)}{k}}.\quad (a=-(n-1),-(n-2),\cdots,n-1;\,\, b=0,1,\cdots,k-1)
\end{align}

\subsubsection*{\underline{Odd $k$}}
For odd $k$, we define $u=e^{\frac{x}{2k}}$ to obtain the following formulas
\begin{align}
\phi_{\ell+1}(u)&=\frac{1}{\pi}\sum_{j\ge 0}\biggl[-\frac{(2\pi i)^{j+1}}{j+1}\sum_{v\in poles}\Res\Bigl[\frac{1}{v^2+\alpha^2 u^2}\frac{v^{k+1}}{v^{2k}+1}\phi_\ell^{(j)}(v)B_{j+1}\Bigl(\frac{\log^{(+)}v}{2\pi i}\Bigr),v\rightarrow w\Bigr]\biggr],
\end{align}
where $\phi_\ell^{(j)}(u)$ are the rational functions given by $\phi_\ell(u)=\sum_{j\ge 0}(\log u)^j\phi_\ell^{(j)}(u)$ and the poles to be collected are
\begin{align}
v&=\pm i\alpha u,\nonumber \\
v&=\alpha^{-2a} e^{\frac{\pi i(2b+1)}{2k}},\quad \Bigl(a=0,1,\cdots,\Bigl[\frac{\ell}{2}\Bigr];\,\,b=0,1,\cdots,2k-1\Bigr)\nonumber \\
v&=\alpha^{-(2a+1)}e^{\frac{\pi ib}{k}}.\quad \Bigl(a=0,1,\cdots,\Bigl[\frac{\ell-1}{2}\Bigr];\,\,b=0,1,\cdots,2k-1\Bigr)
\end{align}
The traces of ${\widehat \rho}^n$ can be computed as
\begin{align}
\Tr{\widehat \rho}^n&=\frac{1}{\pi(\alpha^n-(-1)^n\alpha^{-n})}\sum_{j\ge 0}\biggl[-\frac{(2\pi i)^{j+1}}{j+1}\sum_{w\in poles}\Res\Bigl[\frac{u^{k-1}}{u^{2k}+1}\Psi_n^{(j)}(u)B_{j+1}\Bigl(\frac{\log^{(+)} v}{2\pi i}\Bigr),u\rightarrow w\Bigr]\biggr]
\end{align}
where $\psi(u)=\sum_{\ell=0}^{n-1}(-1)^\ell \phi_\ell(u)\psi_{n-1-\ell}(u)=\sum_{j\ge 0}(\log u)^j\Psi_\ell^{(j)}(u)$ and the poles are
\begin{align}
u&=\alpha^{2a} e^{\frac{\pi i(2b+1)}{2k}},\quad \Bigl(a=-\Bigl[\frac{n-1}{2}\Bigr],-\Bigl[\frac{n-1}{2}\Bigr]+1,\cdots,\Bigl[\frac{n-1}{2}\Bigr];\,\,b=0,1,\cdots, 2k-1\Bigr),\nonumber \\
u&=\alpha^{\pm(2a+1)} e^{\frac{\pi ib}{k}}.\quad \Bigl(a=0,1,\cdots,\Bigl[\frac{n-2}{2}\Bigr];\,\,b=0,1,\cdots, 2k-1\Bigr)\nonumber \\
\end{align}

\section{Exact expressions for $Z(N,k,0,\zeta_2)$}
\label{Z1exact_results}
The technique introduced in sec.~\ref{TWPY} allows us to compute the partition function of the mass deformed ABJM theory $Z(N,k,0,\zeta_2)$ with $\zeta_1=0$ and for small integers $N,k$.
We have computed $Z(N,k,0,\zeta_2)$ for $(k=1 ,N\le 12)$, $(k=2,N\le 9 )$, $(k=3 ,N\le 5)$, $(k=4 ,N\le 5 )$ and $(k=6 ,N\le 4 )$.
Here we display the first few results ($\alpha=e^{2\pi\zeta_2/k}$).
\begin{align}
Z(1,1,0,\zeta_2)&=
\frac{\alpha}{2(1+\alpha^2)},\quad
Z(2,1,0,\zeta_2)=
-\frac{\zeta_2\alpha^3}{(1+\alpha^2)(1-\alpha^4)},\nonumber \\
Z(3,1,0,\zeta_2)&=
\frac{\alpha^5(1+12 \zeta_2 \alpha-\alpha^2)}{8(1+\alpha^2)(1-\alpha^4)(1+\alpha^6)},\quad \cdots
\label{exactk1}
\end{align}
\begin{align}
Z(1,2,0,\zeta_2)&=
\frac{\alpha}{4(1+\alpha^2)},\quad
Z(2,2,0,\zeta_2)=
\frac{\zeta_2^2\alpha^4}
{(1-\alpha^4)^2},\nonumber \\
Z(3,2,0,\zeta_2)&=
\frac{\alpha^7(-1+4 \zeta_2^2+(2+32 \zeta_2^2) \alpha^2+(-1+4 \zeta_2^2) \alpha^4)}{32(1+\alpha^4)(1-\alpha^4)^2(1+\alpha^6)},\quad \cdots
\label{exactk2}
\end{align}
\begin{align}
Z(1,3,0,\zeta_2)&=\frac{\alpha}{6(1+\alpha^2)},\nonumber \\
Z(2,3,0,\zeta_2)&=\frac{\alpha^4(1+(1+4\zeta_2)\alpha+4\zeta_2\alpha^2+(-1+4\zeta_2)\alpha^7\alpha^3-\alpha^4)}{12(1+\alpha)(1+\alpha^2)(1-\alpha^3)(1+\alpha^6)},\nonumber \\
Z(3,3,0,\zeta_2)&=-\frac{\alpha^8}{18\sqrt{3}(1+\alpha^6)^3(1-\alpha^6)}
(2+\sqrt{3}\zeta_2-3\sqrt{3}\alpha+(4-2\sqrt{3}\zeta_2)\alpha^2+3\sqrt{3}\zeta_2\alpha^4\nonumber \\
&\quad +(-4-2\sqrt{3}\zeta_2)\alpha^6+3\sqrt{3}\alpha^7+(-2+\sqrt{3}\zeta_2)\alpha^8)
,\quad \cdots
\label{exactk3}
\end{align}
\begin{align}
Z(1,4,0,\zeta_2)&=\frac{\alpha}{8(1+\alpha^2)},\quad
Z(2,4,0,\zeta_2)=\frac{\alpha^4(1+(-2-8\zeta_2^2)\alpha^2+\alpha^4)}{64(1+\alpha^4)(1-\alpha^4)^2},\nonumber \\
Z(3,4,0,\zeta_2)&=
\frac{\alpha^9}{256(1 - \alpha^2)^2 (1 + \alpha^2)^2 (1 + \alpha^4)^2(1 + \alpha^6) (1 + \alpha^8)}
(5 + 8 \zeta_2 + 4 \zeta_2^2 + (-7 + 8 \zeta_2) \alpha^2\nonumber \\
&\quad + (5 - 8 \zeta_2 + 4 \zeta_2^2) \alpha^4 + (-6 - 32 \zeta_2^2) \alpha^6 + (5 + 8 \zeta_2 + 4 \zeta_2^2) \alpha^8 + (-7 - 8 \zeta_2) \alpha^{10}\nonumber \\
&\quad + (5 - 8 \zeta_2 + 4 \zeta_2^2) \alpha^{12}),\quad \cdots
\label{exactk4}
\end{align}
\begin{align}
Z(1,6,0,\zeta_2)&=\frac{\alpha}{12(1+\alpha^2)},\quad
Z(2,6,0,\zeta_2)=\frac{\alpha^4(1-9\alpha^2+8(2+3\zeta_2^2)\alpha^4-9\alpha^6+\alpha^8)}{432(1-\alpha^4)(1-\alpha^{12})},\nonumber \\
Z(3,6,0,\zeta_2)&=\frac{\alpha^9}{5184 (1 + \alpha^2)^2 (-1 + \alpha^6)^2 (1 + \alpha^6) (1 + \alpha^{12})}
(1 + (-54 - 32 \sqrt{3} \zeta_2 - 24 \zeta_2^2) \alpha^2\nonumber \\
&\quad  + (-15 - 64 \sqrt{3} \zeta_2) \alpha^4 + (30 - 32 \sqrt{3} \zeta_2 + 96 \zeta_2^2) \alpha^6 + (76 + 192 \zeta_2^2) \alpha^8\nonumber \\
&\quad  + (30 + 32 \sqrt{3} \zeta_2 + 96 \zeta_2^2) \alpha^{10} + (-15 + 64 \sqrt{3} \zeta_2) \alpha^{12} + (-54 + 32 \sqrt{3} \zeta_2 - 24 \zeta_2^2) \alpha^{14}\nonumber \\
&\quad  + \alpha^{16}),\quad\cdots.
\label{exactk6}
\end{align}

\subsection{Comparison with saddle point approximation}
\begin{figure}
\begin{center}
\includegraphics[width=12cm]{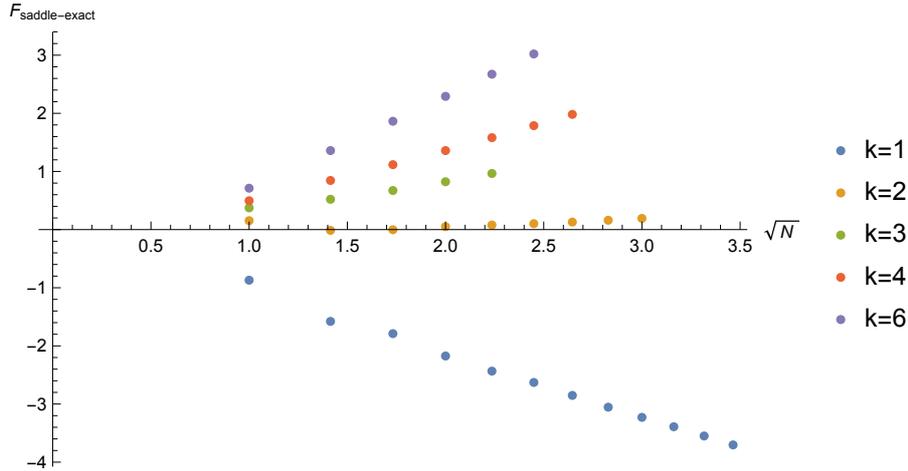}
\end{center}
\caption{
Plot of $F_{\text{saddle}-\text{exact}}$ \eqref{Fsaddle-exact} for $k=1,2,3,4,6$, $\zeta_2=1$.
}
\label{Zexact_vs_saddle}
\end{figure}

Let us compare the exact partition function \eqref{exactk1}-\eqref{exactk6} with the result of the saddle point approximation \eqref{Zinsaddle}.
In figure \ref{Zexact_vs_saddle} we plot the difference between two results
\begin{align}
F_{\text{saddle}-\text{exact}}=\frac{\pi\sqrt{2k}}{3}\sqrt{1+\frac{16\zeta_2^2}{k^2}}N^{\frac{3}{2}}-(-\log Z(N,k,0,\zeta_2))
\label{Fsaddle-exact}
\end{align}
for $k=1,2,3,4,6$ and $\zeta_2=1$.
The plot indicates $F_{\text{saddle}-\text{exact}}\sim \sqrt{N}$ for large-$N$, hence the leading part of the two results ($\sim N^{3/2}$) agree with each other.

\begin{figure}
\begin{center}
\includegraphics[width=8cm]{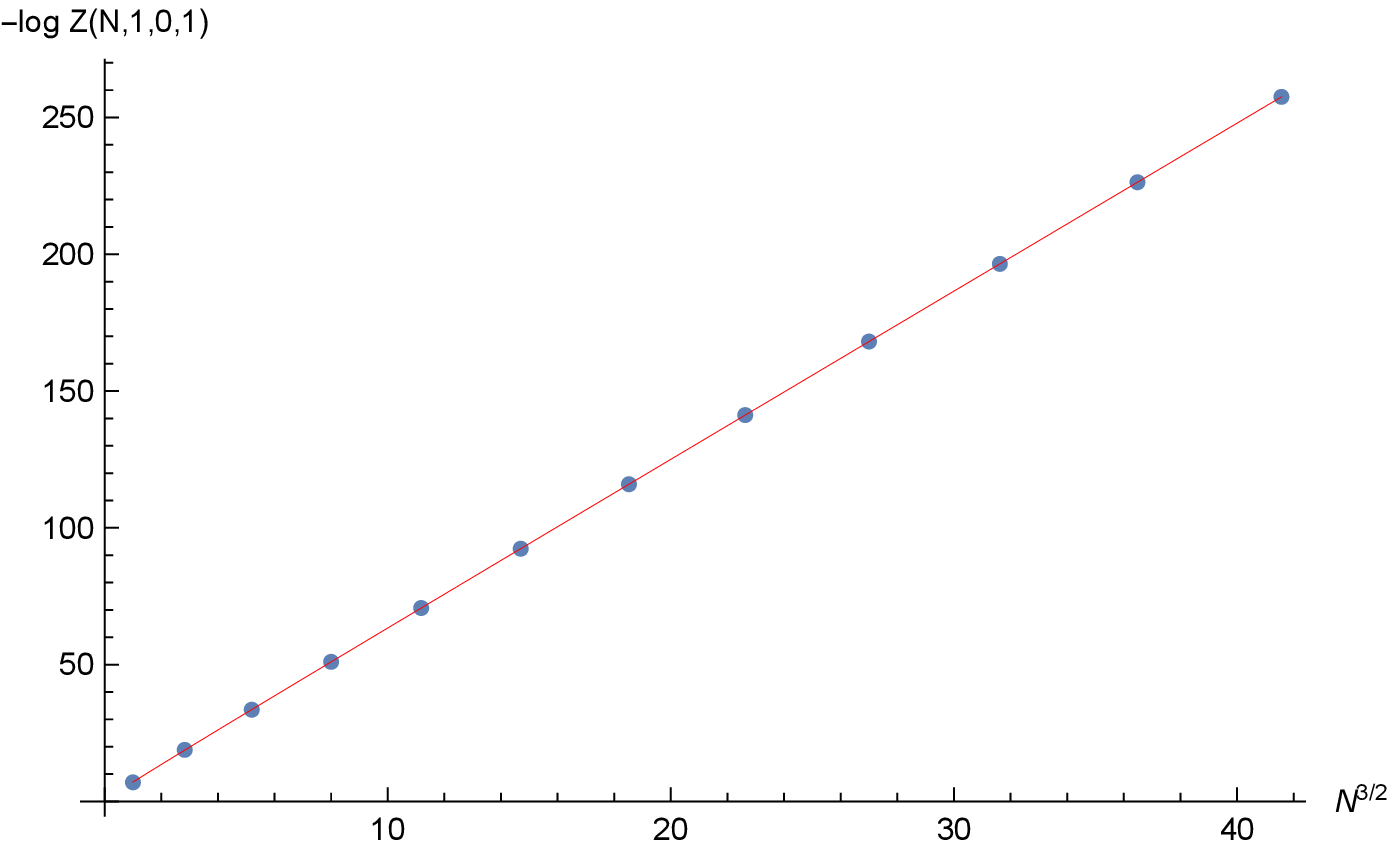}
\includegraphics[width=8cm]{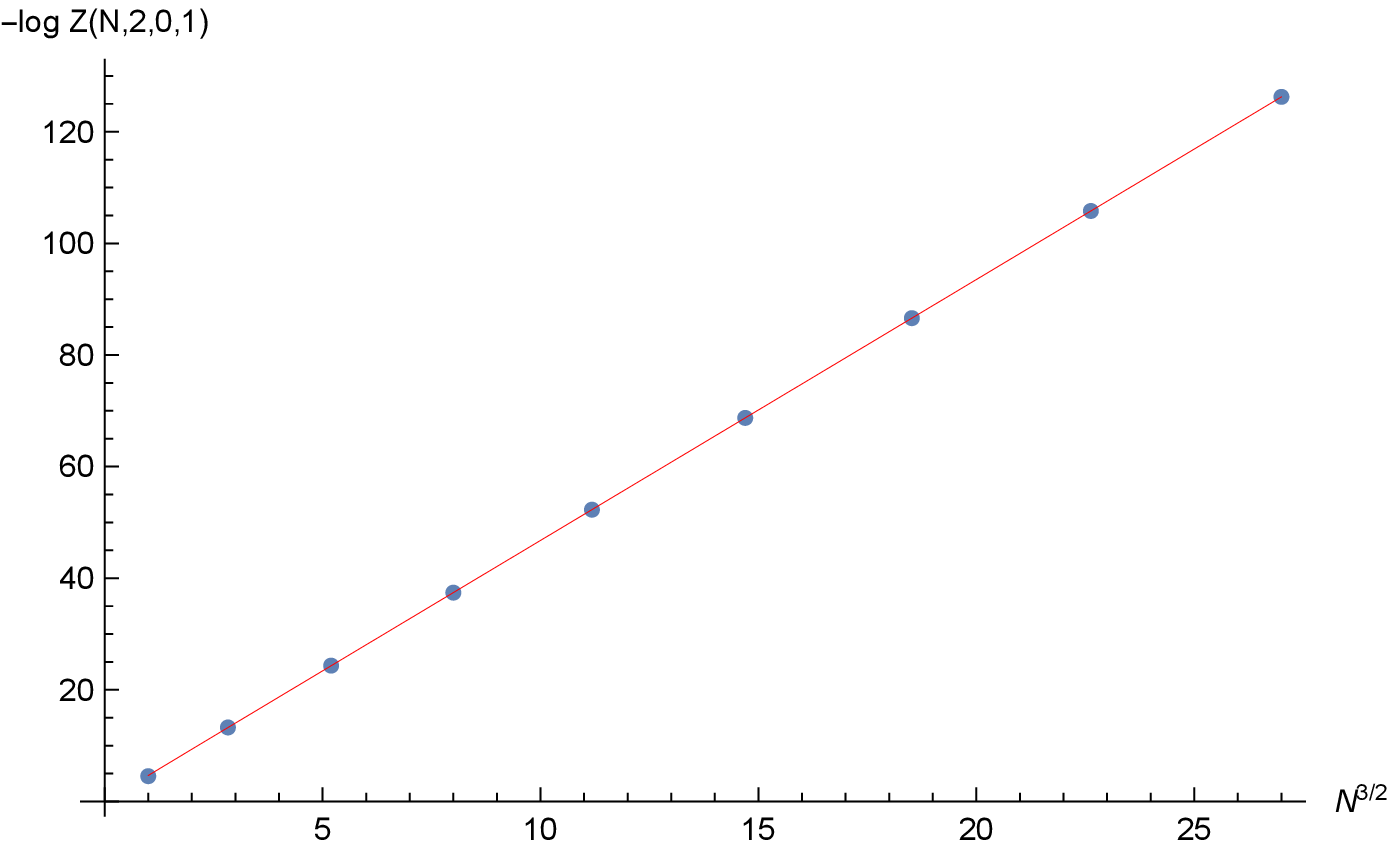}\\
\includegraphics[width=8cm]{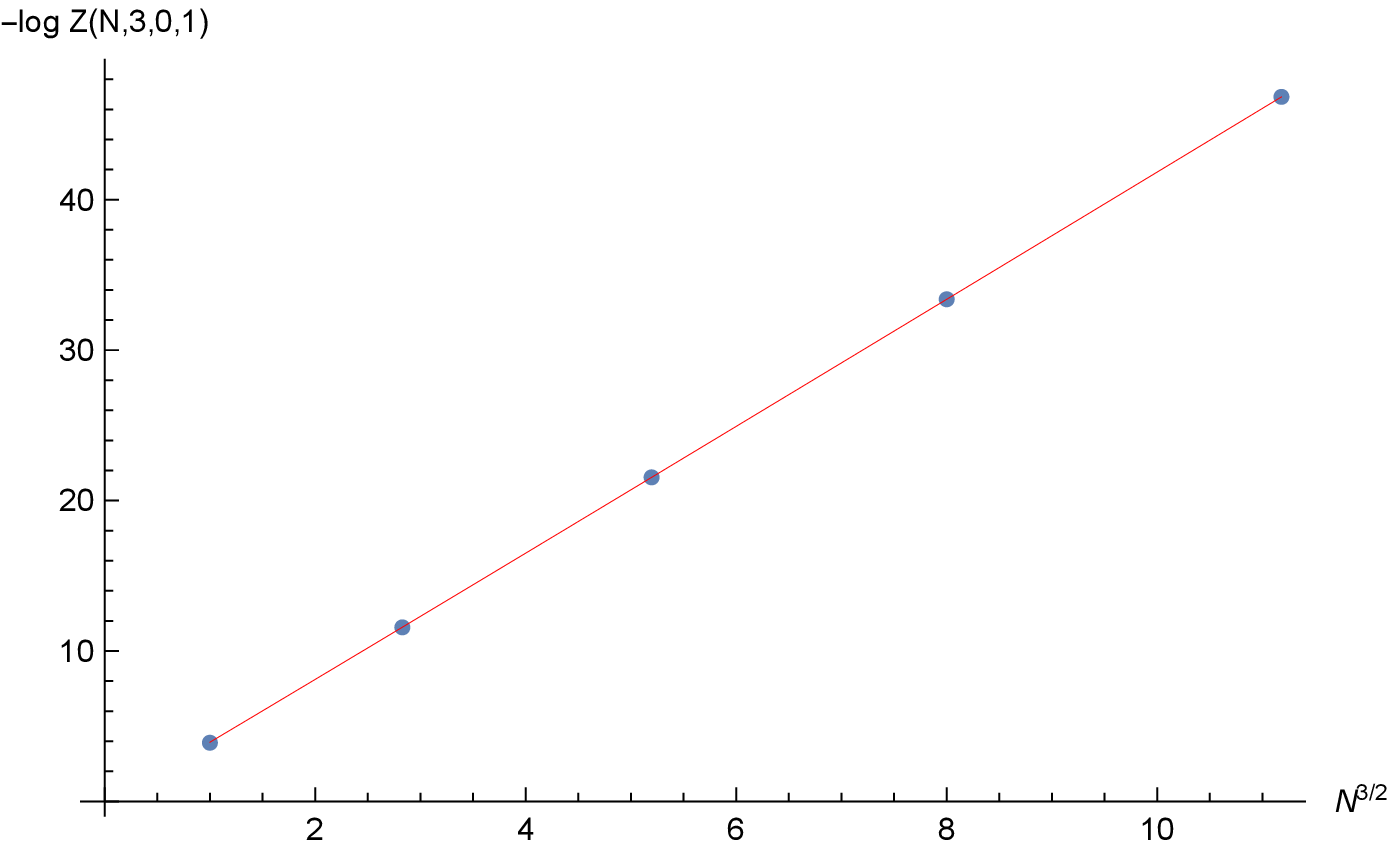}
\includegraphics[width=8cm]{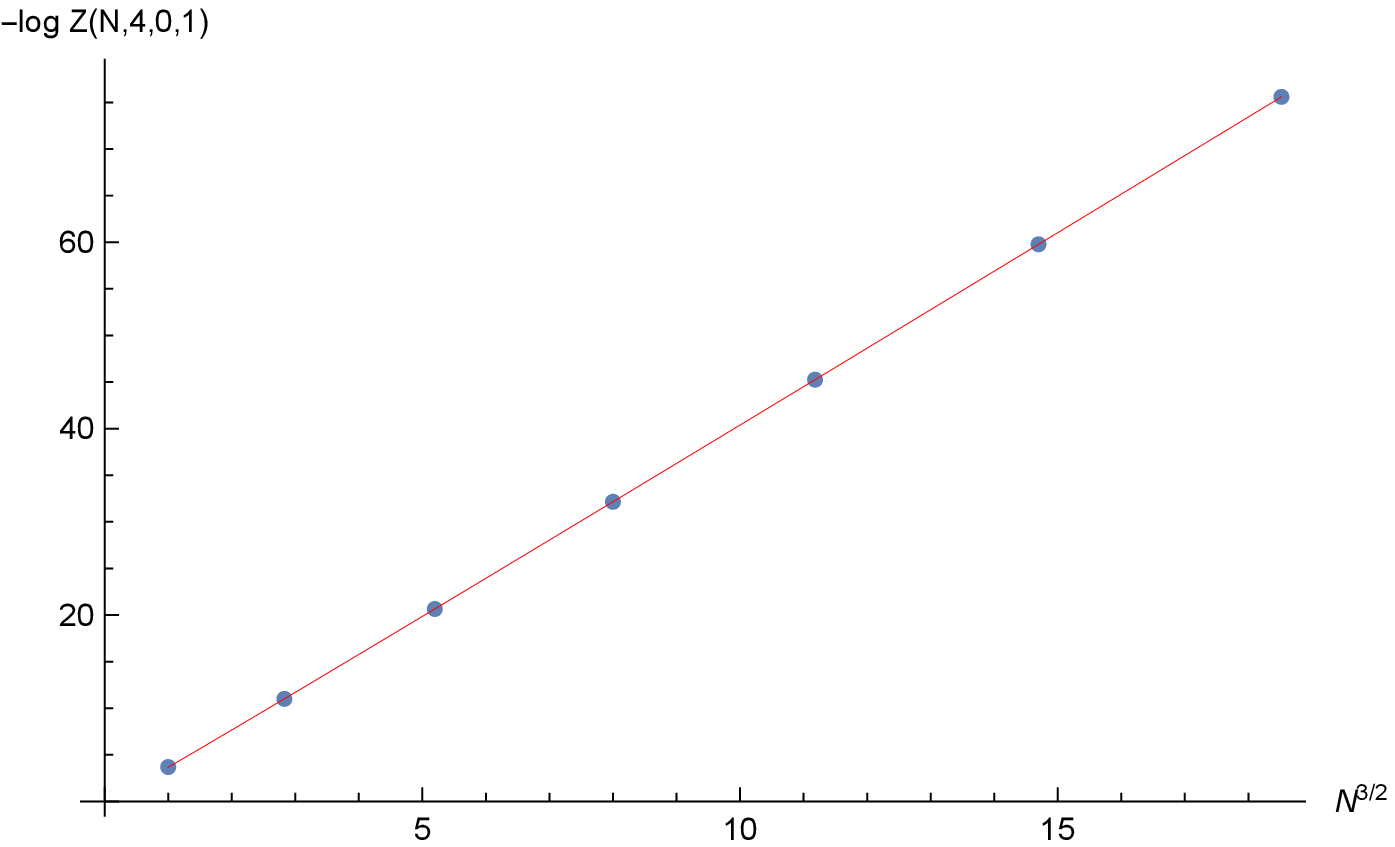}\\
\includegraphics[width=8cm]{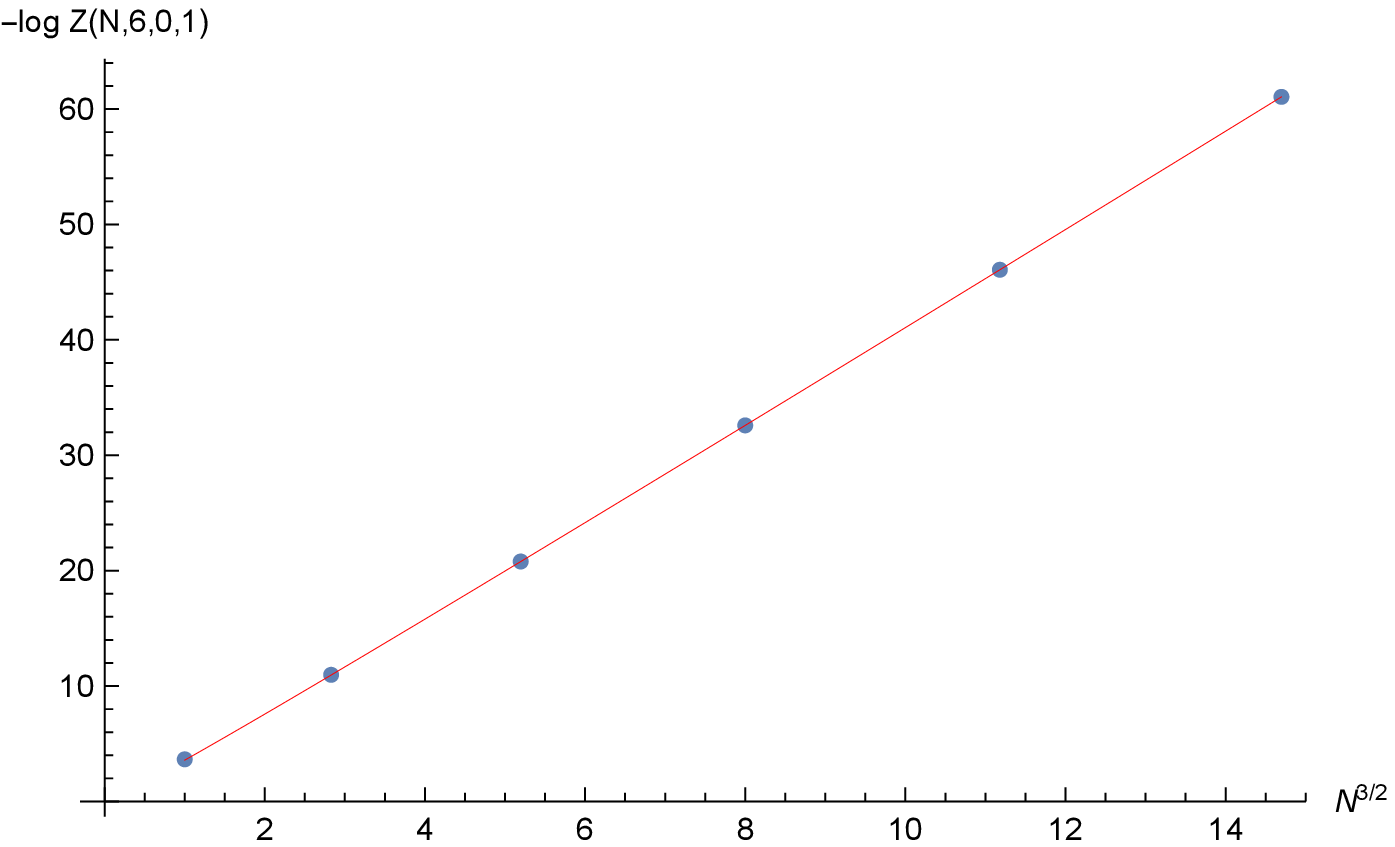}
\end{center}
\caption{
Blue points: exact values of $-\log Z(N,k,0,\zeta_2)$ \eqref{exactk1}-\eqref{exactk6}; Red line: $-\log Z_\text{pert}(N)$ \eqref{Airy}.
}
\label{Zexact_vs_Airy}
\end{figure}

We can make a more refined comparison between the exact results and large-$N$ expansion as follows.
First we notice that the saddle point approximation \eqref{Zinsaddle} agree with the following expression in the large-$N$ limit
\begin{align}
Z_\text{pert}=e^AC^{-\frac{1}{3}}\Ai[C^{-\frac{1}{3}}(N-B)]
\label{Airy}
\end{align}
where
\begin{align}
C&=\frac{2}{k\pi^2(1+\frac{16\zeta_2^2}{k^2})},\quad
B=\frac{k}{24}-\frac{1}{6k}+\frac{1}{2k(1+\frac{16\zeta_2^2}{k^2})},\nonumber \\
A&=\frac{2A_\text{ABJM}(k)
+A_\text{ABJM}(k+4i\zeta_2)
+A_\text{ABJM}(k-4i\zeta_2)
}{4} ,
\end{align}
which is obtained from the partition function of the ABJM theory with R-charge deformation by ignoring the large-$N$ non-perturbative effects ($e^{-\sqrt{N}}$) and replacing the real deformation parameters $\xi,\eta$ formally as $\xi\rightarrow 0$, $\eta\rightarrow 4i\zeta_2/k$ (see eq(1.4) in \cite{Nosaka}).
By comparing the numerical values of \eqref{exactk1}-\eqref{exactk6} and \eqref{Airy} we find good agreement.
As an example, in figure \ref{Zexact_vs_Airy} we display the comparison of the free energy for $\zeta_2=1$.

\subsection{Decoupling limit $\zeta_2\rightarrow\infty$}
\label{comarewithNY}
To compare the mass deformed ABJM theory in the decoupling limit with the $\text{U}(N)_k\times \text{U}(N)_{-k}$ linear quiver superconformal Chern-Simons theory (Gaiotto-Witten theory), it is reasonable to divide the partition function $Z(N,k,0,\zeta_2)$ by $e^{2\pi N^2\zeta_2/k}$, the naive contribution from the massive hypermultiplet.
For $(N,k)=(1,1)$, $(1,2)$, $(1,3)$, $(2,3)$, $(1,4)$, $(2,4)$, $(1,6)$, $(2,6)$, $(3,6)$ the result of $\lim_{\zeta_2\rightarrow \infty} Z(N,k,0,\zeta_2)/e^{-2\pi N^2\zeta_2/k}$ is finite and coincide with the partition function of the Gaiotto-Witten theory with the same $(N,k)$ obtained in \cite{NY1}.

For the other $(N,k)$ we have found the following asymptotic behavior
\begin{align}
Z(N,k,0,\zeta_2)\rightarrow f_{N,k}(\zeta_2)e^{-\frac{2\pi p_{N,k}\zeta_2}{k}},
\end{align}
where $p_{N,k}$ are some integers and $f_{N,k}(\zeta_2)$ are some polynomials of $\zeta_2$, which are listed in the following tables.
\begin{align}
&\begin{tabular}{|c|c|c|}
\multicolumn{3}{l}{$k=1$}\\ \hline
$N$&$p_{N,k}$&$f_{N,k}(\zeta_2)$\\ \hline
 1& 1&$\frac{1}{2}$\\ \hline
 2& 3&$\zeta_2$\\ \hline
 3& 5&$\frac{1}{8}$\\ \hline
 4& 8&$-\frac{1}{32}+\frac{\zeta_2^2}{2}$\\ \hline
 5&11&$-\frac{3}{64}+\frac{\zeta_2^2}{4}$\\ \hline
 6&14&$\frac{1}{64}$\\ \hline
 7&18&$-\frac{5\zeta_2}{192}+\frac{\zeta_2^3}{12}$\\ \hline
 8&22&$\frac{9}{1024}-\frac{\zeta_2^2}{24}+\frac{\zeta_2^4}{12}$\\ \hline
 9&26&$-\frac{11\zeta_2}{768}+\frac{\zeta_2^3}{48}$\\ \hline
10&30&$\frac{1}{1024}$\\ \hline
11&35&$\frac{3}{16384}-\frac{7\zeta_2^2}{1536}+\frac{\zeta_2^4}{192}$\\ \hline
12&40&$-\frac{45}{65536}+\frac{151\zeta_2^2}{36864}-\frac{19\zeta_2^4}{2304}+\frac{\zeta_2^6}{144}$\\ \hline
\end{tabular}
\quad
\begin{tabular}{|c|c|c|}
\multicolumn{3}{l}{$k=2$}\\ \hline
$N$&$p_{N,k}$&$f_{N,k}(\zeta_2)$\\ \hline
 1& 1&$\frac{1}{4}$\\ \hline
 2& 4&$\frac{\zeta_2^2}{2}$\\ \hline
 3& 7&$-\frac{1}{32}+\frac{\zeta_2^2}{8}$\\ \hline
 4&10&$\frac{1}{256}$\\ \hline
 5&15&$-\frac{\zeta_2^2}{192}+\frac{\zeta_2^4}{96}$\\ \hline
 6&20&$-\frac{1}{1024}+\frac{17\zeta_2^2}{4608}-\frac{\zeta_2^4}{144}+\frac{\zeta_2^6}{144}$\\ \hline
 7&25&$\frac{19\zeta_2^2}{9216}-\frac{7\zeta_2^4}{2304}+\frac{\zeta_2^6}{576}$\\ \hline
 8&30&$\frac{3}{65536}-\frac{\zeta_2^2}{3072}+\frac{\zeta_2^4}{6144}$\\ \hline
 9&35&$\frac{1}{262144}$\\ \hline
\end{tabular}\nonumber \\
&\begin{tabular}{|c|c|c|}
\multicolumn{3}{l}{$k=3$}\\ \hline
$N$&$p_{N,k}$&$f_{N,k}(\zeta_2)$\\ \hline
 1& 1&$\frac{1}{6}$\\ \hline
 2& 4&$\frac{1}{12}$\\ \hline
 3& 8&$-\frac{1}{9\sqrt{3}}+\frac{\zeta_2}{18}$\\ \hline
 4& 12&$\frac{1}{432}$\\ \hline
 5& 17&$\frac{1}{2592}$\\ \hline
\end{tabular}
\quad
\begin{tabular}{|c|c|c|}
\multicolumn{3}{l}{$k=4$}\\ \hline
$N$&$p_{N,k}$&$f_{N,k}(\zeta_2)$\\ \hline
 1& 1&$\frac{1}{8}$\\ \hline
 2& 4&$\frac{1}{64}$\\ \hline
 3& 9&$\frac{5}{256}-\frac{\zeta_2}{32}+\frac{\zeta_2^2}{64}$\\ \hline
 4&14&$\frac{1}{1024}-\frac{\zeta_2}{256}+\frac{\zeta_2^2}{512}$\\ \hline
 5&19&$\frac{1}{32768}$\\ \hline
\end{tabular}
\quad
\begin{tabular}{|c|c|c|}
\multicolumn{3}{l}{$k=6$}\\ \hline
$N$&$p_{N,k}$&$f_{N,k}(\zeta_2)$\\ \hline
 1& 1&$\frac{1}{12}$\\ \hline
 2& 4&$\frac{1}{432}$\\ \hline
 3& 9&$\frac{1}{5184}$\\ \hline
 4&16&$\frac{1}{1296}-\frac{\zeta_2}{972\sqrt{3}}+\frac{\zeta_2^2}{7776}$\\ \hline
\end{tabular}
\end{align}

\end{document}